\documentclass[twocolumn,aps,prx,floatfix,superscriptaddress]{revtex4-2}

\usepackage{amsmath,amssymb,amsfonts,graphicx,braket,hyperref}
\hypersetup{colorlinks=true,citecolor=magenta,linkcolor=red}

\graphicspath{{figs/},{figs_SD/},{figs_stability/},{figs_LMC/},{figs_partfunc/},{figs_optman_param/},{./}}

\usepackage{graphicx}
\usepackage[caption=false]{subfig}
\usepackage{physics}
\usepackage{bbold}
\usepackage[normalem]{ulem}
\usepackage{xcolor}

\begin{document}

\title{Toward a Theory of Phase Transitions in Quantum Control Landscapes}

\author{Nicol\`o Beato}
\email{nbeato@pks.mpg.de}
\affiliation{Max Planck Institute for the Physics of Complex Systems, N\"othnitzer Stra{\ss}e 38, 01187 Dresden, Germany}

\author{Pranay Patil}
\email{pranay.patil@iitm.ac.in}
\affiliation{Department of Physics, Indian Institute of Technology Madras, Chennai 600036, India}
\affiliation{Theory of Quantum Matter Unit, Okinawa Institute of Science and
Technology Graduate University, Onna-son, Okinawa 904-0412, Japan}

\author{Marin Bukov}
\email{mgbukov@pks.mpg.de}
\affiliation{Max Planck Institute for the Physics of Complex Systems, N\"othnitzer Stra{\ss}e 38, 01187 Dresden, Germany}

\date{\today}
\begin{abstract}

Control landscape phase transitions (CLPTs) occur as abrupt changes in the cost function landscape upon varying a control parameter, and can be revealed by non-analytic points in statistical order parameters. 
A prime example are quantum speed limits (QSL) which mark the onset of controllability as the protocol duration is increased.
Here we lay the foundations of an analytical theory for CLPTs by developing Dyson, Magnus, and cumulant expansions for the cost function that capture the behavior of CLPTs with a controlled precision. Using linear and quadratic stability analysis, we reveal that CLPTs can be associated with different types of instabilities of the optimal protocol.
This allows us to explicitly relate CLPTs to critical structural rearrangements in the extrema of the control landscape: utilizing path integral methods from statistical field theory, we trace back the critical scaling of the order parameter at the QSL to the topological and geometric properties of the set of optimal protocols, such as the number of connected components and its dimensionality.
We verify our predictions by introducing a numerical sampling algorithm designed to explore this optimal set via a homotopic stochastic update rule. 
We apply this new toolbox explicitly to analyze CLPTs in the single- and two-qubit control problems whose landscapes are analytically tractable, and compare the landscapes for bang-bang and piecewise continuous protocols. 
Our work provides the first steps towards a systematic theory of CLPTs and paves the way for utilizing statistical field theory techniques for generic complex control landscapes. 
\end{abstract}

\maketitle

\section{Introduction} \label{sec:intro}

Ever since quantum systems have become an object of investigation, the question of how to manipulate them optimally has been central to both verifying and understanding the theory behind quantum mechanics, as well as to developing new experimental techniques.
Nowadays, control theory has become an indispensable tool in emerging quantum technologies \cite{dalessandro2021introduction,rabitz2004quantum,boscain2021introduction,ansel2024}: efficient quantum state manipulation is an essential prerequisite 
for the experimental realization of simulations on modern intermediate-noise-scale quantum devices
\cite{koch2022quantum,glaser2015training,acin2017roadmap},
for the improvement of performance in quantum measurement devices  
\cite{ungar2024,li2024,wang2022,rembold2020,degen2017}
or quantum information processing \cite{gough2013quantum,gertler2021protecting,zhou2020optimal,khodjasteh2009dynamical,ahn2002continuous}.
Quantum control theory aims to find protocols that optimize the values of physical quantities of interest: e.g., in state preparation or gate synthesis problems, one is typically interested in minimizing infidelity or expectation values of a certain physical observable such as the energy or the total magnetization~\cite{altafini2012modeling,brif2010control}.

\begin{figure}[t!]
    \centering
    \includegraphics[width=.9\linewidth]{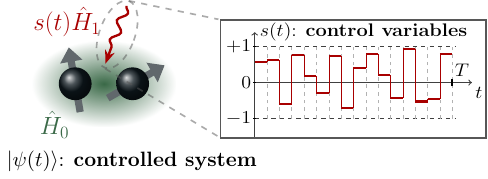}
    \caption{
        In quantum control problems, there are two types of physical degrees of freedom: the \textit{controlled system} (e.g., qubits) and the \textit{control variables} defining the control protocol (e.g., modulation of electric or magnetic fields, or electromagnetic radiation, external forces, etc.). 
        The figure of merit that measures the quality of control protocols defines the control landscape. 
        Control phase transitions refer to critical changes in the structure of the control landscape, as a model parameter (e.g., the protocol duration $T$) is varied. Note that the control landscape can present a many-body problem in the control variables, even when the controlled system consists of as few as a single degree of freedom, see Sec.~\ref{sec:model}.
    }
    \label{fig:qcp}
\end{figure}

In quantum control problems, one distinguishes between two types of physical degrees of freedom: the \emph{controlled system} and the \emph{control variables} (see Fig.~\ref{fig:qcp}).
The controlled system is the quantum system whose state we aim to control.
The control variables correspond to the driving field applied to the controlled system. Time-dependent control protocols have an extensive number of degrees of freedom -- one for each time step when time is discretized, or a single continuous control \textit{field} (in the sense of field theory) composed of infinitely many degrees of freedom.
Consequently, the problem of finding optimal control protocols poses the formidable challenge of searching through high- or infinite-dimensional spaces.

The difficulty in finding analytical solutions arises from the lack of explicit expressions for the solution of the time-dependent Schr\"odinger equation, even for the simplest of quantum systems \cite{boscain2021introduction,ansel2024,koch2022quantum}. This motivated the development of an array of numerical optimal control algorithms. Among them, two classes can be distinguished, namely gradient-based methods (e.g., GRAPE \cite{khaneja2005optimal} and Krotov~\cite{reich2012monotonically}) and gradient-free algorithms (e.g., CRAB \cite{caneva2011chopped,doria2011optimal}). The former class uses the differentiability of the cost function; the latter relies on a clever sampling of the search space and identifies an optimal protocol through direct search methods \cite{mahesh2023quantum,machnes2011comparing}.

The optimization procedure depends strongly on the structure of the underlying quantum control landscape; the latter is defined by an objective that measures the performance of solving the optimization task, and is a functional of the protocol configuration~\cite{chakrabarti2007quantum}. Structural properties of the landscape, such as the number of local minima or saddle-points, depend on both the controlled quantum system and the control variables \cite{ge2022optimization,ge2021}. 
{For instance, the landscape associated with ground state preparation in a chain of qubits exhibits an exponential number of almost-optimal control protocols as local landscape minima, and a connection between quantum control landscapes and glassy systems was established~\cite{Bukov19_Glassy}. 
However, despite the large body of previous work in quantum control, surprisingly little is known about the properties of control landscapes associated with experimentally relevant quantum control problems \cite{van2016optimal,bason2012high,dalgaard2022,matos2021quantifying,caneva2014complexity,ge2022optimization}.
}

The bridge between quantum control landscapes and the physics of glassy-systems becomes apparent once the cost function (defining the landscape) is interpreted as an effective energy function, with control variables playing the role of classical degrees of freedom \cite{hernandezgomez2023optimal,solon2018phase,folena2023introduction,Nishimori01}: from this point of view, optimal protocols correspond to ground states of the effective energy function.
In particular, once the cost function is expanded in terms of the control variables, the $n$th-order expansion coefficient specifies the strength of the $n$-body effective interaction among control degrees of freedom.
So far, the calculation of different terms in the expansion has required the (brute-force) exhaustive exploration of the entire control space \cite{Bukov19_Glassy}, a procedure computationally feasible only for a small number of control parameters. Nonetheless, this approach captures two important features of these effective interactions: they are long-range and multi-body, directly reflecting the non-locality and nonlinearity of the underlying quantum control problem. 

Curiously, as the protocol duration is increased, quantum control landscapes have been shown to exhibit critical structural changes, called ``Control Landscape Phase Transitions" (CLPTs)~\cite{Bukov18_Reinforcement,Bukov18_Broken,Bukov19_Glassy, beato2025_topological}, both in few- and many-body controlled systems.
The name derives from sharp non-analytic behavior arising at the transition point in an order parameter that quantifies the correlations between minima in the optimization landscape. In light of the connection mentioned above between optimization problems and classical statistical physics, CLPTs are naturally associated with phase transitions occurring in the classical effective model. Moreover, since CLPTs are related to structural changes in the optimization landscape, they can affect the complexity of the underlying control problem.
Therefore, an accurate analysis of CLPTs is an important step towards developing a better understanding of quantum control landscapes and shall provide new insights for improving existing optimization algorithms.
Despite the clear numerical evidence for their existence, no analytical characterization of CLPTs has hitherto been developed. 

In this work, we combine analytical and numerical methods to investigate different CLPTs occurring in prototypical quantum control setups.

First, we propose different controlled approximations for the quantum control landscape, and analyze to what extent they reproduce the CLPTs present in the exact landscape. 
In particular, the central object is the evolution operator, which we expand using different analytical expansions (Dyson, Magnus, and cumulants); the cost function landscape (i.e., the infidelity in the case of state-preparation or a physical observable in the case of observable optimization) is then obtained via projection onto the initial and target quantum states. 
Remarkably, the analytical expansions decouple the controlled quantum degrees of freedom from the control fields; hence, the framework remains general and valid for arbitrary quantum systems. From this result, we conjecture that CLPTs are ubiquitous in optimal control and can arise in both quantum and classical dynamical systems.
Finally, our method treats piecewise continuous and bang-bang protocols on equal footing: the coefficients in the landscape expansions do not depend on the family of control protocols, but the landscapes do.

Second, we use the analytical expansion of the landscape to characterize CLPTs in single- and two-qubit setups. By working in a continuous protocol space, we make a precise connection between non-analytic behaviors of order parameters and structural rearrangements in the control landscape.
In particular, we keep track of optimal protocols while increasing {adiabatically} the duration of the quantum evolution; this reveals CLPTs to be associated with different kinds of instabilities of the optimal protocols present in the landscape. The method is independent of the controlled quantum system and provides additional means to detect and characterize CLPTs.

Third, the landscape expansions allow us to estimate analytically the critical scaling behavior of order parameters at the quantum speed limit phase transition. To this end, we first frame the problem in the context of statistical field theory by defining an appropriate partition function. Then, we evaluate the partition function, defined using a path integral, within a Gaussian approximation.

Our work is a first attempt to apply methods from statistical field theory, designed to investigate generic complex optimization landscapes \cite{ros2023high,fyodorov2022,gamarnik2022disordered,urbani2023,zdeborova2016statistical,hu2012stability,morampudi2017clustering,krzakala2007landscape,mezard2005landscape,monasson1997statistical,mezard2002analytic,mezard1987spin,Nishimori01}, to the study of quantum control landscapes \cite{koch2022quantum,ge2022optimization,ge2021}. 
We directly address the many-body character of the time-dependent \emph{control fields}.
Although we focus on one- and two-body models, our methods are formally independent of the properties of the controlled quantum system, and we discuss potential applications to many-body controlled systems. 
For this reason, the framework we present provides a starting point for analyzing more complex systems.

This paper is organized as follows.
In Sec.~\ref{sec:model} we introduce the control problems we later use to benchmark our theory and discuss the corresponding control phase diagrams.
In Sec.~\ref{sec:expansions} we derive the analytical expansions for the quantum control landscape and discuss their interpretation from the point of view of classical spin-models in statistical physics. 
In Sec.~\ref{sec:stability} we discuss how CLPTs are connected to structural changes in the optimal level set, \text{i.e.} the set of control protocols that minimize the cost function. In particular, using linear and quadratic stability analysis we detect and classify different CLPTs.
In Sec.~\ref{sec:LMC} we introduce a sampling algorithm based on the Metropolis-adjusted Langevin dynamics, revisit the QSL phase transition for piecewise continuous protocols, and study the properties of the optimal level set beyond the QSL. 
In Sec.~\ref{sec:Tqslcritical} we estimate the critical scaling of the order parameters at the QSL phase transition, in both cases of continuous and bang-bang protocols.
Finally, in Sec.~\ref{sec:manybody}, we comment on the challenges in extending the methods we develop to controlled many-body quantum systems.
We summarize the main results and discuss their physical significance in Sec.~\ref{sec:outro}.


\section{Toy models for quantum control landscape phase transitions (CLPTs)} 
\label{sec:model}

Consider the generic control problem 
\begin{equation}
    \label{eq:H(t)}
    \hat H(t) = \hat H_0 + s(t) \hat H_1,
\end{equation}
where the Hamiltonian is divided between a constant ``drift" term $\hat H_0$ and a time-varying ``control" term $s(t)\hat H_1$. The control parameter $s$, regarded as a function of time $t$, is called control ``protocol" (or ``schedule"). In the qubit systems we consider below, the protocol modulates the external magnetic field in time, but more generic controls are also frequently used in experiments.
To stay away from the trivial adiabatic limit, we consider a fixed protocol duration $T$ of the order of the inverse energy scale in the drift term, \textit{i.e.}, $T\sim ||{\hat{H}_0}^{-1}||$, and track evolution times $t\in [0,T]$. Consistently with typical limitations in present-day experiments \cite{Bukov19_Geometric}, we assume a bounded control term $\hat H_1$ and restrict the problem to the class of bounded functions $\abs{s(t)}{\le}1,\,{\forall} t {\in} [0,T]$.

We are interested in CLPTs emerging in state-preparation problems. Let us denote the initial state of the system by $\ket{\psi_0}$, and the target state (\textit{i.e.}, the state we want to prepare) -- by $\ket{\psi_\ast}$. Notice that the control term $s(t)\hat H_1$ modifies the bare drift evolution of the system. The optimal control problem then consists in finding those protocols $s_\ast(t)$ that achieve a maximum overlap between the target $\ket{\psi_\ast}$ and the time-evolved state $\ket{\psi(t)}{=}\hat U_{s}(t,0)\ket{\psi_0}$ at the final time $t{=}T$; in other words, given $T$ we want to find (all) the protocols $s(t)$ that minimize the so-called ``infidelity",
\begin{equation}
\label{eq:infid-def}
    I(T)[s] = 1-\abs{\bra{\psi_*} \hat U_s(T,0) \ket{\psi_0}}^2,
\end{equation}
where $\hat U_s(T,0) {=} \mathcal T \mathrm e^{-i \int_0^T \dd t \hat H(t)}$ is the time-ordered evolution operator, dependent on the time-varying protocol $s$.

The infidelity defines a functional over protocol space (\textit{i.e.}, the space of all control functions), $s\mapsto I(T)[s]$, henceforth referred to as the ``quantum control landscape"; the latter depends both on the specifics of the Hamiltonian $\hat H(t)$ and the details of the controlled system. Each point $s$ of the infidelity landscape corresponds to a unique protocol $s(t)$; optimal protocols by definition reside in the (possibly many) global infidelity minima.  
By introducing both analytical and numerical techniques, we will analyze changes in the structure of minima in the control landscape in few-qubit systems, as we vary the protocol duration $T$. 

The quantum state preparation problem associated with few-qubit Hamiltonians of the form~\eqref{eq:H(t)} was first studied in Refs.~\cite{Bukov18_Broken} and \cite{Bukov19_Glassy}. In particular, the search for optimal protocols was restricted to the so-called `bang-bang' class, where $|s(t)|$ takes its maximum allowed value. This restriction is formally motivated within the framework of Pontryagin's Maximum Principle which, under certain conditions, guarantees the existence of at least one optimal protocol within the bang-bang class \cite{SchaettlerLedzewicz12}; that said, some control problems have optimal protocols mixing bang-bang and continuous parts~\cite{baldwin2021optimal}. Concretely, the time axis $[0,T]$ is divided in $N$ equidistant time steps $\{t_1,\dots,t_N\}$  and hence $s(t_i){=}s_i{=}{\pm}1$ at each time step $t_i$. Optimal control algorithms, such as Stochastic Descent, can then be used to collect a set of $M$ (locally) optimal protocols, $\mathcal S {=} \{s^{(1)},\dots,s^{(M)}\}$ as a probe for the structure of minima in the quantum control landscape. 

The piecewise-constant bang-bang protocol $s{=}\{s_i\}$ can be regarded as classical Ising spin degrees of freedom while the infidelity landscape, $I(T)[s]$, defines an effective Ising energy function. The corresponding spin model
\begin{equation}
\label{eq:I_s}
    I(T)[s] {=} c(T) {+} \frac1N\sum_{j=1}^N b_j(T) s_j {+} \frac{1}{2N^2} \sum_{i,j}^N J_{ij}(T) s_i s_j {+} \cdots
\end{equation}
can exhibit CLPTs, numerically investigated in Ref.~\cite{Bukov18_Reinforcement}. Note that, despite the discretization of time, there is no underlying regular lattice structure, \textit{i.e.},~the control landscape is effectively an infinite-dimensional system (although discretized time induces a natural ordering of the lattice sites).

CLPTs can be detected using the order parameter
\begin{equation}
\label{eq:q}
    q(T) = \frac1N \sum_{i=1}^N \qty[\expval{s_i^2}_{\mathcal S} - \expval{s_i}^2_{\mathcal S}], 
\end{equation}
which is closely related to the Edwards-Anderson order parameter used to measure spin-glass order \cite{Nishimori01,mezard1987spin}. Here, $\expval{s_i}_{\mathcal S},\expval{s_i^2}_{\mathcal S}$ represent the average on site $i$ over the set of (locally) optimal protocols $\mathcal S$. By definition, $q(T)$ quantifies fluctuations among the set $\mathcal S$ with respect to the average protocol $\expval{s}_{\mathcal S}$. 
Within the bang-bang family, we can further simplify $q(T)$ to
\begin{equation}
\label{eq:q_BB}
    q_\text{BB}(T) = 1 - \frac1N\sum_{i=1}^N \expval{s_i}^2_{\mathcal S}. 
\end{equation}
For convex landscapes, the set $\mathcal S$ contains a single protocol, and we have $q_\text{BB}(T){\equiv} 0$. On the other hand, whenever protocols in $\mathcal S$ are uncorrelated, we have $\expval{s_i}^2_{\mathcal S}{=}0$, and $q_\text{BB}(T){\equiv}1$. Intermediate values of $q(T)$ correspond to correlations between protocols in $\mathcal S$.

Little is known about the nature and origin of control phase transitions in general. Despite their formal resemblance to $k$-SAT problems and spin glasses \cite{Nishimori01}, their analytical study is hampered by the lack of techniques to determine the form of the coupling constants $c, b_j, J_{ij}$ in the energy function $I(T)[s]$. A primary reason for this is the infinite character of the energy function in Eq.~\eqref{eq:I_s}, comprising both long-range and multi-body terms, which pose a daunting challenge for analytical studies. In turn, although numerical methods allow us to sample the control landscape and map out the corresponding phase diagram, they prove insufficient when it comes to determining and understanding the detailed mechanisms that drive these phenomena. 

In this paper, we introduce methods formally independent of the specific quantum system and focus our analysis on two few-qubit toy models.
Before diving into the details, we briefly revisit their control phase diagrams and summarize the properties of the existing control phases.

\subsection{Single-qubit control landscape}

Consider the single-qubit system
\begin{equation}
    \hat H(t) = h_z \hat S^z + s(t) h_x \hat S^x,
    \label{eq:1q-H}
\end{equation}
where $\hat S^\alpha,\,\alpha{=}x,y,z$ are the spin-1/2 operators. The control problem we investigate aims to prepare the target state $\ket{\psi_\ast} {=} \ket{\text{GS}(h_x/h_z{=}{-}2)}$, starting from the initial state $\ket{\psi_0} {=} \ket{\text{GS}(h_x/h_z{=}2)}$; $\ket{\text{GS}(h_x/h_z)}$ denotes the ground state of the Hamiltonian specified by the ratio $h_x/h_z$ (with $s{=}1$). In addition, we fix the protocol duration $T$ and set, throughout the evolution, $h_z{=}{-}1,h_x{=}{-}\sqrt{5}$ (\textit{i.e.}~$s(t)$ is the only time-dependent parameter). Even though precise values of the control problem parameters affect the CLPTs' locations, the qualitative picture remains unaffected.

\begin{figure}
\centering
\includegraphics[width=.48\textwidth]{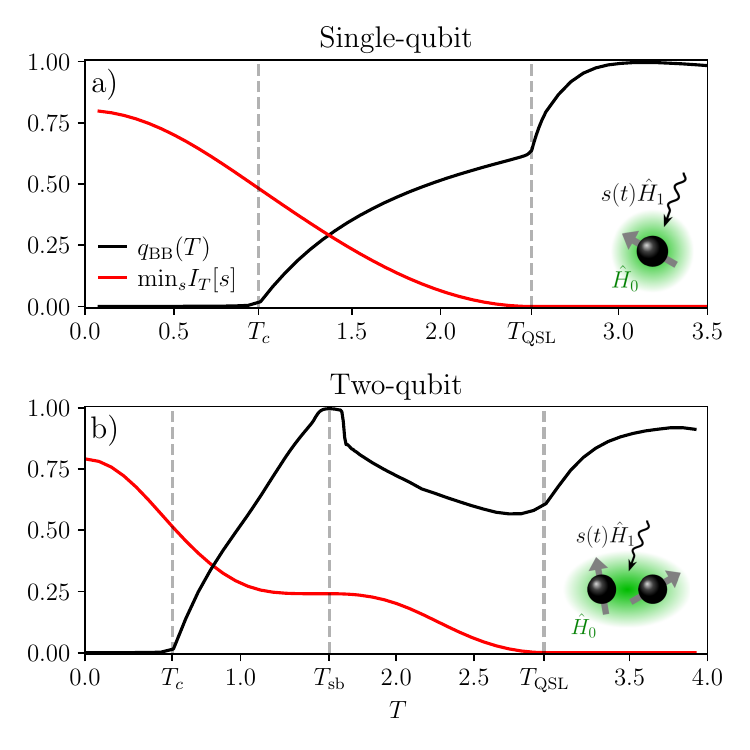}
\caption{
Control Landscape Phase Transitions (CLPTs) in the single- (a) and two-qubit (b) control problems. The curves shown are obtained from Stochastic Descent (SD) algorithm in the exact infidelity landscape, searching in the space of \emph{bang-bang} protocols. The minimum infidelity $\min_s I(T)[s]$ and the order parameter $q_\text{BB}(T)$ (cf.~Eq.~\eqref{eq:q_BB}) are shown as a function of the protocol duration $T$ for a fixed number of bang-bang steps. Non-analytic points in the order parameter $q_\text{BB}(T)$ mark control landscape phase transitions (dashed vertical lines). Similar transitions appear in both control problems: at $T{=}T_c$, and at $T{=}T_\text{QSL}$ (where the infidelity vanishes). The two-qubit problem possesses an additional symmetry-breaking transition at $T{=}T_\text{sb}$ related to the separation of the set of optimal protocols into two separated subsets which break the $s(t) {\leftrightarrow} {-}s(T{-}t)$ symmetry of the control problem \cite{Bukov18_Broken}.
}
\label{fig:phase_diags}
\end{figure}

In Fig.~\ref{fig:phase_diags}a we show the corresponding bang-bang control phase diagram as a function of the parameter $T$. For $T{<}T_c$, the order parameter vanishes ($q_\text{BB}{=}0$) and hence there exists a single optimal protocol. At $T{=}T_c$ the bang-bang infidelity landscape suddenly acquires additional local minima, associated with a cusp in $q_\text{BB}$. While a simple physical picture shows that at the critical duration $T_c$ a \emph{singular arc} appears in the optimal protocol  \cite{boscain2021introduction,ansel2024,Bukov18_Reinforcement},
little is known about how the underlying microscopic spin model $I(T)[s]$ determines the properties of $q_\text{BB}(T)$, including the critical exponent at the transition.  
For $T_c{\leq} T{\leq} T_\text{QSL}$, the bang-bang infidelity landscape exhibits, in addition to a unique global minimum, multiple almost-optimal local minima; the correlations among these minima give rise to a finite value $0{<}q_\text{BB}{<}1$. At the quantum speed limit, $T{=}T_\text{QSL}$, the landscape undergoes a second phase transition where the number of global minima proliferates; a second non-analytic point in $q_\text{BB}(T)$ emerges. It is currently unclear how the presence of a finite quantum speed limit modifies the structure of the energy function $I(T)[s]$, nor how these changes affect the behavior of the order parameter $q_\text{BB}(T)$. 

\subsection{Two-qubit control landscape}

While the control landscape of the single-qubit problem in Eq.~\eqref{eq:1q-H} is not fully understood, finding the optimal protocol is rather straightforward due to the simplicity of the qubit system \cite{Hegerfeldt13_Driving}. 
Therefore, we also investigate a two-qubit model
\begin{equation}
\label{eq:2q-H}
    \hat H(t) = J\hat S^z_{1}\hat S^z_{2} +h_z(\hat S^z_1 + \hat S^z_2) 
    +s(t) h_x(\hat S^x_1 + \hat S^x_2)
\end{equation}
with $h_z,h_x$ as in the single-qubit problem and $J{=}{-}2$.The two qubits are coupled via an Ising interaction $J$, and are exposed to a constant global $z$-field, and a time-varying global $x$-field. The initial and target states read as $\ket{\psi_0} {=} \ket{\text{GS}(h_x/h_z{=}2,J/h_z{=}1)}$ and $\ket{\psi_\ast} {=} \ket{\text{GS}(h_x/h_z{=}{-}2,J/h_z{=}1)}$.
To the best of our knowledge, the exact solution to the corresponding constrained optimal control problem in Eq.~\eqref{eq:2q-H} is unknown.
This two-qubit model will allow us to study the wider applicability of the techniques we develop. 

For $J{=}0$, we recover the single-qubit problem discussed above; in particular, the control phase diagram exhibits an overconstrained phase ($T{<}T_c$), and a fully controllable phase $T{>}T_\text{QSL}$, separated by a correlated phase $T_c{\leq} T{\leq} T_\text{QSL}$ (cf.~Fig.~\ref{fig:phase_diags}a and Ref.~\cite{Bukov18_Broken}).
However, for finite values of $J$, additional phases appear in the control landscape. 

In Ref.~\cite{Bukov18_Broken} it was shown that the infidelity function features an Ising $\mathbb{Z}_2$ symmetry:
\begin{equation}
\label{eq:Z2_symm}
    I_T[s(t)] = I_T[-s(T-t)],
\end{equation}
corresponding to the simultaneous application of a protocol time-reversal symmetry $s(t) {\to} s(T{-}t)$ and global flip of the $x/y$-projection $\hat S^{x/y}_j {\to} {-}\hat S^{x/y}_j$. On the level of the effective classical spin model, the same $\mathbb{Z}_2$ symmetry manifests itself as a global spin flip, combined with a reflection about the center of the lattice: $s_j {\to} {-}s_{N-j}$. The presence of this symmetry hinges on the choice of initial and target states. 

Interestingly, symmetry breaking occurs within the correlated phase of the two-qubit bang-bang control phase diagram (cf.~Fig.~\ref{fig:phase_diags}(b)), at a critical protocol duration $T_\text{sb}$. This leads to a doubly degenerate optimal level set that features two global minima, whose protocols can be transformed into one another by the symmetry operation. 
Moreover, the order parameter $q_\text{BB}$ exhibits a discontinuity at the symmetry-breaking transition $T_\text{sb}$. 
The same $\mathbb{Z}_2$ symmetry is present in the single-qubit control problem~\eqref{eq:1q-H}, as well as in the corresponding multi-qubit generalization \cite{Bukov18_Reinforcement}; nevertheless, rather mysteriously, the corresponding landscapes do not feature a symmetry-broken phase \cite{Bukov19_Glassy}. While this phenomenon was discovered and studied numerically, at present, it is not known what the necessary and sufficient conditions for it to occur are; it is also unclear what its relation to the other control phase transitions is.

In the following, we develop and discuss techniques to derive analytical expressions for the quantum control landscape. This allows us to go beyond bang-bang protocols and investigate the corresponding landscapes defined over the piecewise continuous protocol space. 

\section{Perturbative expansions for the control landscape}  \label{sec:expansions}

We now proceed to present a systematic derivation of the analytical expansion for the infidelity landscape, cf.~Eq.~\eqref{eq:I_s}, using three different controlled techniques: the Dyson, Magnus, and cumulant expansions. 
We test their accuracy when truncated to the leading few orders, and compare and contrast their ability to capture the transitions in the control phase diagrams of the single- and two-qubit toy models.

\subsection{Analytical expansions for the infidelity} 
\label{ssec:expansion0}

Consider the general case of piecewise continuous protocols $s(t)$.
We can write down a general expansion of the infidelity $I(T)[s]$ functional as
\begin{align}
\label{eq:infid-exp}
I(T)[s] &{=} c(T) {+} \int_0^T\dd{t} b_t(T) s_{t} {+} \frac12 \int_0^T\dd^2  t\, J_{t_1t_2}(T) s_{t_1} s_{t_2} {+} \cdots 
\end{align}
with some a priori unknown coefficients $c(T)$, $b_t(T)$, $J_{t_1t_2}(T),\dots$ that depend parametrically on the protocol duration $T$. At the level of the underlying effective spin model, as the notation suggests, they play the role of a constant shift, an external field, and a two-body interaction strength, respectively.

To determine these unknown coefficients, we rewrite Eq.~\eqref{eq:infid-def} in terms of the density matrix associated with the quantum state $\hat U_s(T,0) \ket \psi$. In an $N$-qubit system, a generic density matrix can be expanded as $\rho {=} \dyad{\psi}{\psi}{=} d^{-1} + \hat {\vec S} \cdot \vec n$ where $d{=}2^N$ is the Hilbert space dimension and the vector $\hat{\vec S}$ contains all traceless spin operators acting on the Hilbert space (e.g., Pauli or Gell-Mann matrices for a qubit or qutrit system, respectively; Pauli strings for a qubit chain) and we choose the normalization convention $\Tr{\hat S^i\, \hat S^j} {=} \delta_{ij}/2$ with $i,j{=}1,\dots,d^2{-}1$. Notice that to each quantum state $\ket{\psi} {\in} \mathbb{C}^{d}$ corresponds a real ``dual" vector $\vec n {\in} \mathbb{R}^{d^2{-}1}$; in terms of the dual vector $\vec n$, the infidelity becomes 
\begin{align}
\label{eq:fid-real}
I(T)[s] &= 1-
\abs{\bra{\psi_*} \hat U_s(T,0) \ket{\psi_0}}^2 \notag \\
&= 1-\bra{\psi_*} \hat U_s(T,0)\; \rho_0\; \hat U_s(T,0)^\dagger \ket{\psi_*} \notag \\
&= 1-\bra{\psi_*} \left( d^{-1} + \hat U_s(T,0)\; \hat {\vec S}\; \hat U_s(T,0)^\dagger \cdot \vec{n}_0 \right) \ket{\psi_*} \notag \\
&= 1 - d^{-1} - \bra{\psi_\ast} \hat {\vec S} \ket{\psi_\ast}  M_s(T,0) \, \vec n_0 \notag \\
&= 1 - d^{-1} -(1/2)\qty(\vec{n}_* \cdot \, M_s(T,0) \, \vec n_0).
\end{align}
The rotation matrix $M_s(T,0)$, which we refer to as control propagator, defines the representation of the unitary evolution group on the dual space $\mathbb R^{d^2{-}1}$; in other words, the quantum evolution operator is replaced by the orthogonal evolution operator over the space of real $(d^2{-}1)$-dimensional vectors $\vec n$. Explicitly, $(M_s(T,0))_{ij} {=} 2\Tr(\hat S^i \hat U_s(T,0) \hat S^j \hat U_s(T,0)^\dagger)$.

A few remarks are in order: 
(i), note that by going to the dual description, we manage to swap the original quantum expectation values for an ``expectation" in the vectors $\vec{n}_{0,\ast}$; in a sense, the quantum nature of the original problem is now hidden in the dual space dimension which grows faster than the qubit Hilbert space size. This indicates that any transitions in the control landscape are not quantum by nature, but merely classical; 
(ii), on a technical level, the above procedure allows us to eliminate the absolute value square in the definition of the infidelity; this is useful since it will allow us to more easily find the expression for the landscape from Eq.~\eqref{eq:infid-exp};
(iii), notice that this procedure is generic and applies to control problems where the cost function is defined in terms of observables (e.g., energy minimization) that possess easy-to-identify expansions in operator space: one simply has to replace $\rho_0$ by the corresponding observable. 
Last but not least, (iv), we have managed to reduce the problem of computing the infidelity landscape to the problem of expanding the control propagator $M_s(T,0)$ in terms of the protocol $s(t)$. 

To make further progress analytically, let us write the control propagator $M_s(T,0)$ as 
\begin{equation}
M_s(T,0) = \mathcal T \mathrm e^{\int_0^T m_s(t)dt},
\end{equation}
where $m_s(t)$ is the anti-symmetric matrix generating the rotation $M_s(T)$; assuming the relation $[\hat S^i, \hat S^j]{=}if_{ijk} \hat S^k$ with structure constants $f_{ijk}$, $m_s(t)$ can be computed directly from the Hamiltonian as $(m(t))_{ij} {=} \sum_k f_{ijk} 2\Tr(\hat S^k \hat H(t))$.

For a general Hamiltonian of the form in Eq.~\eqref{eq:H(t)}, we can separate the time dependence in the generator as $m_s(t) {=} m_0 {+} s(t) m_1$.
However, this additive structure of the generator does not yet allow for a straightforward expansion in terms of the protocol $s$ (see App.~\ref{app:expansions}).
To circumvent this problem, we now perform a change-of-frame transformation, which leads to an effective re-summation of all relevant subseries \cite{bukov2015universal}. To this end, we use a rotating frame transformation on the level of the generator, resulting in
\begin{align}
    m_s(t) \mapsto m_s'(t) &= M_0(t,0)^\mathrm{t} [m_s(t) - \partial_t] M_0(t,0),
\end{align}
with $M_0(t,0) {=} \exp(t m_0)$ and $(\cdot)^\mathrm{t}$ denotes the matrix transpose.
Note that physical quantities, such as the infidelity, remain invariant. Thus, in the primed reference frame, the infidelity can be written as
\begin{align}
I_s(T) &= 1-d^{-1} -(1/2)\left(\vec n_* \cdot M_0(T,0) M'_s(T,0) \,\vec n_0 \right),
\end{align}
where $M'_s(T,0)$ is the transformed evolution operator associated with the generator $m_s'(t)$ in the new reference frame. We can also absorb the $s$-independent rotation in the target state: $\vec n'_*(T) {=} \vec n_* M_0(T,0)$, to simplify notation. 
The only remaining protocol-dependent quantity is now the control propagator $M'_s(T,0)$; hence, in the following, we seek a matrix-valued expansion in powers of $s$. 

\textit{Dyson expansion.} The Dyson series associated with $M'_s(T,0)$ reads as
\begin{align}
M'_s(T,0) &= \mathbb 1 + \int_0^T\dd{t}\, \partial_s m'_s(t)s(t)  \notag\\ 
 &+ \frac12 \int_0^T\dd^2 t\, \mathcal T \qty[\partial_s m_s'(t_1)\partial_s m_s'(t_2)]s(t_1)s(t_2) \notag \\
 &+ \dots \qquad.
\end{align}
We can now easily read off the infidelity expansion coefficients in Eq.~\eqref{eq:infid-exp}:
\begin{align}
\label{eq:infid-exp-0-coeffs}
c(T) &= 1-d^{-1} -(1/2)\left(\vec n_*'(T) \cdot \vec n_0\right) \\
b_t(T) &= -(1/2) \left( \vec n_*'(T) \cdot \, \partial_sm'_s(t)\, \vec n_0 \right) \notag \\
J_{t_1t_2}(T) &= -(1/2) \left( \vec n_*'(T) \cdot \, \mathcal T \qty[\partial_sm'_s(t_1)\partial_sm'_s(t_2)]\,  \vec n_0 \right) \notag
\end{align}
and similarly for higher-order terms. 
We have thus obtained explicit closed-form expressions for all coefficients in the infidelity expansion, which define the control landscape. Remarkably, the whole series expansion can be explicitly written once the quantity $\partial_s m_s'(t)$ is known.

As a side note, notice that when restricted to bang-bang protocols, a second-order truncation of the series expansion yields an effective classical spin model that bears resemblance to the well-known Hopfield network \cite{Nishimori01}.

It now becomes clear that the (matrix-valued) control propagator $M'_s(T,0)$ generates the control landscape. In particular, sandwiching it between $\vec{n}_\ast'$ and $\vec{n}_0$ projects to a specific landscape for \textit{any} choice of initial and target states, and ultimately produces the corresponding infidelity expansion coefficients in Eq.~\eqref{eq:infid-exp}. Moreover, by choosing $\vec{n}'_\ast$ appropriately, we can obtain the corresponding control landscape also for observables, e.g., when minimizing the energy, or maximizing the expectation of some spin component, etc. Therefore, the control propagator is a fundamental object of control landscape theory. 

As we will see below, the Dyson series is one example of a functional expansion for the infidelity. In particular, Eq.~\eqref{eq:infid-exp-0-coeffs} corresponds to the infidelity functional-Taylor expansion \emph{centered} around the protocol $s{=}0$ (written in terms of the orthogonal generator $m_s'(t)$). In App.~\ref{app:expansions} we generalize this result to an arbitrary center $s{\ne}0$ and a generic reference frame; moreover, we derive the expansion for both unitary and orthogonal evolution operators. In Sec.~\ref{sec:stability} we use these general expressions to study the near-optimal region of the control landscape. 

A well-known deficiency of the \emph{truncated} Dyson series for the evolution operator is the violation of orthogonality of the resulting approximate operator; as a result, the corresponding Dyson-expanded infidelity landscape may not be restricted to the interval $[0,1]$. 

\textit{Magnus expansion.} A possible remedy to the unitarity/orthogonality loss is given by the Magnus expansion \cite{Magnus54,blanes2009magnus,blanes2010pedagogical} for the control propagator $M'_s(T,0)$:
\begin{equation}
    M_s'(T,0) = \exp[\sum_{n=1}^\infty \Omega_{n,s}(T,0)]
    \label{eq:magnus_exp}
\end{equation}
where the first few terms read as 
\footnote{
Here, we use the shorthand notation $s_t=s(t)$ and $m'_t=m'_s(t)$.
}
\begin{eqnarray*}
    \Omega_{1,s}(T,0) &=& \int_0^T\! \dd t\,  \partial_s m'_{t} s_{t}, \\
    \Omega_{2,s}(T,0) &=& \frac12 \int_0^T\! \int_0^{t_1}\! \dd^2\,  t\, [\partial_s m'_{t_1},\partial_s m'_{t_2}]  s_{t_1} s_{t_2},\\
    \Omega_{3,s}(T,0) &=& \frac16 \int_0^T\! \int_0^{t_1}\! \int_0^{t_2}\!\dd^3 t\, \bigg( [\partial_s m'_{t_1},[\partial_s m'_{t_2},\partial_s m'_{t_3}]]  \\
    && + [[\partial_sm'_{t_1},\partial_s m'_{t_2}],\partial_s m'_{t_3}]] s_{t_1} s_{t_2} s_{t_3} \bigg).
\end{eqnarray*}

Owing to its structure, truncating the Magnus series at finite order yields an orthogonal operator, and the resulting fidelity respects the $[0,1]$ interval bound. Notice that the exponential map acting on the truncated expansion generates infinitely many terms that are absent in the Dyson expansion. However, the Magnus expansion does not allow for a direct scalar effective energy interpretation since its terms are matrix-valued (\textit{i.e.} generators of the rotation group $SO(N^2{-}1)$).

\textit{Cumulant expansion.} Substituting the Magnus expansion in Eq.~\eqref{eq:fid-real}, we find
\begin{equation*}
    I(T)[s] = 1-d^{-1} -(1/2) \vec n_*' \cdot \exp(\Sigma) \vec n_0
\end{equation*}
where $\Sigma$ compactly denotes the Magnus expansion. Rewriting $I(T)[s] {=} 1 {-} e^{\log(1-I(T)[s])}$, we can expand the logarithm using the so-called cumulant expansion as
\begin{equation}
    \log(1-I(T)[s]) = \log\frac{2d^{-1} + \vec n_*' \cdot \exp(\Sigma) \vec n_0}{2} = \sum_{m=0}^\infty \frac{\kappa_m}{m!}.
\end{equation}
This expansion is common in statistical mechanics, where the free energy $F$, linked to the partition function $Z$ through the identity $-\beta F {=} \log Z$, is typically expanded in cumulants. In the present case, the fidelity $1 {-} I(T)[s]$ plays the role of the partition function. The first few terms of the cumulants expansion read as
\begin{align*}
    \kappa_0 &= \log\frac{2d^{-1} + \vec n_*' \cdot \vec n_0}{2}  \\
    \kappa_1 &= \frac{\vec n_*' \cdot \Sigma \vec n_0}{2d^{-1} + \vec n_*' \cdot \vec n_0} \\
    \kappa_2 &= \frac{\vec n_*' \cdot \Sigma^2 \vec n_0}{2d^{-1} + \vec n_*' \cdot \vec n_0} - \qty(\frac{\vec n_*' \cdot \Sigma \vec n_0}{2d^{-1} + \vec n_*' \cdot \vec n_0})^2.
\end{align*}

We now turn to the quality of the three infidelity expansions presented. 
Since we want to develop a technique to better understand the near-optimal region of the quantum control landscape, it is important to quantify the extent to which the infidelity expansions reproduce the behavior of already known quantities, such as the non-analytic points in the order parameter $q_\text{BB}(T)$ across control phase transitions. For this reason, in the following, we compare the three expansions using the two toy models introduced in Sec.~\ref{sec:model}.

\subsection{Single-qubit problem}
\label{ssec:SD_1Q}

So far the discussion has been carried out for a generic quantum system. In the case of Hamiltonian \eqref{eq:1q-H} one obtains
\begin{align}
m'_s(t) &= h_xs(t)[T_x \cos(h_z t)+ T_y \sin(h_z t)],
\label{eq:1q-m}
\end{align}
where $\{T_i\}_1^3$ are the generators of the group $SO(3)$ defined by $(T_i)_{jk} {=} \epsilon_{ijk}$, with $\epsilon_{ijk}$ the $SO(3)$ structure constants. We emphasize that having a closed-form expression for $m'_s(t)$ is sufficient to generate \textit{any} (multi-body) coupling term in the Dyson expansion for the infidelity landscape.

\begin{figure}
\centering
\includegraphics[width=.235\textwidth]{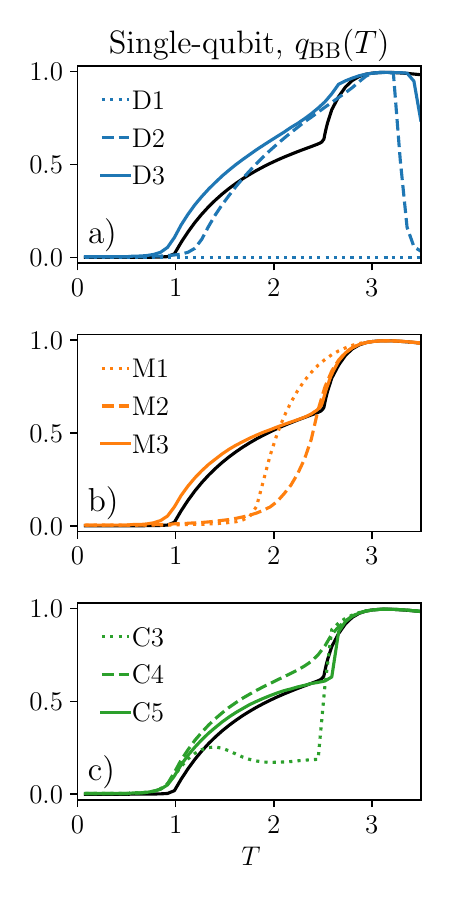}
\includegraphics[width=.235\textwidth]{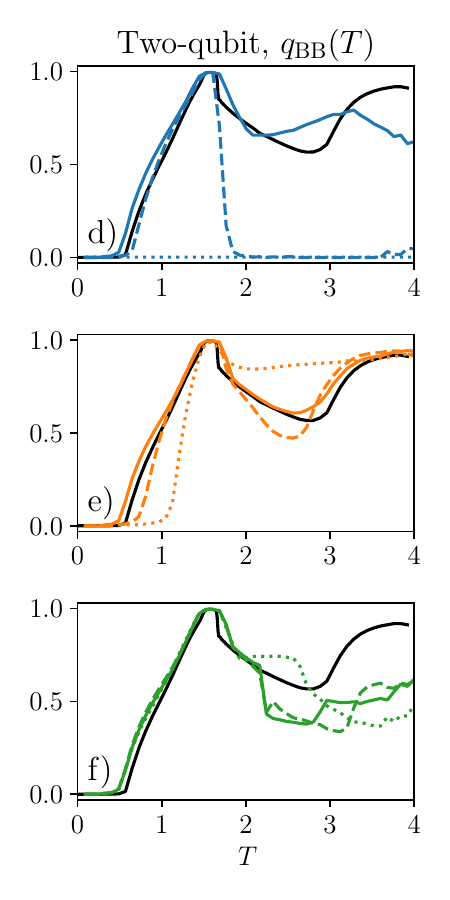}
\caption{
Comparison of the order parameter $q_\text{BB}(T)$ (cf.~Eq.~\eqref{eq:q_BB}) obtained from Stochastic Descent (SD) in the landscape defined by the infidelity expansions introduced in Sec.~\ref{sec:expansions}, in the single- and two-qubit control problem. 
Control protocols are here restricted to the bang-bang class.
The black curve refers to the exact infidelity landscape in ~Eq.~\eqref{eq:infid-def}. The letters D, M, and C in the legend stand for Dyson, Magnus, and Cumulant expansion while the adjacent integer specifies the order of truncation.
\textit{Single-qubit:}
The $T{=}T_c{\simeq}0.98$ transition is approximately captured by all three methods. The second transition at $T{=}T_\text{QSL}{\simeq}2.51$ is only visible in the Magnus and Cumulants results.
\textit{Two-qubit:}
The $T{=}T_c{\simeq}0.56$ and $T{=}T_\text{sb}{\simeq}1.57$ transitions are approximately captured by all three methods. The non-analytic point at $T{=}T_\text{QSL}{\simeq}2.95$ is less sharp than the corresponding point in the single-qubit problem and only the Magnus expansion truncated at third order is able to reproduce the qualitative behavior of the exact curve.
}
\label{fig:SD}
\end{figure}

In Fig.~\ref{fig:SD} we evaluate the performance of the three infidelity expansions using Stochastic Descent (see App.~\ref{app:SD}). Since our motivation to introduce the expansions is the study of the quantum landscape transitions, we benchmark the three methods using the order parameter $q_\text{BB}(T)$, introduced in Eq.~\eqref{eq:q_BB}. 
In our simulations, we truncate Dyson and Magnus expansions up to third order. 

The first transition around $T_c$ is visible in all three expansions within the truncation orders considered. While the Dyson expansion is not able to capture the $T_\text{QSL}$ transition at third-order, the Magnus and cumulants expansions results follow remarkably well the exact curve $q_\text{BB}(T)$ (at third- and fifth-order, respectively). 
Thus, we conclude that higher than third-order terms, generated in the Magnus expansion from the exponential map but completely absent in the other two expansions, are essential to reproduce CLPTs in the exact infidelity landscape.

\subsection{Two-qubit problem}
\label{ssec:SD_2Q}

Let us apply our theory to the two-qubit problem from Eq.~\eqref{eq:2q-H}. The rotating frame is defined by removing the drift term $\hat H_0$ from the Hamiltonian. In this case, the drift term has two contributions $\hat H_0 = \hat H_{0,1} + \hat H_{0,2}$,
\begin{equation*}
\hat H_{0,1} = h_z \hat S^z_\text{tot},\qquad \hat H_{0,2} = J S_1^z S_2^z.
\end{equation*}
Since $[\hat H_{0,1}, \hat H_{0,2}]=0$, the reference frame transformation factorizes as $\hat U_0 = \hat U_{0,1} \hat U_{0,2}$. A straightforward calculation shows that 
\begin{equation*}
    \hat H'(t) = s(t) h_x\hat U_{0,2}^\dagger \qty[\cos(h_z t) \hat S^x_\text{tot} - \sin(h_z t) \hat S_\text{tot}^y] \hat U_{0,2}
\end{equation*}
where $\hat U_{0,2}$ is a diagonal operator in the computational $z$-basis.

In this case, the two-qubit quantum control problem allows for an orthogonal decomposition of the Hamiltonian operator into the triplet and singlet subspaces; moreover, the quantum states $\ket{\psi_{0,*}}$ have no component in the singlet subspace. Hence, we may restrict $\vec S {=} \vec \lambda/2$ to the $SU(3)$ basis defined by Gell-mann matrices $\vec \lambda$. Explicitly, in the dual description we have:
\begin{align}
    m_s'(t) = s(t) \sqrt2 h_x 
                (
                &T_1 \cos(h_z t + J t/2) \ \notag\\
                -&T_2 \sin(h_z t + J t/2) \ \notag\\
                +&T_6 \cos(h_z t - J t/2) \ \notag\\
                -&T_7 \sin(h_z t - J t/2)
                )
                \label{eq:2q-m}
\end{align}
where the generators $\{T_i\}_1^8$ are given by $(T_i)_{jk} = f_{ijk}$ with $f_{ijk}$ the $SU(3)$ structure constants.

Stochastic Descent results for the two-qubit problem are reported in Fig.~\ref{fig:SD}. Both the $T_c$ and $T_\text{sb}$ transitions are captured by the three expansions. In the latter case, the jump discontinuity of $q_\text{BB}(T)$ is smoothened out by the approximated landscape. On the contrary, the $T_\text{QSL}$ transition is partially captured only by the Magnus expansion truncated at third order.

A final remark is in order. In this section, we used the infidelity expansions, centered around $s_0{=}0$, to detect CLPTs of the exact control landscape. The choice of the center $s_0{=}0$ is convenient since it allows us to make no extra assumption regarding the shape of the optimal protocol(s) of the landscape. Nonetheless, as we will see in the next section, better results can be achieved by centering the expansion around carefully chosen protocols ($s_0{\ne}0$). This can be done whenever additional information regarding the properties of the optimal protocol(s) is known.

\subsection{Landscape expansions in generic quantum control problems}

Let us now comment on the potential applicability of landscape expansions to study CLPTs in other quantum control problems. 

In App.~\ref{app:expansions}, we discuss the convergence properties of the landscape expansions introduced in this section and show that they depend on the norm of the generator integrated over the total evolution, namely
\begin{align}
&\int_0^T \dd t\,\norm{H(s(t))} & \text{or}& & &\int_0^T \dd t\,\norm{m(s(t))}.
\label{eq:integrated-norm}
\end{align}
In the case of bounded protocols, the integral can be bounded by the product $T \max_t(\norm{H(s(t))})$. 
The applicability of the techniques developed in this section for generic quantum control problems depends on this quantity. Physically, $\norm{\hat H(t)}$ depends on the size of the quantum system, the interactions between its components, and the amplitude of the external control fields.

In particular, if one is interested in approximating the exact landscape within a given error (uniformly in the space of control protocols), the order of truncation of the expansion grows with the duration of evolution and the norm of the Hamiltonian $\norm{\hat H(t)}$. 

Nevertheless, when one is interested in studying CLPTs, it is not necessary to work with a uniform approximating functional. To see this, we observe that (i) CLPTs are determined by a subset of all the allowed protocols, namely, minima of the infidelity, and (ii) the numerical approximation of the infidelity is not required, as we are interested in the relative location of minima to capture the relevant physics of CLPTs, not necessarily in their infidelity values.
Therefore, the ultimate limit for the applicability of landscape expansions in the study of CLPTs depends on the time-integrated norms in Eq.~\eqref{eq:integrated-norm}, restricted to the near-optimal region of the control landscape.

\section{Stability analysis and control landscape phase transitions}
\label{sec:stability}

In the previous section, we introduced three different analytical expansions for the control landscape, centered around the protocol $s{=}0$. In App.~\ref{app:expansions} we discuss the relation with the functional Taylor expansion. This section uses the infidelity Taylor expansion to analyze CLPTs in the single- and two-qubit control problem introduced in Sec.~\ref{sec:model}. Our purpose is twofold. On the one hand, critical points in the control phase diagram are identified with different types of instability relative to the minimum of the quantum control landscape. On the other, we consider the Taylor expansion as a function of the parameter $T$ to identify and follow the landscape minimum. In particular, we first use the $T{\to}0$ limit to analyze the landscape in a linear approximation, and to identify the structure of the optimal protocol in this limiting case. Subsequently, we track how the optimal protocol changes as we increase $T$ by enforcing ``linear stability". 

\begin{figure}
    \centering
    \includegraphics[width=.35\textwidth]{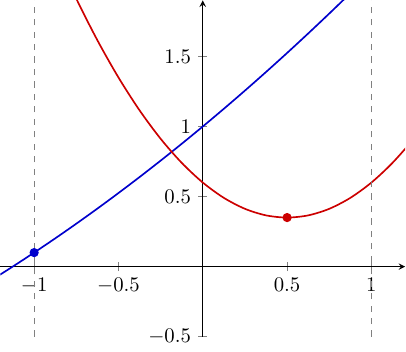}
    \caption{
    Two different one-dimensional functions (blue and red curves) illustrating the stability conditions for optimal points in Eq.~\eqref{eq:lin_stab_cond}. 
    The minimum of the red curve at $x{=}+0.5$ (red dot), lying inside the allowed domain $[-1,1]$, satisfies $f'(x){=}0,f''(x){>}0$.
    The minimum of the blue curve at $x{=}{-}1$ (blue dot), lying at the edge of the allowed domain $[-1,1]$, satisfies $f'(x){>}0$.
    }
    \label{fig:schem_stability}
\end{figure}

The idea can be understood by using a one-dimensional analogy (see Fig.~\ref{fig:schem_stability}). Let $I{=}[-1,1] {\subset} \mathbb R$ and $f {:} I {\to} \mathbb R$ be a convex function with minimum $x_0 {\in} I$. The Taylor expansion centered in $x{=}x_0$ and truncated at second-order reads $f(x) {=} f(x_0) + \dv{f}{x}\eval_{x_0} (x{-}x_0) {+} \frac{1}{2!} \dv[2]{f}{x}\eval_{x_0}(x{-}x_0)^2$. If $x_0$ does not belong to the boundary $\{-1,1\}$, then the linear term vanishes $\dv{f}{x}(x_0){=}0$ and the second-order term is positive $\dv[2]{f}{x}{(x_0)} {>} 0$. Otherwise, if $x_0$ belongs to the boundary $\{-1,1\}$, the linear term may be finite but then, either $f'(x) {>} 0$ if $x_0{=}{-}1$ or $f'(x) {<} 0$ if $x_0{=}1$; notice that the two cases are summarized by the condition $\mathrm{sgn}(f'(x_0)) {=} {-}\mathrm{sgn}(x_0)$, with $\mathrm{sgn}$ the sign function.
Finally, if we let $f(x){=}f_T(x)$ to also continuously depend on a parameter $T$, we may then study the behavior of its minimum $x_0$ as a function of $T$. 

The control landscape (represented by the function $f_T(x)$ in the above analogy) is a functional of the protocol $s$ and depends on the duration of the quantum evolution $T{\in}[0,\infty)$. Our strategy is based on the continuous dependence of infidelity on the parameter $T$, and relies on two ideas. First, once an optimal protocol at a given $T$ is found, it can be used as an ansatz to search for optimal protocol(s) at $T'{=}T{+}\Delta T$. We exploit this idea in two independent ways: (i) analytically, by formulating a variational ansatz for the optimal shape of the protocol at $T'$ using the optimal shape at $T$, and (ii) numerically, by using the numerical optimum at $T$ as a starting point for the new run of the optimization algorithm at $T'$.
Second, at a fixed $T$, we use the infidelity expansion around a given optimal protocol $s_0$ to study its stability under generic perturbations $s_0{+}\delta s$. In this case, the stability (instability) of the protocol $s_0$ depends on whether the infidelity increases (decreases) after the perturbation $s{+}\delta s$ is applied. 

We demonstrate that this procedure (which we dub ``adiabatic tracing" below) is sufficient to map out the control phase diagram up to $T_\text{QSL}$ in the one- and two-qubit problems, respectively (cf.~Fig.~\ref{fig:phase_diags}).
In addition, notice that the control phase diagram was originally obtained by searching for optimal configurations in the space of bang-bang protocols~\cite{Bukov18_Reinforcement,Bukov18_Broken}.
Here, we go beyond this restriction and study the landscape over the space of bounded piecewise continuous protocols, $s(t) {\in} [-1,1]$. Interestingly, the phenomenology observed in the bang-bang case extends over to this larger space, but with a few caveats: in Sec.~\ref{ssec:partfunc_qBB}, we comment on the relation between the control problem defined over bang-bang and bounded piecewise continuous protocol spaces.

\subsection{Adiabatic tracing in quantum control landscapes} 
\label{ssec:stability_tracing}

Observe that, when expanding the infidelity around a generic protocol $s_0(t)$,  the coefficients $c(T)$, $b_t(T)$, $J_{t_1t_2}(T)$ in Eq.~\eqref{eq:infid-exp} depend on $s_0$ itself: in this section we make this dependence explicit but also keep track of the parametric dependence on the protocol duration $T$:
\begin{equation*}
    c[s](T),\quad b_t[s](T),\quad J_{t_1t_2}[s](T), \quad \text{etc.}
\end{equation*}

We define the linear stability of the protocol $s_0$ with respect to the infidelity landscape $I[s](T)$ analogously to the one-dimensional example $f_T(x)$ discussed above. The protocol $s_0$ is ``linearly unstable" if
\begin{align}
  b_t[s_0](T) &\ne 0 & \abs{s_{0,t}}&<1 \notag \\
  -\mathrm{sgn}\,(b_t[s_0](T)) &\ne \mathrm{sgn}\, (s_{0,t}) & \abs{s_{0,t}}&=1 
  \label{eq:lin_stab_cond}
\end{align}
for some $t {\in} [0,T]$. The two cases distinguish the situations where $s(t)$ is inside $[-1,1]$ or at its boundary (see Fig.~\ref{fig:schem_stability}).

As in the one-dimensional example for non-convex functions, one needs in general to define also the quadratic stability of the protocol $s_0$. In this case, the second-order coefficient $J_{t_1t_2}[s_0](T)$ plays the role of the Hessian matrix (\textit{i.e.} the second-order term in the Taylor expansion) for multi-dimensional functions. In particular, the stability of $s_0$ depends on the set of eigenvalues $\{\lambda_n\}$, defined by 
\footnote{
Observe that, by construction, $J_{tt'}$ is the real and symmetric kernel of a Hilbert-Schmidt integral operator; \textit{i.e.} it satisfies the following properties: (i) $J_{tt'}{\in}\mathbb R$, (ii) $J_{tt'}{=}J_{t't}$, (iii) $\int_0^T \dd^2t\, \abs{J_{tt'}}^2 {<} \infty$.
Then, if we search for eigenfunctions $\{f^{(n)}_{t}\}$ in the functional space $L^2([0,T])$, the associated eigenvalues $\{\lambda_n\}$ exist well-defined, are real, have a finite range and an accumulation point at zero.
In this case, we may estimate the eigenvalues $\{\lambda_n\}$ by first discretizing the time domain $[0,T]$ in $L$ steps and, subsequently, consider the limit $L{\to}\infty$.
}
\begin{equation}
     \frac1T \int_0^T \dd t'\, J_{tt'} f^{(n)}_{t'} = \lambda_n f^{(n)}_{t}, 
     \label{eq:hess-eigeq}
  \end{equation}
with $\{f^{(n)}_{t}\}_n$ the eigenfunctions (the dependence on $s_0$ and $T$ is omitted for simplicity). Then, the protocol $s_0$ is ``quadratically unstable'' if there are negative eigenvalues in the spectrum of the ``Hessian operator" $J_{t_1t_2}[s_0](T)$. In practice, we can perform the diagonalization numerically: discretizing the time axis in $L$ equal-size intervals we find the matrix eigenvalue equation:
 \begin{equation*}
     \frac1L \sum_{j=1}^L J_{ij} f^{(n)}_{j} = \lambda_n f^{(n)}_{i},\quad n=1,2,\dots, L.
 \end{equation*}
In this case, the number of eigenvalues is controlled by the number of discrete segments $L$ we divide our protocol into, and is not strictly physical. 
We report in App.~\ref{app:quadratic_scaling} the finite-$L$ scaling analysis.

We are now equipped and in a position to introduce each step of the method and discuss the corresponding results for the single- and two-qubit control problems.
As a starting point, we consider the limit $T {\to} 0^+$ where the infidelity landscape can be approximated by a Taylor series truncated to first order (throughout our work we use the ${+},{-}$ superscripts to distinguish the right- or left-limit, respectively). Expanding around the protocol $s{=}0$,
\begin{equation*}
    I(T)[s] \approx I_T[0] + \int_0^T \dd t\, b_t[0](T) s(t)
\end{equation*}
one obtains the infidelity minimum $s_0(t) {\equiv} {-}\text{sgn}[b_t[0](T)]$. In fact, the above infidelity expansion can be interpreted as an effective classical energy $E(s)\equiv I(T)[s]$ of the field $s(t)$ in which $b_t[0](T)$ plays the role of an external field applied to the variable $s(t)$. Since this is a non-interacting problem, energy is trivially minimized when $s(t)$ has maximum absolute value and is anti-aligned with $b_t[0](T)$ 
\footnote{
The choice of $s(t){\equiv}0$ as a starting protocol is arbitrary but convenient. For other choices $s{\ne}0$, finding the optimal protocol requires more attention.
}.
In this way, in both the one- and two-qubit problems in Eq.~\eqref{eq:1q-H} and \eqref{eq:2q-H}, for $T{\to}0$ we find the optimal protocol
\begin{equation}
s_0(t) \equiv \begin{cases}
+1 & t<T/2 \\
-1 & t>T/2, \\
\end{cases}.
\label{eq:stab_s0}
\end{equation}

\begin{figure*}
\centering
\includegraphics[width=\textwidth]{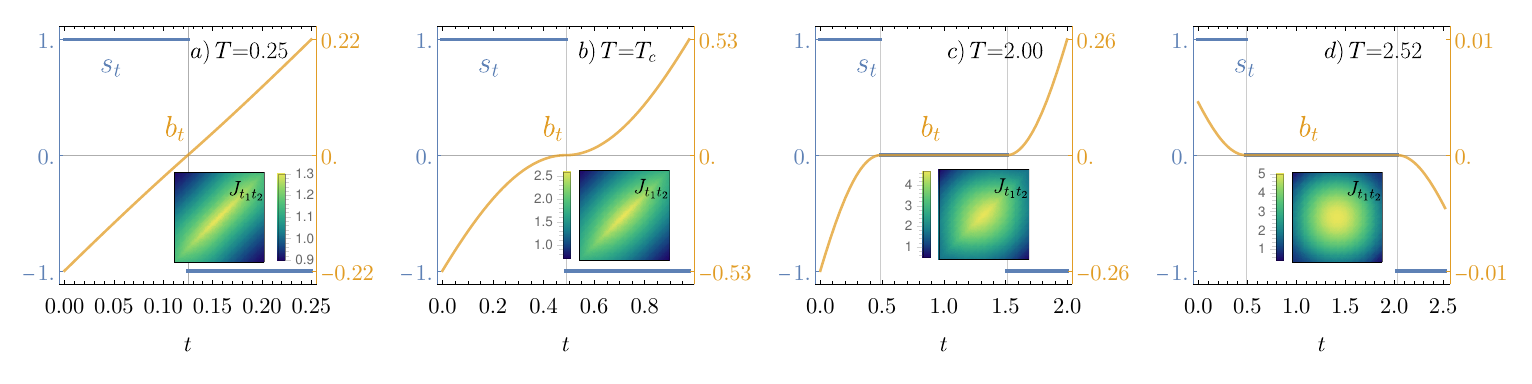}
\caption{
Stability analysis in the single-qubit control problem. The plots show the linear ($b_t$, orange curve) and second-order terms ($J_{t_1t_2}$, colormap insets) of the infidelity Taylor expansion centered around optimal piecewise continuous protocols ($s_t$, blue curve) found by the adiabatic tracing method (cf.~Sec.~\ref{sec:stability}), for different value of the duration $T$ of the quantum evolution (CLPTs in $T_\text{c}{\simeq}0.98,\,T_\text{QSL}{\simeq}2.51$).
\textit{Linear term:} $b_t$. For $T{\in}[0,T_c]$ the optimal protocol is $s_0$ in Eq.~\eqref{eq:stab_s0}. (\textbf{\textit{a}}) At $T{=}0.25$, $s_0$ is anti-aligned with $b_t$ and is linearly stable. (\textbf{\textit{b}}) For $T{=}T_\text{c}$, $s_0$ becomes linearly unstable around $t{=}T/2$ as shown by the horizontal inflection of $b_t$ around $t{=}T/2$. (\textbf{\textit{c}}) For $T{\in}[T_c,T_\text{QSL}]$ the optimal protocol is $s_{\Delta_0(T)}$ in Eq.~\eqref{eq:stab_s_Delta}. At $T{=}2.0$, $s_{\Delta}$ is stable provided $\Delta$ is chosen as Eq.~\eqref{eq:stab_s_Delta} (cf.~Fig.~\ref{fig:stab_1q_Delta}a). (\textbf{\textit{d}}) At $T{=}T_\text{QSL}$, the first-order term is exactly zero and it flips sign for $T{=}T_\text{QSL}^+$: $s_{\Delta_0(T)}$ becomes linearly unstable.
\textit{Quadratic term:} $J_{t_1t_2}$. In the language of statistical mechanics, the second-order term represents the effective two-body interactions of the field $s_t$. In this case, interactions are antiferromagnetic (AFM) and long-range: the system is frustrated. For $T{=}T_c$ (panel (\textbf{\textit{b}})), the protocol $s_0$ becomes linearly unstable around $t{=}T/2$ and the local AFM interactions around the instability point dictate the new optimal protocol structure $s_{\Delta(T)}$ (cf.~Eq.~\eqref{eq:stab_s_Delta}).
}
\label{fig:stab_1q}
\end{figure*}

\subsection{Single-qubit problem}

The starting point for our analysis is the protocol $s_0$ defined in Eq.~\eqref{eq:stab_s0}. As a first step, we check its linear stability. In Fig.~\ref{fig:stab_1q} we show the first order term $b_t[s_0](T)$ of the infidelity Taylor series centered at $s_0$, for $T{=}0.25$ and $T{=}T_c$. In particular, we observe that $s_0$ is linearly stable for $T{\in}[0,T_c]$. Exactly at $T{=}T_c$, $b_t[s_0](T)$ (regarded as a function of the variable $t$) inverts its sign around the point $t{=}T/2$: $s_0$ has a linear instability  for $T{>}T_c$ around the point $t{=}T/2$. In this way, we associate the $T_c$ control landscape phase transition with linear instability of the protocol $s_0$, optimal for $T{\in}[0,T_c]$.

Since the instability only affects a local neighborhood of $t{=}T/2$, it is suggestive to look for the new infidelity minimum in the region $T{=}T_c^+$ by locally (\textit{i.e.}, around $t{=}T/2$) modifying the previous optimal protocol $s_0(t)$. 
For $T{=}T_c$, in the neighborhood of $t{=}T/2$, the first-order term $b_t[s_0](T)$ has a negligible contribution to the infidelity compared to the second-order term $J_{t_1t_2}[s_0](T)$. Moreover, using the language of statistical physics, we can interpret $J_{\frac{T_c}{2},\frac{T_c}{2}}[s_0](T_c){>}0$ as an antiferromagnetic interaction (cf.~$T{=}T_c$ case in Fig.~\ref{fig:stab_1q}b). This observation allows us to make an educated guess for the optimal protocol in the parameter region $T{>}T_c$:
\begin{equation}
s_\Delta(t) \equiv \begin{cases}
0 & t \in \left[\frac{T}{2}-\Delta, \frac{T}{2}+\Delta\right] \\
s_0(t) & \text{elsewhere},
\end{cases} 
\label{eq:stab_s_Delta}
\end{equation}
where $\Delta {\in} [0,T/2]$ is a free parameter in the ansatz that may depend on the protocol duration $T$ itself
\footnote{
This argument should not be considered as rigorous. In our case, it is sufficient to guess the structure of the optimal protocol for $T{>}T_c$. In general, a more quantitative analysis is required to obtain the optimal protocol structure past a given CLPT (see \cite{hernandezgomez2023optimal} for an example).
}.

\begin{figure}
\centering
\includegraphics[width=.23\textwidth]{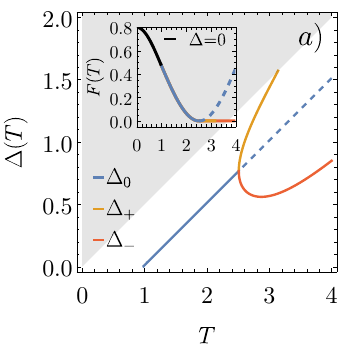}\,
\includegraphics[width=.24\textwidth]{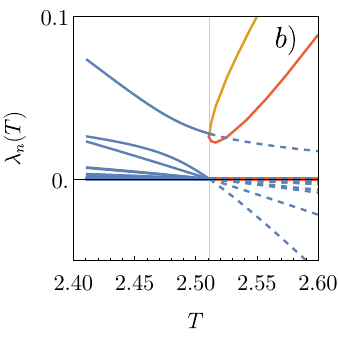}
\caption{
Stability analysis in the single-qubit control problem.
(\textbf{\textit{a}}) $\Delta$ parametrizes the family of protocols $s_{\Delta}$ in Eq.~\eqref{eq:stab_s_Delta}. $\Delta(T)$ is defined by linear stability of the corresponding protocol. For $T{\in}[0,T_\text{QSL}]$, $\Delta_0(T)$ is single-valued: within the family $s_{\Delta}$ there is a single optimal protocol. The bifurcation $\Delta(T){=}\Delta_\pm(T)$ occurs at $T{=}T_\text{QSL}$ so that for $T{>}T_\text{QSL}$ $\Delta(T)$ is double-valued: within the family $s_{\Delta}$ there are two optimal protocols. The gray area corresponds to $\Delta{>}T/2$ and is inaccessible by definition.
(\textbf{\textit{b}}) Eigenvalues of the quadratic operator $J_{t_1t_2}$ from the infidelity Taylor expansion centered around the optimal protocol $s_{\Delta(T)}$. For $T{\to}T_\text{QSL}$, $L{-}2$ eigenvalues (blue solid curves) vanish: the protocol $s_{\Delta_0(T)}$ becomes quadratically unstable. The largest positive eigenvalue, approximately constant and with magnitude ${\sim}1$, is shown in App.~\ref{app:quadratic_scaling}.
For $T{\ge}T_\text{QSL}$, the two protocols $s_{\Delta_\pm(T)}$ are optimal ($I_T{=}0$); the associated quadratic operator has $L{-}2$ vanishing and $2$ positive eigenvalues. We deduce the existence of $L{-}2$ orthogonal deformations of the optimal protocols $s_{\Delta_\pm(T)}$ that preserve the vanishing infidelity.
}
\label{fig:stab_1q_Delta}
\end{figure}

Past the critical point $T_c$, the family of protocols $s_\Delta$ parametrized by $\Delta{\in}[0,T/2]$, contains the optimal protocol. 
Searching for linearly stable protocols within the family $s_\Delta$ fixes $\Delta$ for each $T$. In fact, this constraint defines the curves $\Delta_0(T),\Delta_\pm(T)$, shown in Fig.~\ref{fig:stab_1q_Delta}a and selects the stable protocols among the family $s_\Delta$.
For $T{\in}[T_c,T_\text{QSL}]$, we find only one linearly stable protocol within this family. On the other hand, a bifurcation in the curve $\Delta(T)$ occurs which coincides exactly with the critical point at $T_\text{QSL} {\simeq} 2.51$, marking the end of the single-valued regime. More precisely, we find perfect agreement with the analytical behavior
\begin{equation}
    \Delta(T) = \begin{cases}
        0                    & T \in [0,T_c] \\
        \frac 1 2 (T{-}T_c) & T \in [T_c,T_\text{QSL}] \\
        \frac 1 2 (T{-}T_c) + \delta_\pm(T) & T \gtrsim T_\text{QSL}
    \end{cases}
    \label{eq:Delta}
\end{equation}
where $\delta_\pm(T) {\sim} {\pm} (T{-}T_\text{QSL})^{1/2}$ as $T{\to}T_\text{QSL}^+$. 
Hence, for later convenience, let us define 
\begin{align}
    \Delta_0(T)     &{=} (T{-}T_c)/2, &
    \Delta_\pm(T)   &{=} \Delta_0(T) {+} \delta_\pm(T).
    \label{eq:Delta0_Deltapm}
\end{align}
The presence of the bifurcation at $T_\text{QSL}$ implies that the infidelity landscape possesses two distinct linearly stable protocols $s_{\Delta_\pm}$. For $T{\ge}T_\text{QSL}$ the infidelity associated with $s_{\Delta_\pm}$ vanishes (\textit{i.e.} they are global optimal protocols) as it is visible in the inset of Fig.~\ref{fig:stab_1q_Delta}a. Consistent with the global optimality condition, we also find $b_t[s_{\Delta_\pm}](T){=}0$, $\forall t{\in}[0,T]$. Therefore, for $T{>}T_\text{QSL}$ the infidelity landscape possesses at least two (and possibly more) globally optimal protocols $s_{\Delta_\pm}$. This indicates a non-trivial structure emerging in the near-optimal region in the control landscape. 

A better understanding of the $T_\text{QSL}$ control phase transition is provided by the quadratic stability analysis, and in particular by the eigenvalues of the Hessian operator. The results for the optimal protocol discussed so far \{$s_0, s_{\Delta_0}, s_{\Delta_\pm}$\} are reported in Fig.~\ref{fig:stab_1q_Delta}b (see also App.~\ref{app:quadratic_scaling}) and can be summarized as follows: 
(i) $s_0$ and $s_{\Delta_0}$ are quadratically stable in $[0,T_\text{QSL}]$. 
(ii) For $s_{\Delta_0}$, all except two eigenvalues of $J_{t_1t_2}[s_{\Delta_0}](T)$ change sign from positive to negative exactly at the quantum speed limit $T_\text{QSL}$. 
(iii) For $s_{\Delta_\pm}$, all except two eigenvalues of $J_{t_1t_2}[s_{\Delta_\pm}](T)$ vanish as $T {\to} T_\text{QSL}^-$ and they remain zero for $T{>}T_\text{QSL}$; the remaining two non-vanishing eigenvalues are positive.
Thus, the control phase transition at $T_\text{QSL}$ is associated with linear and quadratic instability.

It is interesting to notice that the presence of infinitely many zero eigenvalues is associated with infinitely many deformations of the protocols $s_{\Delta_\pm}$ preserving zero infidelity. This fact has important implications. 
First, we deduce that at $T_\text{QSL}$ the number of optimal protocols may grow from one to \emph{infinite}.
Second, quadratic stability analysis relative to the optimal protocols $s_{\Delta_\pm}$ suggests the existence of an infinite-dimensional ``optimal level set", defined implicitly by the equation $I(T)[s]{=}0$, containing optimal and homotopically equivalent protocols; in this case, different optimal protocols (in the same connected component) could be continuously deformed into one another while preserving optimality. Such a property can be valuable in experimental applications: one can select a subset of optimal protocols based on desired criteria, such as robustness against experimental noise \cite{ge2021}.

The local character of the linear and quadratic stability analysis does not allow for a direct study of the global properties of the optimal level set. 
For example, the constraint $\abs{s(t)}{\le}1$ limits the allowed deformations around the two protocols $s_{\Delta_\pm}$, potentially preventing the exploration of the optimal level set. Hence, motivated by the stability analysis and keeping in mind the limitations of our local analytical analysis, in Sec.~\ref{sec:LMC}, we investigate the existence of a level set of continuously connected optimal protocols using Metropolis-adjusted Langevin dynamics.

\subsection{Two-qubit problem}
\label{ssec:stability_2q}

\begin{figure}
\centering
\includegraphics[width=.241\textwidth]{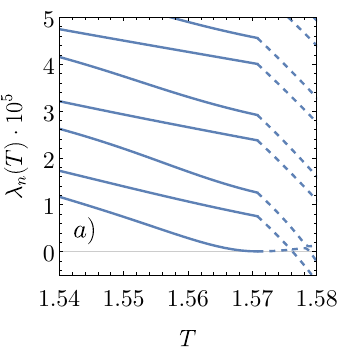}
\includegraphics[width=.229\textwidth]{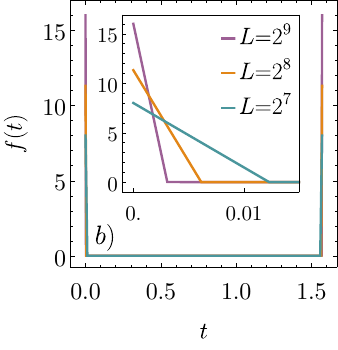}
\caption{
Stability analysis in the two-qubit control problem.
(\textbf{\textit{a}}) Eigenvalues of the quadratic operator $J_{t_1t_2}$ from the infidelity Taylor expansion centered around the protocol $s_{\Delta(T)}$, optimal for $T{\le}T_\text{sb}$.  
In the plot, we discretize the $[0,T]$ interval in $L{=}64$ steps; as $L$ grows, the smallest eigenvalue decreases but remains positive for $T{<}T_\text{sb}{\simeq}1.57$. The CLPT in $T_\text{sb}$ is associated with a single vanishing eigenvalue in the second-order term. 
(\textbf{\textit{b}}) Eigenfunction $f(t)$ associated with the vanishing eigenvalue, plotted for $T{\simeq}T_\text{sb}$. As $L$ grows, $f(t)$ approximates a sum of two delta distributions peaked at the interval boundary, $t{=}0,T$. Notice the even parity $f(t){=}f(T{-}t)$: this property is associated with the breaking of the symmetry of the quantum control problem, $s(t){\leftrightarrow}{-}s(T{-}t)$, occurring at $T{=}T_\text{sb}$.
}
\label{fig:stab_2q}
\end{figure}

The two-qubit control problem follows the same qualitative behavior of the single-qubit case up to $T{=}T_\text{sb}$. Following the same steps as in Sec.~\ref{ssec:stability_tracing}, we correctly identify the optimal protocol $s_0$ for $T{\to}0$ and detect, by linear stability analysis, the critical point $T_c{\simeq}1.57$. For $T{\ge}T_c$, only one protocol satisfies linear stability condition among the family \eqref{eq:stab_s_Delta} parametrized by $\Delta{\in}[0,T/2]$ (see Fig.~\ref{fig:stab_2q}). 

The $T_\text{sb}$ control phase transition presents new features. 
(i) As $T {\to} T_\text{sb}$, $\Delta(T) {\to} T/2$, so that $\Delta(T)$ reaches the boundary of its domain $[0,T/2]$ and $s_{\Delta(T)} {\to} 0$. 
(ii) For $T{=}T_\text{sb}$, the linear-order term $b_t[s_{\Delta(T)}](T)$ vanishes: $s{=}0$ is a saddle point at $T{=}T_\text{sb}$.
(iii) For $T{=}T_\text{sb}$, the lowest eigenvalue of the Hessian operator $J_{t_1t_2}[s_{\Delta(T)}](T)$ vanishes (cf.~Fig.~\ref{fig:stab_2q}).
These properties are reminiscent of the behavior of control phase transitions in the one-qubit problem $T_\text{QSL}$, where the vanishing eigenvalues in the Hessian operator mark the presence of a branching phenomenon in the low-infidelity region of the landscape. However, for the two-qubit problem at $T{=}T_\text{sb}$ there is only one vanishing eigenvalue. 

In Ref.~\cite{Bukov18_Broken} the $T_\text{sb}$ control phase transition was related to a bifurcation process. It was shown that the unique optimal protocol (valid in $T {\le} T_\text{sb}$) bifurcates into two optimal protocols (valid in $T {\ge} T_\text{sb}$) individually violating the symmetry $s(t) {\leftrightarrow} {-}s(T{-}t)$ of the quantum control problem.
Interestingly, in the quadratic stability analysis, the eigenfunction associated with the vanishing eigenvalue has even parity, namely, it violates the symmetry $f(t) {=} -f(T{-}t)$ (cf.~Fig.~\ref{fig:stab_2q}).
This characteristic instability signals the presence of a bifurcation of a single optimal protocol into a pair of symmetry-violating optimal protocols.

For $T{>}T_\text{sb}$, piecewise continuous optimal protocols can be found using numerical optimization algorithms (for example, the LMC algorithm introduced in Sec.~\ref{sec:LMC}).
In App.~\ref{app:quadratic_scaling}, we report the quadratic stability analysis at $T{=}T_\text{QSL}$ for the two-qubit problem, where the infidelity is expanded around optimal protocols found by the LMC algorithm.
The results are analogous to the single-qubit case, except that there are now $L-4$ vanishing eigenvalues, for $T{\to}T_\text{QSL}$.

\section{Stochastic dynamics within the optimal level set beyond the quantum speed limit}
\label{sec:LMC}

In Sec.~\ref{sec:stability}, the control phase transition occurring at $T_\text{QSL}$ has been associated with the formation of infinitely many global minima in the quantum control landscape, branching out from isolated optimal protocols. Past $T_\text{QSL}$, stability analysis suggests that the quantum control landscape possesses a level set of distinct protocols satisfying $I[s]{=}0$ (\textit{i.e.} ``optimal") and that can be transformed into each other via continuous deformations. In this section, we introduce a simple stochastic dynamics algorithm to explore the low-infidelity region of the landscape and probe the properties of the optimal level set $\{s(t) : I(T)[s]{=}0\}$. Despite its simplicity, this approach can uncover important properties of the level set, such as connectivity, the local structure, and bulk size.

Let us introduce the algorithm we use to explore the optimal level set. First, we discretize the time domain $[0,T]$ in $L$ steps $(t_1,\dots,t_L)$ and approximate the piecewise continuous protocol $s(t)$ with a piecewise-constant protocol $\vb s {=} (s(t_1),\dots,s(t_L))$ where $s(t_i) {\in} [-1,1]$. Then, we introduce a stochastic map $\vb s_n {\mapsto} \vb s_{n+1}$ by combining the gradient-free Langevin update rule,
\begin{align}
\label{eq:LMC_update_update} 
s_{n+1}(t_i) &= s_n(t_i) + \xi(t_i)
\\
\expval{\xi(t_i)} &= 0,\quad 
\expval{\xi(t_i) \xi(t_j)} = \sigma^2 \delta_{ij}, \notag
\end{align}
with the standard Metropolis acceptance rule,
\begin{align}
P_\text{acc}(s(t_i)_{n+1}) 	&= \min(1, \exp(-\beta L \Delta I_{n+1})) 
\label{eq:LMC_update_metropolis}\\
\Delta I_{n+1}						&= I_T[s_{n+1}] - I_T[s_{n}]. \notag
\end{align}
Here, the subscript $n$ labels the iteration of the stochastic dynamics, $\xi(t_i)$ is a Gaussian random variable of zero mean and variance $\sigma^2$, and $\beta$ has the intuitive meaning of (dimensionless) inverse-temperature. This dynamics is a gradient-free version of the standard Langevin-Monte Carlo algorithm (also known as Metropolis-adjusted Langevin algorithm \cite{Roberts02}); we thus refer to this algorithm as ``LMC". The reasons behind such a simple algorithm are two-fold. On the one hand, we neglect the gradient term contribution (typically present in Langevin dynamics) because we use the algorithm to explore the optimal level set, characterized by a vanishing gradient term. On the other, we do not exploit information from the Hessian analysis in order to have an unbiased exploration algorithm.

\begin{figure*}[t!]
\centering
\includegraphics[width=.3\textwidth]{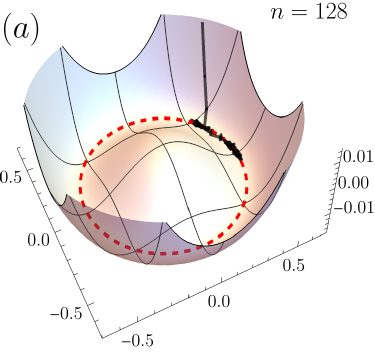}\,
\includegraphics[width=.3\textwidth]{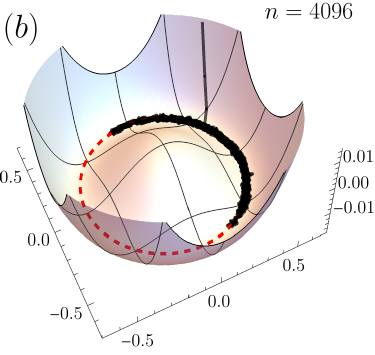}\,
\includegraphics[width=.3\textwidth]{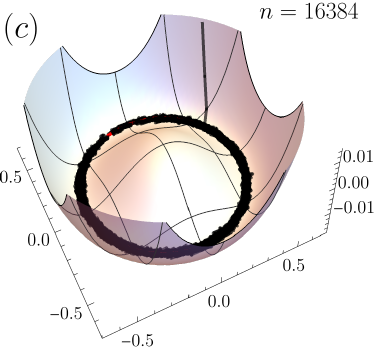}
\caption{
Illustration of LMC dynamics in a two-dimensional toy-model system $\vb s_n {=} (x_n,y_n) {\in} \mathbb R^2$ subject to the landscape $V(\vb s_n) {=} {-}(1/8) x_n^2 {+} (1/4) x_n^4$, (``Mexican-hat") with $\beta{=}10^4,\, \sigma{=}10^{-3/2}$ (cf. Eq.~\eqref{eq:LMC_update_update}). The trajectory of $\vb s_n$ in a single LMC run is plotted up to three different iterations. The stochastic motion along the optimal level set (the landscape valley denoted by the red-dashed curve) becomes visible.
In general, LMC dynamics can be separated into three consecutive stages. (\textbf{\textit{a}}) \textit{Thermal relaxation stage.} The system $\vb s_n$ starts from the random configuration $\vb s_0$ and reaches the region surrounding the optimal level set $I_T{=}0$. (\textbf{\textit{b}}) \textit{Diffusion stage.} By stochastic moves, $\vb s_n$ diffuses in the optimal level set; this motion cannot be suppressed by increasing $\beta$ (provided the system has enough time to diffuse). (\textbf{\textit{c}}) \textit{Equilibrium sampling stage.} Asymptotically, $\vb s_n$ explores the optimal level set. Using LMC dynamics, we sample different configurations from the optimal level set and study their statistical properties.
}
\label{fig:LMC_sketch}
\end{figure*}

LMC possesses two free parameters: $\beta,\sigma{>}0$. In principle, they are independent of one another and need to be tuned carefully for the dynamics to explore efficiently the space of low-infidelity protocols. On one side, the dynamics is confined to a volume around the $I(T)[s]{=}0$ hypersurface that shrinks as $\beta {\to} \infty$ and $\sigma {\to} 0$. On the other, as $\sigma {\to} 0$ the LMC dynamics slows down, and more time is needed for the optimal level set exploration.  
In what follows, we explore the optimal level set at $\beta {=} 10^{4}, 10^{5}, 10^{6}$ and choose the value of $\sigma$ maximizing the acceptance ratio for each $\beta$. In these cases, the LMC dynamics possesses three distinct stages: 
\begin{enumerate}
    \item Thermal relaxation stage: $\vb s_n$ descends the infidelity landscape, from the initial random configuration to the low-infidelity region, reaching the bottom of one (of possibly many) landscape ``valley" (Fig.~\ref{fig:LMC_sketch}a);
    \item Diffusion stage: $\vb s_n$ moves stochastically along the bottom of the valley since the Metropolis acceptance rule constrains the size of infidelity fluctuations $\Delta I {\sim} L\beta$ (Fig.~\ref{fig:LMC_sketch}b);
    \item Equilibrium sampling stage: $\vb s_n$ explores the landscape valley and samples a statistically representative set of optimal protocols (Fig.~\ref{fig:LMC_sketch}c). This happens on a time scale that is much larger than that of the diffusion stage, and thus ensures that individual samples are uncorrelated.
\end{enumerate}
Although the main motivation for the LMC algorithm is the study of the optimal level set, we find that the control phase transition at $T_\text{QSL}$ affects each of the three stages. As a consequence, next we separately present the main results from the three stages of LMC dynamics.

\subsection{Thermal relaxation stage}
\label{ssec:LMC_T}

Upon initialization of the LMC algorithm, each component of the initial protocol $\vb s_0$ is drawn randomly from a uniform distribution in the interval $[-1,1]$. Subsequently, the stochastic dynamics evolves the protocol $\vb s_n$ decreasing the corresponding infidelity $I[\vb s_n]$, thus ``descending" the landscape. By studying how the infidelity changes as the protocol evolves from the initial random configuration, we obtain information about the structure of the quantum control landscape above the low-infidelity region.

For $\beta {\gtrsim} 10^3$ the behavior of the algorithm during thermal relaxation is largely independent of $\sigma {\lesssim} 10^{3/2}$. In general, during this stage the non-equilibrium dynamics can be divided into two substages: one characterizes the motion at the beginning of thermal relaxation while the other appears subsequently and persists until the system thermalizes. Interestingly, the whole behavior can be captured by an effective one-dimensional model that admits a closed-form solution and provides a better intuition on the structural changes occurring in the landscape across the different control phase transitions (cf.~App.~\ref{app:LMC_T}). In particular, the CLPT affects the non-equilibrium dynamics in the thermal relaxation stage of LMC and one may then detect the presence of the CLPT without sampling optimal protocols. We remark that so far CLPTs have been characterized exclusively in terms of statistical properties of the locally optimal protocols, whereas here we provide a dynamical characterization.

\subsection{Diffusion stage}
\label{ssec:LMC_D}

Once the protocol $\vb s_n$ reaches the bottom of a valley in the infidelity landscape, it moves stochastically along the flat directions of the landscape. If the infidelity minimum is unique (or the infidelity minima are isolated from each other), fluctuations of $\vb s_n$ are suppressed by the effective inverse temperature $\beta$, cf.~Eq.~\eqref{eq:LMC_update_metropolis}. Otherwise, if different infidelity minima are continuously connected with each other, protocol fluctuations within the flat directions are not suppressed by taking $\beta {\to} \infty$. Therefore, if the $T_\text{QSL}$ control phase transition is associated with the creation of infinitely many flat directions in the infidelity valley, then the LMC dynamics after thermal relaxation should be strongly affected by this control phase transition.

For a quantitative description, it is useful to fix the parameter  $\beta$ (in the Metropolis acceptance rule) and vary the size of the stochastic jumps $\sigma {\ll} \sqrt\beta$. In this limit, we approximate the LMC dynamics with an $L$-dimensional diffusive motion subject to a constraining potential given by the infidelity. For short times, LMC explores the neighborhood of the initial point, and its motion can be decomposed in an approximately free motion along the valley (the tangent space) and fluctuations along the orthogonal directions (cf.~Fig.~\ref{fig:LMC_sketch}b). At long times the LMC dynamics diffuses around the optimal level set, ideally exploring this set in a relatively dense way (\textit{i.e.} such that the mean values of the observables we sample using LMC dynamics converge to constant values). 

In more concrete terms, during the diffusive stage, we consider the covariance matrix related to the protocol fluctuations $\vb s_n$. Using this quantity, for each $\beta$ we estimate the minimum time required to diffuse through the optimal level set (cf.~App.~\ref{app:LMC_D}). In addition, eigenvalues and eigenvectors of the covariance matrix can be used to study the local structure (during early-diffusion time) or the bulk size (for asymptotic diffusion time) of the optimal level set.

\subsection{Equilibrium sampling stage}
\label{ssec:LMC_E}

Ultimately, the $T_\text{QSL}$ control phase transition affects the statistical properties of protocols lying in the low-infidelity region of the landscape. Accordingly, we primarily use LMC to sample protocols continuously connected with each other through sufficiently small stochastic jumps, cf. Eq.~\eqref{eq:LMC_update_update}. In particular, we use LMC's diffusive motion to move along the infidelity valleys and store the protocol configuration $\vb s_n$ every $\Delta n{=}2^{14}$ iterations (see App.~\ref{app:LMC_D}) such that, between two samples, LMC explores a global portion of the optimal level set. For each set of LMC parameters, we collect $R$ LMC runs acquiring $M$ optimal protocols each. Notice that, in each run, LMC starts from a different random protocol $\vb s_0$ and the system first relaxes to the global optimal level set before the sampling of the $M$ optimal protocols begins.

\begin{figure}
\centering
\includegraphics[width=.45\textwidth]{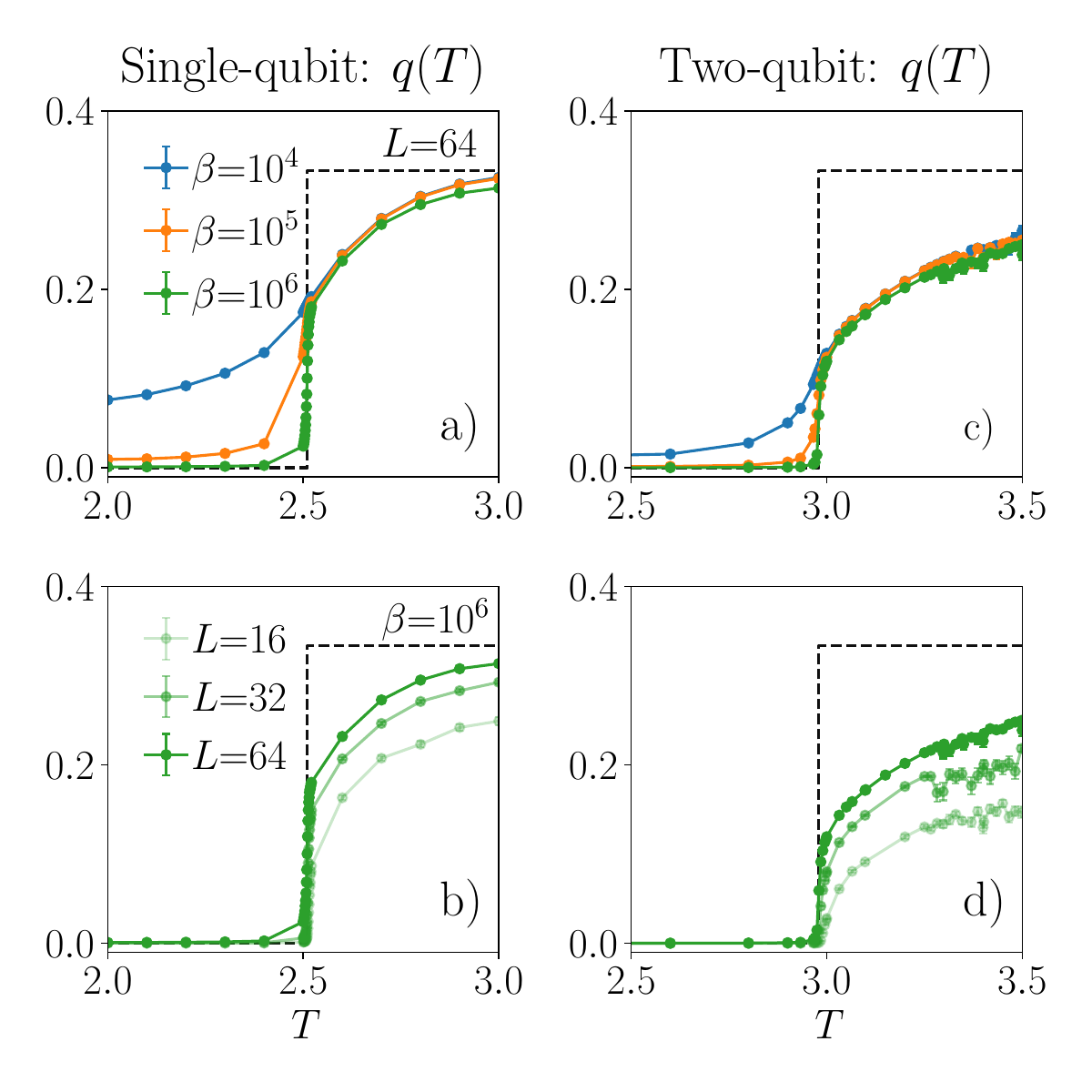}
\caption{
Single- and two-qubit control problems. Jump-discontinuity of the order parameter $q(T)$ in Eq.~\eqref{eq:q} at the quantum speed limit; $q(T)$ quantifies fluctuations of the piecewise continuous protocol as it explores the optimal level set under LMC dynamics. A non-vanishing value of $q(T)$ (as $\beta {\to} \infty$) implies that the optimal level set possesses a family of optimal protocols connected to each other via continuous transformations (homotopy).
The black dashed curve shows the analytical prediction discussed in Sec.~\ref{ssec:partfunc_q}.
In the two-qubit case, the errors associated with the mean value of $q(T)$ are relatively small for $T {\lesssim} 3.3$ but they show a sudden increase for $T{\gtrsim}3.3$ for the same number of LMC steps. As discussed in Ref.~\cite{beato2025_topological}, this is related to the presence of a new disconnected component in the optimal level set arising at $T{\simeq}3.3$.
}
\label{fig:LMC_E_q}
\end{figure}

As a first observable, consider the order parameter $q(T)$ defined in Eq.~\eqref{eq:q} which measures the size of fluctuations around the mean protocol.
The average $\expval{\cdot}$ is here performed over each LMC run separately, in order to separate contributions from different connected components of the optimal level set.
The existence of an optimal level set composed of isolated points is related to a vanishing $q(T)$ in the limit $\beta {\to} \infty$. In Fig.~\ref{fig:LMC_E_q} we show $q(T)$ for different values of the inverse-temperature $\beta$ and the system size $L$, for the single- and two-qubit problems. In the single-qubit case, as $\beta$ grows we find $q(T) {\to} 0$ for $T{\in}[0,T_\text{QSL}]$: this confirms that for $T{<}T_\text{QSL}$ the infidelity minimum is unique for piecewise-constant protocols. By contrast, for $T{>}T_\text{QSL}$ the minimum is not unique since $q(T)$ is independent of $\beta$. Remarkably, the scaling analysis with $L$ reveals the $T_\text{QSL}$ transition as a first-order transition: namely, $q(T)$ has a jump discontinuity in $T{=}T_\text{QSL}$.  
In the two-qubit case, similarly we find $q(T) {\to} 0$ (as $\beta$ is increased) for $T {\in} [0,T_\text{QSL}]$. Notice that this does not contradict the presence of two isolated optimal protocols in the interval $[T_\text{sb},T_\text{QSL}]$: each LMC run is confined to one of the two valleys which are not continuously connected to each other. Therefore, the protocol fluctuations still depend on the inverse-temperature $\beta$. For $T{=}T_\text{QSL}$, $q(T)$ acquires a finite value with a jump-discontinuity (as in the single-qubit case). Interestingly, for $T{\gtrsim}3.3$ in the two-qubit case the presence of multiple connected components in the optimal level set gives rise to a larger standard deviation of the mean value $q(T)$ for the same number of LMC samples (error-bar in Fig.~\ref{fig:LMC_E_q}c,d) \cite{beato2025_topological}.

\begin{figure}
\centering
\includegraphics[width=.45\textwidth]{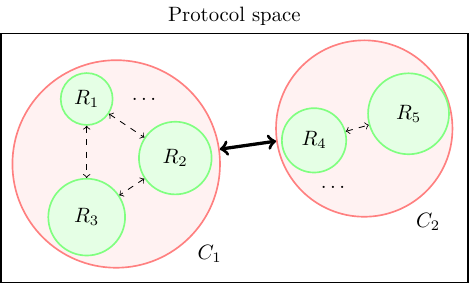}
\caption{
    Schematic visualization of different sets of protocols $R_1,R_2,\dots$ sampled by independent LMC runs. 
    As $\beta{\to}\infty$, stochastic jumps between different connected components of the optimal level set, represented by the red sets $C_1,C_2$, become ever more unlikely during LMC dynamics. In this limit, the sets $R_1,R_2,\dots$ are confined in either one of the two components. The distance between two sets, defined in Eq.~\eqref{eq:LMC_run_distance_set} and represented by the arrows in the figure, can vanish, as the number of sampled sets increases, only when the two sets are located in the same connected component. Whenever two sets are located in different connected components, the distance reaches asymptotically (in LMC iterations) a non-vanishing value.    
}
    \label{fig:LMC_distance_sketch}
\end{figure}

The number of connected components in the optimal level set can be estimated by computing the distance between pairs of LMC runs. The $M$ protocols sampled within each LMC run are connected with each other by the LMC dynamics (see, e.g., Fig.~\ref{fig:LMC_sketch} for a schematic cartoon). Assuming that LMC approximates homotopic transformations between optimal protocols, we regard the $M$ protocols collected during each LMC run as continuously connected to one another. On the contrary, protocols in different runs may belong to different connected components of the optimal level set since LMC thermalizes to different protocols in different runs. Therefore, by computing the distance between pairs of protocols belonging to different LMC runs, we can estimate the number of connected components in the optimal level set. In Fig.~\ref{fig:LMC_distance_sketch} we show a schematic representation of the idea behind the distance analysis, involving protocols within and across runs and we discuss details in App.~\ref{app:LMC_E}. In the single-qubit control landscape, our results suggest the presence of a single connected component for all $T{\in}[T_\text{QSL},3.5]$: as the number of protocols per run is increased, the distance steadily decreases to zero following an algebraic decay. 
The two-qubit problem, where the situation is more complex and multiple connected components are present, is discussed in detail in Ref.~\cite{beato2025_topological}.

\section{Critical behavior at the quantum speed limit} 
\label{sec:Tqslcritical}

In Sec.~\ref{sec:LMC}, we presented numerical results showing the relation between the control landscape phase transition at $T{=}T_\text{QSL}$ and the formation of an infinite dimensional optimal level set lying in the zero-infidelity hyperplane. So far this analysis was purely numerical, and little intuition about the general structure of this infinite-dimensional manifold was given.

In this section, we complement the numerical study, limited by construction to finite-size ($L{<}\infty$) systems, with an analytical characterization of the optimal level set and the associated $T_\text{QSL}$ transition. In particular, we shed new light on the nature of the non-analytic points in the order parameters $q_\text{BB}(T),\, q(T)$ visible in Figs.~\ref{fig:SD} and \ref{fig:LMC_E_q}.

In what follows, we focus on the properties of the infidelity landscape that are independent of the controlled quantum system at the quantum speed limit transition.
Similar to Ginzburg-Landau's theory of phase transitions, we set up an analytical theory and relate the scaling in $\Delta T {=} T{-}T_\text{QSL}$ of coefficients in the infidelity expansion (carrying information about the local geometry of the landscape) to the critical behavior of the order parameters $q(T),q_\text{BB}(T)$, in the limit $\Delta T\to0^+$.

\subsection{Critical behavior of the order parameter \texorpdfstring{$q(T)$}{q(T)}} 
\label{ssec:partfunc_q}

The order parameters $q(T),\,q_\text{BB}(T)$, introduced in Eqs.~\eqref{eq:q} and \eqref{eq:q_BB}, capture statistical properties of local minima in the quantum control landscape.
For their evaluation, a method to compute the expectation values $\expval{s(t)}$ and $\expval{s(t)^2}$ has to be introduced. 
So far, we used numerical algorithms (SD and LMC) to sample control protocols from the low-infidelity region and we computed the expectation values by averaging over the samples. 
Now, as a first step towards an analytical approach, we shall specify a rule for the computation of the expectation values.

\subsubsection{Introducing the partition function}
\label{sssec:partfunc-intro}

Given the set of admissible control protocols $\mathcal X$, expectation values can be analytically evaluated once a set of weights $w(x),\, x {\in} \mathcal X$ is introduced. 
Then, the expectation value of a generic observable $f{:}\mathcal X {\to} \mathbb R$ is computed as the ratio
\begin{align*}
    \expval{f}_{\mathcal X} &= \frac{\sum_{x\in\mathcal X}f(x)w(x)}{\sum_{x\in\mathcal X}w(x)}.
\end{align*}
Defining the partition function $Z[f] {=} \sum_{x\in\mathcal X}f(x)w(x)$, we can compactly rewrite $\expval{f}_{\mathcal X} {=} Z[f]/Z[1]$.
In our problem, $\mathcal X$ is the infinite-dimensional space of bounded protocols $\{s(t): \abs{s(t)}{\le}1 \}$ and the sum is a functional integral (path integral) formally written as
\begin{align*}
    Z[f] &= \int_{-1}^{+1} \prod_t \dd s_t\, f[s]w[s].
\end{align*}

Since we are interested in the order parameters $q(T),\,q_\text{BB}(T)$ and because they are related to a structural change of the low-infidelity region, we can for example choose the Boltzmann-Gibbs weights $w_\beta[s]{=}\exp({-}\beta I[s])$, with $I[s]{=}I(T)[s]$ the infidelity and $\beta$ is a free parameter
\footnote{Note that here $\beta$ is different from the corresponding hyperparameter of the LMC algorithm in Sec.~\ref{sec:LMC}.}.
In this way, as $\beta {\to} \infty$, the dominant contributions to the partition function have the lowest possible infidelity, and the weight $w_\beta[s]$ concentrates on globally optimal protocols: 
\begin{equation*}
    Z[f] = \int_{-1}^{+1} \prod_t \mathrm ds_t\, f[s] \delta(I[s]-I_\text{min})\ ,
\end{equation*}
where $I_\text{min}=I_\text{min}(T)$.
For $T \ge T_\text{QSL}$, by definition $I_\text{min}=0$ and the partition function reduces to
\begin{equation}
    Z[f] = \int_{-1}^{+1} \prod_t \mathrm ds_t\, f[s] \delta(I[s]),\qquad T>T_\text{QSL}.
    \label{eq:partfunc-exact}
\end{equation}

In the next sections, we explicitly evaluate the partition function and relate, for $T{-}T_\text{QSL}{=}\Delta T{\to}0^+$, the non-analytic behavior of the order parameters $q(T)$ and $q_\text{BB}(T)$ to the structural changes occurring in the low-infidelity region of the landscape.

\subsubsection{Optimal level set parametrization}
\label{sssec:optset-param}

Starting just before the quantum speed limit $T_\text{QSL}$, by definition, a subset of control protocols reaches the minimum value of the infidelity, as $T$ is increased. 
In what follows, we assume a \emph{finite} number of locally optimal protocols, which we denote $\{s_0^{(i)}\}_{i=1}^{N_0}$. 
This is the case, for example, in the single- and two-qubit problems introduced in Sec.~\ref{sec:model} (cf.~Secs.~\ref{sec:stability} and \ref{sec:LMC}), and we expect the same situation in generic controlled quantum systems.

Let us focus on the structure of the control landscape around a single locally optimal protocol $s_0$ from the set $\{s_0^{(i)}\}$. Our goal is to use the infidelity expansion centered at the protocol $s_0$ to obtain a parametrization of the optimal level set around $s_0$, for $T\ge T_\text{QSL}$.

The infidelity expanded around a given protocol $s_0$ reads as
\begin{align}
    I(T)[s_0+\delta s] &= c + \int \dd t\, b_t \delta s_t + \frac12 \int \dd ^2 t\, J_{t_1t_2} \delta s_{t_1} \delta s_{t_2} + \notag\\
    &+ \frac{1}{3!} \int \dd ^3 t\, d_{t_1t_2t_3} \delta s_{t_1} \delta s_{t_2} \delta s_{t_3} + \notag\\
    &+ \frac{1}{4!} \int \dd ^4 t\, g_{t_1t_2t_3t_4} \delta s_{t_1} \delta s_{t_2} \delta s_{t_3} \delta s_{t_4} + \dots
    \label{eq:infidelity-expansion-t}
\end{align}
where $\delta s$ represents the deviation from the protocol $s_0$. The $T$ and $s_0$ dependence in the expansion coefficients $c,b_t,J_{t_1t_2},\dots$ is omitted for notational simplicity. 
We expand $\delta s_t$ in the eigenbasis $\{f^{(n)}_t\}_n$ of the Hessian operator $J_{t_1t_2}$ as $\delta s_t = \sum_{n=1}^\infty \delta s_n f^{(n)}_t$, with $\delta s_n$ the new degrees of freedom.
Using the orthonormality of the eigenbasis, $\int_0^T f^{(n)}_t f^{(m)}_t \dd t / T = \delta_{nm}$, we obtain
\begin{align}
    I(T)[s_0+\delta s] &= c 
    + T\sum_{n=1}^\infty b_n \delta s_n 
    + \frac{T^2}{2} \sum_{n=1}^\infty \lambda_n \delta s_n^2 + \notag\\
    &+ \frac{T^3}{3!} \sum_{n,m,k}^\infty d_{nmk}\delta s_{n}\delta s_{m}\delta s_{k} + \notag\\
    &+ \frac{T^4}{4!} \sum_{n,m,k,l}^\infty g_{nmkl}\delta s_{n}\delta s_{m}\delta s_{k}\delta s_{l} + \dots
    \label{eq:infidelity-expansion-n}
\end{align}
where $\{\lambda_n\}$ are the eigenvalues of $J_{t_1t_2}$. Notice that we defined $b_n{=}\int_0^T \mathrm dt\, f^{(n)}_t b_t / T$ and analogously for higher-order terms: $J_{nm}{=}\delta_{nm}\lambda_n,d_{nmk},g_{nmkl},\dots$ .

Next, we restrict the analysis to a neighborhood of the quantum speed limit. We study the behavior of the coefficients of the expansion in Eq.~\eqref{eq:infidelity-expansion-n}, in the limit $\Delta T = T-T_\text{QSL}\to0^-$.
In this regime, there are important constraints on these coefficients, originating from the infidelity lower bound being attained at zero and the assumption of local optimality of the protocol $s_0$. 

Since $I_T[s]\in[0,1]$, odd-order terms $b_n,d_{nmk},\dots$ vanish at least as fast as $\Delta T$; otherwise, the infidelity bound $I_T[s]\ge0$ would be severely violated at $\Delta T {=} 0$.
The zeroth-order term, $c\in[0,1]$, has a zero at $\Delta T=0$ by assumption, so that necessarily $c\sim\Delta T^2$.
The eigenvalues of the second-order term, $\lambda_n$, separate into two subsets: a number $n_+$ of them remain positive while the rest vanish: $\lambda_{n\le n_+}{\sim}1$ and $\lambda_{n>n_+}{\sim}\Delta T$. In general, $n_+$ depends on the controlled quantum system \cite{chakrabarti2007quantum}; for example, in the single- and two-qubit cases introduced in Sec.~\ref{sec:model}, we found $n_+{=}2$ and $n_+{=}4$, respectively (cf.~App.~\ref{app:quadratic_scaling}).
Finally, consistent with the optimality for $T\le T_\text{QSL}$ of the protocol $s_0$, we necessarily have $b_{n>n_+}\sim\Delta T^2$; otherwise, the protocol $s_0$ would not be a minimum of the infidelity for $\Delta T {=}0$.

In summary, the most general scaling behaviors in $\Delta T$ of the infidelity expansion coefficients are
\begin{align}
    c &\sim \Delta T ^2 && \notag\\
    b_{n{\le}n_+} &\sim \Delta T & b_{n{>}n_+} &\sim \Delta T^2
    \label{eq:scalings-rest} \\
    d_{nmk} &\sim \Delta T & g_{nmkl} &\sim 1 \notag 
\end{align}
and 
\begin{align}
    \lambda_{n{\le}n_+} &\sim 1 & \lambda_{n{>}n_+} &\sim \Delta T,
    \label{eq:scalings-lambda}
\end{align}
with $n_+$ a positive integer. 

We obtain an approximate parametrization of the optimal level set $\{s:I(T)[s]{=}0\}$ around $s_0$ by truncating the infidelity expansion in Eq.~\eqref{eq:infidelity-expansion-t} to a given order, and consider the analytical continuation of the scaling in Eqs.~\eqref{eq:scalings-lambda}, \eqref{eq:scalings-rest}, for $\Delta T\to0^+$.

\subsubsection{Hard-boundary constraint approximation}

The hard-boundary constraint $\abs{s(t)} {\le} 1$ imposed on the protocol space amounts to a restriction of the integration domain for each coordinate $s_t$. Unfortunately, this constraint does not allow for an explicit evaluation of the partition function, so we resort to a suitable approximation. 

As a first step, we rewrite the constraint as
\begin{align}
    \int_{-1}^{1} \dd s_t &= 
     \int_{-1}^{1} \dd  \lambda_t \int_{\mathbb R} \dd s_t \delta(s_t - \lambda_t) \notag\\
    &= \int_{-1}^{1} \dd  \lambda_t \int_{\mathbb R} \dd s_t \int_{\mathbb R} \frac{\dd \nu_t}{2\pi}\, e^{i \nu_t (s_t - \lambda_t)},
    \label{eq:hb-1}
\end{align}
where we introduced the Fourier representation of the Dirac-delta distribution $\delta(s_t {-} \lambda_t)$. Integrating over $\lambda_t$ gives
\begin{align*}
    \int_{-1}^{1} \dd  \lambda_t \, e^{-i \nu_t \lambda_t} &= 2\frac{\sin(\nu_t)}{\nu_t} \approx 2e^{- \nu_t^2/6}.
\end{align*}
The above approximation trades the $\sin(x)/x$ function for a Gaussian displaying the same behavior for $\nu_t {\to} 0$. Notice that both functions are peaked around $\nu_t{=}0$ and vanish for $\abs{\nu_t}{\to}\infty$ 
\footnote{
More formally, we rewrite
$\exp(i \nu_t s_t) \sin(\nu_t)/\nu_t = \exp[i \nu_t s_t + \log(\sin{\nu_t}/\nu_t)]$ and expand the exponent in the right-hand side at second-order in $\nu_t$.
}.
Using this approximation, we finally integrate over $\nu_t$ in Eq.~\eqref{eq:hb-1}. 
Eventually, we obtain,
\begin{equation}
    \int_{-1}^{1} \dd s_t \approx \int_{\mathbb R} \dd s_t\, e^{-(3/2)s_t^2}, 
    \label{eq:hb-approx}
\end{equation}
which approximates the constraint $\abs{s(t)}{\le}1$ (we neglect the overall prefactor, as it will not affect expectation values).

Equation \eqref{eq:hb-approx} is the Gaussian approximation of the hard-boundary constraint, $\abs{s(t)}\le1$.
We observe the following important fact. If we relax the infidelity constraint, this approximation of the hard-boundary constraint does not affect the first two moments $\expval{s_t},\,\expval{s_t^2}$. Namely,
\begin{align}
     \frac{\int_{-1}^{+1}\dd s_t s_t}{\int_{-1}^{+1}\dd s_t} &= \frac{\int_{\mathbb R} \dd s_t\, e^{-(3/2)s_t^2} s_t}{\int_{\mathbb R} \dd s_t\, e^{-(3/2)s_t^2}} = 0 
     \notag \\
    \frac{\int_{-1}^{+1}\dd s_t s_t^2}{\int_{-1}^{+1}\dd s_t} &= 
    \frac{\int_{\mathbb R} \dd s_t\, e^{-(3/2)s_t^2} s_t^2}{\int_{\mathbb R} \dd s_t\, e^{-(3/2)s_t^2}} = \frac13.
    \label{eq:expval_s}
\end{align}
However, higher moments will differ as the two distributions are distinct.

\subsubsection{Evaluating \texorpdfstring{$q(T)$}{q(T)}}
\label{sssec:partfunc_evaluating_q}

Over the space of piecewise continuous protocols, the order parameter $q(T)$, defined in Eq.~\eqref{eq:q}, reads as
\begin{equation*}
    q(T) = \frac1T  \int_0^T\dd t\, \qty(\expval{s_t^2} - \expval{s_t}^2).
\end{equation*}
As we ultimately want to compare with the numerical results shown in Sec.~\ref{sec:LMC}, we perform the averages appearing in $q(T)$ over each local minimum in $\{s_0^{(i)}\}_1^{N_0}$, separately. That is, 
\begin{align}
    q(T) &= \frac{1}{N_0}\sum_{i=1}^{N_0} q_i(T), \\
    q_i(T) &= \frac1T  \int_0^T\dd t\, \qty(\expval{s_t^2}_i - \expval{s_t}_i^2),
\end{align}
where $\expval{\cdot}_i$ is the average relative to the minimum $s_0^{(i)}$. The remainder of this section is dedicated to the calculation of $q_i(T)$. Therefore, except for $q_i(T)$, we will omit the dependence on the label $i$, for notational convenience.

As we saw in Sec.~\ref{sssec:optset-param}, the parametrization of the optimal level set $\{s:I[s]=0\}$ is conveniently written in terms of the local Hessian eigenbasis $\{f^{(n)}\}$. In this basis, the order parameter becomes
\begin{equation}
    q_i(T) = \sum_{n=1}^\infty \qty(\expval{s_n^2} - \expval{s_n}^2).
    \label{eq:q_nspace}
\end{equation}
Thus, to estimate $q_i(T)$ we need to compute $\expval{s_n} = Z[s_n]/Z[1]$ and $\expval{s_n^2} = Z[s_n^2]/Z[1]$. 

Considering the approximations introduced in the last two sections, the exact partition function in Eq.~\eqref{eq:partfunc-exact} is now replaced by
\begin{align}
    Z[f] &\approx \int_{\mathbb R} \prod_t^L \dd s_t\, e^{-(3/2)(L/T)\int_0^T dt s_t^2} \delta(I(T)[s]) f[s].
    \label{eq:partfunc_approx}
\end{align}
We notice that the effective system size $L$ represents the dimensional cutoff in the path integral, namely, $\prod_t^L \dd s_t {\equiv} \dd s_{t_1} \dd s_{t_2} \dots \dd s_{t_L}$. As we are interested in the case of piecewise continuous protocols, we will ultimately consider the $L\to\infty$ limit.

We give the evaluation of the partition function in App.~\ref{app:q} to keep the discussion clear, and present here only the main results. Eventually, from Eq.~\eqref{eq:partfunc_approx}, we obtain the simple expression
\begin{align}
    q_i(T) &\sim \frac{1}{3 L}\sum_{n=1}^L (1+y_*\lambda_n)^{-1}, &  L &\to \infty.
    \label{eq:q_partfunc}
\end{align}
The quantity $y_*$ depends on $\Delta T$ and is restricted to the interval $(-\lambda_+^{-1},-\lambda_-^{-1})$, with $\lambda_{+,-}$ the positive and negative eigenvalues of largest absolute value. For $\Delta T \le 0$, we find that the eigenvalues $\{\lambda_n\}$ are positive and $y_*=\infty$ so that $q_i(T){=}0$.

From this result, we deduce that the jump discontinuity of the order parameter $q_i(T)$ essentially depends on the existence of the negative eigenvalue(s); in this situation, the quantity $y_*$ acquires a finite value and the set of optimal protocols $\{s:I[s]=0\}$ becomes degenerate.
To see this, we observe the two possible behaviors
\begin{align}
    (1+y_*\lambda_n)^{-1} &\to 0, & y_*\lambda_n \to \infty 
    \notag\\
    (1+y_*\lambda_n)^{-1} &\to 1, & y_*\lambda_n \to 0.
    \notag
\end{align}
Since $\abs{y_*} < \abs{\lambda_-^{-1}}$ and we have $\abs{\lambda_n} \to 0$ as $n,L\to\infty$ from general properties of the spectrum of the Hessian operator $J_{nm}$ (cf.~Sec.~\ref{sec:stability}), we deduce that $q_i(T)$ acquires a finite value once $\abs{\lambda_-^{-1}} < \infty$ exists (otherwise, $y_*=\infty$). In this case, we obtain $q(T) \to 1/3$ as $L\to\infty$ (see  App.~\ref{app:q} for details).

The above result has the following physical interpretation. 
The requirement of vanishing infidelity imposes a number $n_+$ of constraints on the $L$ degrees of freedom of the control protocol, when the quantum duration $T$ is larger than the quantum speed limit $T_\text{QSL}$. 
Therefore, in the $L\to\infty$ limit, the extra $L-n_+$ degrees of freedom introduced are redundant -- the control protocol is underconstrained. 
In the control landscape, these extra degrees of freedom are associated with \emph{soft modes}, deformation of optimal control protocols that preserve the infidelity within second-order. 
The order parameter $q(T)$ measures the average size of allowed deformations between different optimal protocols, so that its value grows with $L$, for $T>T_\text{QSL}$.
In particular, the limit value $1/3$ arises from the freedom left by the infidelity expansion truncated at the second-order on these extra degrees of freedom, as $L\to\infty$ (cf.~Eqs.~\eqref{eq:expval_s} and \eqref{eq:q_nspace}).

Considering higher-order terms in the infidelity expansion introduces additional constraints on the $L$ degrees of freedom. In a path-integral language, higher-order terms introduce interactions between the different modes that ultimately affect the value of the jump-discontinuity of $q(T)$. Thus, we understand that higher-order terms in the infidelity expansion are necessary to accurately predict the size of the jump-discontinuity of $q(T)$ at the quantum speed limit.

In conclusion, our analysis shows that the universal feature of the order parameter $q(T)$ is the presence of a jump-discontinuity, whereas the size of the jump depends on the details of the controlled quantum system.

\begin{figure}
\centering
\includegraphics[width=.5\textwidth]{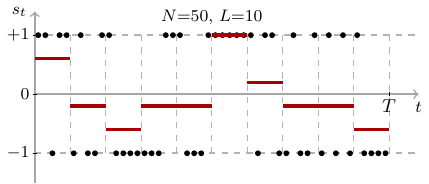}
\caption{
Schematic representation of the coarse-graining average performed on a bang-bang protocol with $N$ steps, into a piecewise-constant protocol with $L$ time steps. The average value in each bin depends on the number of $+1$ and $-1$ that it contains; in the limit $N{\to}\infty$, different bang-bang protocols can approximate the same coarse-grained continuous protocol. This redundancy is responsible for the density of state factor $\rho[s]$ appearing in partition function \eqref{eq:partfunc-exact-BB} and estimated in \eqref{eq:BB_coarsegrainaverage_dos}.
}
\label{fig:coarse_grain_average}
\end{figure}

\begin{figure}
\centering
\includegraphics[width=.5\textwidth]{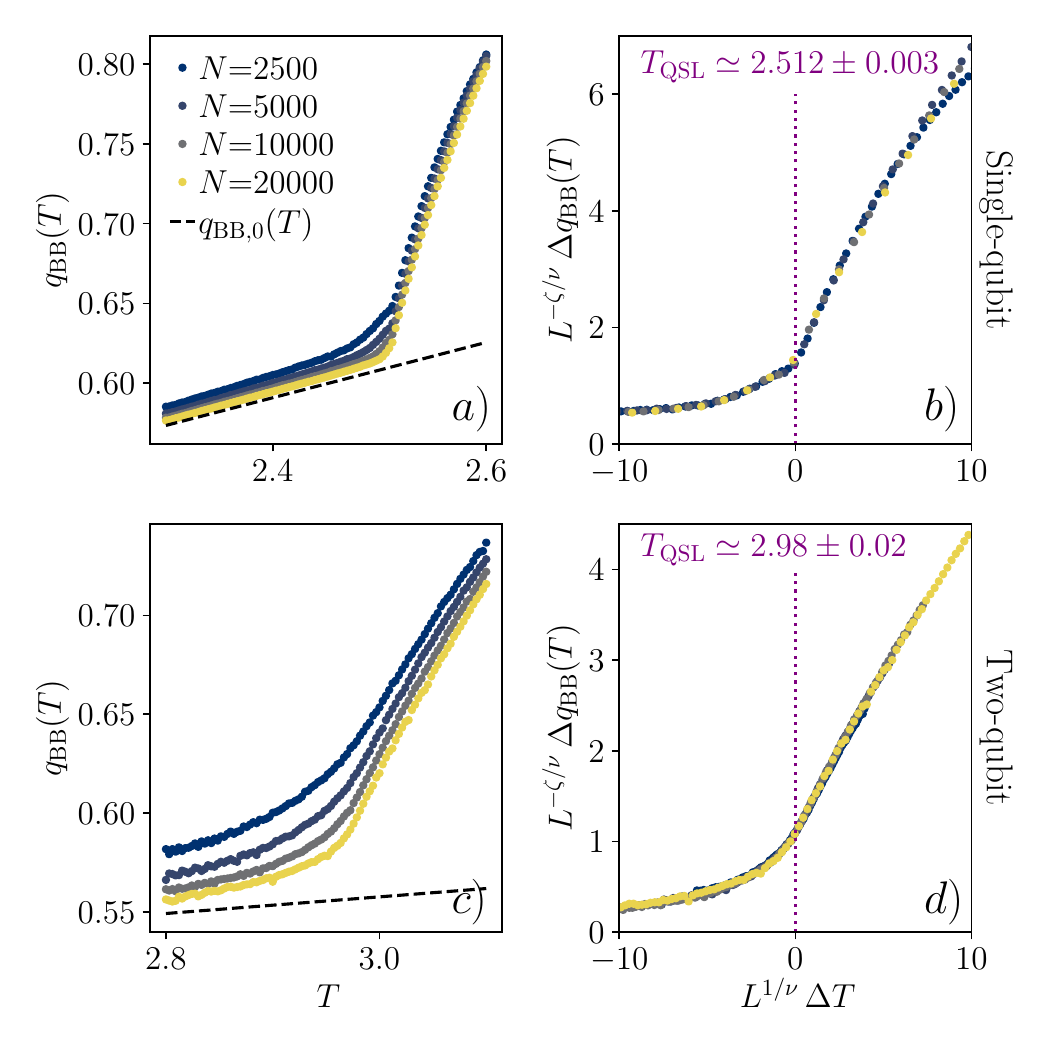}
\caption{
    Numerical estimate of the critical behavior of the order parameter $q_\text{BB}(T)$, for the single- and two-qubit system.
    Panels (\textbf{\textit{a}}) and (\textbf{\textit{c}}) show the curves $q_\text{BB}(T)$ obtained from Stochastic Descent with a finite number of bang-bang steps, near the quantum speed limit transition in $T_\text{QSL}$. 
    Panels (\textbf{\textit{b}}) and (\textbf{\textit{d}}) shows the collapsed curves $L^{\zeta_n/\nu}\Delta q_\text{BB}(L^{1/\nu}\Delta T)$, obtained by fitting $T_\text{QSL},\zeta_n,\nu$ and $q_\text{BB,0}(T)$ ($\Delta T {=} T {-} T_\text{QSL}$).
    The critical exponents $\zeta_n$ extracted in the two cases are in reasonable agreement with the analytical result in Eq.~\eqref{eq:qBB-partfunc-quartic} (see text).
    The scaling collapse locates the quantum speed limit $T_\text{QSL}$ with $\sim 0.1{-}1\,\%$ precision.
}
\label{fig:q_BB_critical}
\end{figure}

\subsection{Critical behavior of the order parameter \texorpdfstring{$q_\text{BB}(T)$}{qBB(T)}}
\label{ssec:partfunc_qBB}

In this section, we show that the partition function  introduced in the last section, defined over the continuous protocol space, can be used for the analytical estimation of the critical exponent of $q_\text{BB}(T)$, defined within the space of bang-bang control protocols.

The average $\expval{s_i}$, contained in Eq.~\eqref{eq:q_BB} defining the order parameter $q_\text{BB}(T)$, can be formally written as
\begin{equation*}
    \expval{s_i} = \sum_{\{s_i\}} w(\{s_i\}) s_i,
\end{equation*}
where the sum runs over the bang-bang control protocols space and $w(\{s_i\})$ are the statistical weights.

Nonetheless, in the limit of an infinite number of bang-bang steps $N{\to}\infty$,  a coarse-grain average of bang-bang protocols would give
\begin{equation*}
    \expval{s_i} = \int_{-1}^{+1} \prod_t \dd s_t\,  w[s] \rho[s] s_{t_i}
\end{equation*}
where the density of states $\rho[s]$ keeps track of how many bang-bang protocols are mapped to the same continuous protocol by the coarse-grain average procedure  (cf.~Fig.~\ref{fig:coarse_grain_average}). 

Therefore, we now consider the partition function
\begin{equation}
    Z_\text{BB} = \int_{-1}^{+1} \prod_t \dd s_t\, \delta(I[s]) \rho[s]
    \label{eq:partfunc-exact-BB}
\end{equation}
and estimate $\rho[s]$ in the limit $N{\to}\infty$ by subdividing the bang-bang protocol $\{s_i\}$ in $L$ bins, each containing $N/L$ consecutive bang-bang steps. Assuming the spins $s_i{=}{\pm}1$ in each bin are uncorrelated, each $s_i$ becomes an independent random variable, and the number of equivalent configurations within each bin only depends on the number density of up or down spins (cf.~Fig.~\ref{fig:coarse_grain_average}). 

In the limit $N,L{\to}\infty,N/L{\to}\infty$ we obtain
\begin{align}
    \rho[s]     &= \prod_t\rho(s_t) = e^{\sum_t S(s_t)}
    \label{eq:BB_coarsegrainaverage_dos} \\
    S(s_t)      &\sim -(N/L) \qty[s_t^+\log s_t^+ + s_t^-\log s_t^-],
    \notag
\end{align}
where $t$ labels the bin, $s_t^\pm{=}(1{\pm}s_t)/2$ is the density of up and down spins in the bin $t$ and $S(s_t)$ is the entropy associated with the variable $s_t$. Consequently, for a given protocol $s$, we obtain the density of states $\rho[s] {=} e^{S[s]}$ associated with the Shannon entropy
\begin{align}
    S[s] &= -\frac NT \int_0^T \dd t\,\qty[s_t^+\log s_t^+ + s_t^-\log s_t^-].
    \label{eq:shannon}
\end{align}

This result is independent of the particular binning subdivision: changing the number of bang-bang steps $s_i$ contained in each bin $s_t$ yields the same density of states provided that spins in each bin remain decorrelated in the limit $N{\to}\infty$.

\subsubsection{Shannon entropy approximation}

We would like to explicitly evaluate the partition function in Eq.~\eqref{eq:partfunc-exact-BB}. Unfortunately, the Shannon entropy makes the integral intractable by means of elementary analytical tools. Since we are interested in the simplest possible analytical estimate of the order parameter $q_\text{BB}(T)$, we search for an approximation of the Shannon entropy that allows the partition function to be evaluated by elementary means. 

To this end, we replace the nonlinear Shannon entropy with the quadratic function
\begin{equation}
    S[s] \mapsto S_2[s] = \frac{N\alpha}{2} \qty(1-\frac1T\int_0^T \dd t\,s_t^2)
    \label{eq:shannon-approx}
\end{equation}
where $\alpha$ is a positive free parameter. $S_2[s]$ is commonly known as the Tsallis entropy with entropic-index 2 \cite{tsallis1988possible}.
Notice that $\alpha$ can be fixed by imposing a further constraint (for example, $\alpha{=}1$ approximates the Shannon entropy density around $s_t{=}0$ up to second-order). Still, its precise value will not affect the critical scaling of $q_\text{BB}$, as we explain below.

Finally, following the Shannon entropy approximation, we extend the integration domain of $\{s_t\}$ from $[-1,1]$ to $\mathbb R$.

\subsubsection{Evaluating \texorpdfstring{$q_\text{BB}(T)$}{qBB(T)}}
\label{sssec:partfunc_evaluating_qBB}

Evaluating the partition function in Eq.~\eqref{eq:partfunc-exact-BB} within a second-order approximation of the infidelity expansion, we obtain the critical behavior (see App.~\ref{app:qBB})
\begin{align}
     \Delta q_\text{BB}(T) &= q_{\text{BB}}(T) - q_{\text{BB},*}(T) \sim c_0 + c_1\Delta T,
     \label{eq:qBB-partfunc-quadratic}
\end{align}
with $q_{\text{BB},*}(T)$ given by the analytical behavior of $q_{\text{BB}}(T)$ for $T<T_\text{QSL}$, and $c_{0,1}$ constant in $\Delta T$.
This result deviates from the continuous curve observed in Fig.~\ref{fig:q_BB_critical} by the presence of the constant term $c_0$. 
In App.~\ref{app:qBB}, we show that this jump discontinuity in the analytical prediction of $\Delta q_\text{BB}(T)$ is an artifact of the second-order truncation of the infidelity expansion.
In addition, we show that including quartic order terms yields (cf.~App.~\ref{app:qBB})
\begin{align}
     \Delta q_\text{BB}(T) = q_\text{BB}(T) {-} q_\text{BB,0}(T) &\sim c_2 \Delta T^{\zeta_a} + c_3 \Delta T^{2\zeta_a},
     \label{eq:qBB-partfunc-quartic}
\end{align}
with $c_{2,3}$ constant in $\Delta T$ and an undetermined critical exponent in the allowed interval $\zeta_a\in[0,1/2]$.
We notice that this interval for $\zeta_a$ is obtained within the approximation of the coarse-grain Shannon entropy, so that small deviations are possible.
Our analysis shows that the precise value of $\zeta_a$ depends on the details of the landscape, around the optimal control protocol(s) at the quantum speed limit. More precisely, $\zeta_a$ depends on the fourth-order terms in the infidelity expansion.

For comparison, in Fig.~\ref{fig:q_BB_critical} we perform a numerical scaling collapse around the quantum speed limit $T_\text{QSL}$ of the curves $q_\text{BB}(T)$ obtained in the single- and two-qubit example discussed in Sec.~\ref{sec:model} (see App.~\ref{app:qBB-scalcoll}). 
The results return the numerical critical scaling $\Delta q_\text{BB}(T) \sim \Delta T^{\zeta_n}$, with $\zeta_n\approx0.7$ and an estimated error of order $\sim 0.1$, in both cases.

Overall, we find the numerical critical exponent $\zeta_n$ in reasonable agreement with the analytically predicted range for $\zeta_a$.
The possible deviation between the two results may have different origins. 
Our analytical result relies on the approximation of the entropy associated with the coarse-grain average, where we neglected correlations at different times $t$ and adopted a quadratic approximation for the Shannon entropy.
Our numerical results are affected by systematic errors, which limit the precision of the estimated critical exponent (cf.~App.~\ref{app:qBB-scalcoll}).
Finally, in practice, the value of coefficients $c_2,c_3$ may be important for a precise comparison with the numerical results.

From our theoretical analysis, we deduce that the universal feature of the order parameter $q_\text{BB}(T)$ is the presence of a cusp, $\Delta q_\text{BB}(T) \sim \Delta T^{\zeta}$, with a fractional critical exponent $\zeta$, whose precise value depends on the details of the controlled quantum system.

In conclusion, let us reemphasize the simplicity and generality of our analysis of the landscape at the quantum speed limit transition.
Similar to Ginzburg-Landau's theory of phase transitions, our method focuses on the generic scaling behavior of the coefficients in the landscape expansion and therefore applies to generic controlled quantum systems.
Using this approach, we connect the changes in the geometry of the landscape to the non-analytic behavior of order parameters and explain their universal properties at the quantum speed limit transition.

\section{Extension to generic many-body controlled quantum systems}
\label{sec:manybody}

Although the formal extension of our methods to many-body controlled quantum systems is straightforward, the exponentially growing Hilbert space dimension (in the number of controlled degrees of freedom) affects their practical implementation. Therefore, we now briefly discuss the broader applicability of our methods to generic quantum control problems, highlighting some potential challenges and presenting useful ideas on how to tackle them.

As discussed in Sec.~\ref{sec:expansions}, the convergence of the landscape expansions is controlled by the norm of the Hamiltonian and the total protocol duration.
Therefore, for many-body quantum systems in particular, we anticipate a larger order of truncation in the landscape expansions to reliably capture the transitions in the control landscape.
Nevertheless, there exist classes of many-body problems possessing polynomially large algebras \cite{orozcoruiz2024}. In this case, the matrices associated with the generators of the dynamics are sparse, and higher-order terms of the expansion can be efficiently computed.
In addition, when interested in a particular region of the control protocols space, we can reduce the order of truncation by changing the center of the analytical expansion using, e.g., an educated guess for a (close-to-optimal) protocol.
This is particularly relevant for the study of CLPTs, as they originate from sudden changes in the subset of locally-optimal protocols. 

The adiabatic tracing method and the numerical exploration algorithm, used to study the structure of the optimal level set in Secs.~\ref{sec:stability} and \ref{sec:LMC}, require that (i) the evolution of the quantum system can be computed and (ii) locally-optimal protocols can be obtained through optimization procedure. 
This limits the applicability of these methods to quantum systems whose evolution can be efficiently performed; in addition, the optimization of control protocols has to be feasible, within numerical precision. 
Recently, it was demonstrated that both conditions are satisfied in many-body control problems with polynomially large algebras (in the number of controlled degrees of freedom): this allows one to numerically compute almost-optimal quantum control protocols that prepare macroscopically entangled Greenberger-Horne-Zeilinger states, and the topologically ordered ground state of the cluster Ising model without the need to store the quantum state \cite{orozcoruiz2024}. 
We also notice that a similar idea to adiabatic tracing was recently applied as an efficient optimization strategy for the Quantum Approximate Optimization Algorithm -- a parametrized ansatz used to control the evolution in quantum many-body circuits \cite{sack2023recursive,medina2024analytical}.
Finally, we remark that our methods are formally independent of the method used to evolve the quantum system: the evolution does not necessarily have to be \emph{exact} or evaluated on a \emph{classical} computer.
For instance, the quantum dynamics could also be evaluated experimentally on a quantum simulator, via circuit dynamics in a quantum computer, or using approximate numerical methods on a classical computer (such as time-dependent density matrix renormalization group or time-evolving block decimation for one-dimensional chains).
In summary, experimentally relevant many-body control problems exist where the techniques we develop can be applied. 

The analytical framework based on path integrals, introduced in Sec.~\ref{sec:Tqslcritical}, does not require detailed information about the quantum control problem.
Its predictions apply to control landscapes, where a finite number of locally-optimal protocols become global optima, as the protocol duration $T$ crosses the quantum speed limit.
Note that this requires a finite quantum speed limit~\footnote{The quantum speed limit may be infinite, e.g., when a large number of local minima preclude any practical optimization procedure from finding optima with an arbitrarily small infidelity.}. 

In conclusion, extending the methods developed in this work to many-body controlled quantum systems presents an exciting and challenging open direction for future research.

\section{Conclusion}
\label{sec:outro}

\subsection{Summary of the main results}

In this work, we combined different analytical and numerical techniques that shed new light on quantum control phase transitions. At its core, our contribution is fourfold. 

First,
we derive three different analytical expansions that approximate the infidelity functional (or similar objective functions) of the single- and two-qubit control problems and capture the underlying control phase transitions. In particular, the Dyson and cumulant expansions admit an effective energy interpretation where the protocol configuration plays the role of a spin- or field-configuration (depending on the constraints imposed on the protocol; cf.~bang-bang vs.~piecewise continuous protocol space).
Control phase transitions are captured by the approximate landscape, provided the order of truncation is large enough.
Nonetheless, we show that changing the center of the expansion allows one to reduce the order of truncation. 

Second, 
we introduce the adiabatic tracing method and study the near-optimal region of the control landscape as the protocol duration $T$ is varied.
Starting from an infinitesimal $T$ and incrementing its value adiabatically (\textit{i.e.}, while optimizing the control protocol), we relate control phase transitions to precise changes in the structure of the landscape around the optimal protocol(s).
This analysis relies on the knowledge of optimal protocol(s) of the landscape; in our case, we find them using either analytical ans\"atze (motivated by the landscape expansions) or numerical optimization algorithms (simulated annealing with LMC dynamics).
More generally, available tools in the quantum optimal control community (such as gradient-based or gradient-free optimizers or machine-learning methods) can be adapted to trace optimal protocols as the protocol duration is varied.
In this context, notice that the adiabatic tracing method can be regarded as a simulated annealing procedure, where, instead of the temperature $\beta^{-1}$, the annealing parameter is the control protocol duration $T$.
This suggests a new direction for improving currently available optimization procedures (both analytical and numerical) in quantum control problems.

Third, 
we use a gradient-free version of the Langevin-Monte Carlo algorithm (LMC) to explore the optimal level set, defined in the infinite-dimensional protocol space by the equation $I(T)[s]{=}0$. Using LMC, we characterize its topological/geometrical properties (connectedness and bulk sizes) and we obtain a numerical estimate of the order parameter $q(T)$. 
Notice that LMC explores the optimal level set by considering stochastic fluctuations at finite inverse-temperature $\beta$, in the space of piecewise-constant functions with $L$ steps. To overcome these numerical limitations, we explicitly check for finite-size effects by studying the results of the simulations while varying $L$ and $\beta$.

Fourth,
inspired by statistical field theory methods, we rephrase the optimal control problem in path integral language, and use this new formulation to estimate the critical behavior of order parameters $q(T)$ and $q_\text{BB}(T)$ for $T {\to} T_\text{QSL}$. 
This framework does not require detailed information about the specific quantum control problem. Instead, similar to the Ginzburg-Landau treatment of phase transitions, the method takes advantage of the analyticity of the cost function, and considers the most general scaling behavior for the coefficients in the landscape expansion.
This method allows us to connect the critical behavior of the statistical order parameters with both the changes occurring in the geometry of the control landscape and the constraints imposed on the protocol space. 
In addition, it captures the jump-discontinuity of $q(T)$ and the fractional critical exponent of $q_\text{BB}(T)$, raising the question about universality in the behavior of the order parameters at the quantum speed limit transition, irrespective of the controlled quantum system.

\subsection{Physical Significance of Control Landscape Phase Transitions}

Let us now turn to the physical implications of CLPTs for both the control of quantum systems and the optimization of control parameters.
In what follows, we discuss the three transitions at $T_c,T_\text{QSL},T_\text{sb}$ separately.

The $T_c$ transition signals the lack of local controllability of the quantum system.
At $T_c$, the shape of the optimal control protocol changes: from a bang-bang to a singular type (see Fig.~\ref{fig:stab_1q}, for an example).
The singular arcs occurring in the control protocol reflect the experimental limitations in selecting the control fields; in particular, they appear when the applied fields cannot fully steer the system's evolution 
\cite{dalessandro2021introduction,
chakrabarti2007quantum,
kirchhoff2018}.
Depending on the accessible family of control protocols, the $T_c$ transition may have a sizable effect on the optimization landscape and introduce a large degeneracy of locally-optimal protocols.
For the class of bang-bang control protocols we investigated, the $T_c$ transition sets the separation between a convex landscape and a landscape exhibiting multiple local minima.

The quantum speed limit transition ($T_\text{QSL}$) sets the boundary between an unsolvable (overconstrained) and a solvable (underconstrained) optimization problem; its existence determines whether the target state can be reached in a finite time, given the accessible controls.
Physically, the existence of a finite quantum speed limit depends on the physical constraints of the applied control fields, such as limited amplitude (as in our case), bandwidth, or total power output.
For a generic controlled quantum system, no efficient way to determine the quantum speed limit is known, which makes structural changes in the control landscape a viable alternative.
In our work, we show that order parameters capturing correlations between different optimal protocols exhibit universal features (independent of the controlled quantum system) at $T_\text{QSL}$.
Therefore, even though the quantum speed limit is typically characterized by the minimal cost achieved by optimal control protocols, here we precisely quantify how it affects correlations between optimal control protocols.
In addition, we show that a finite number of local minima, present before the transition, generate a multitude of optimal control protocols.
This observation allows us to approximately reconstruct the set of optimal protocols for $T>T_\text{QSL}$, using information from optimal protocols at $T\le T_\text{QSL}$; the procedure can have useful practical applications, as some minima within the optimal level set may have preferable properties (e.g., robustness to experimental noise), and can find application in experiments.

Finally, the $T_\text{sb}$ transition marks the onset of a control-symmetry broken phase. Here, optimal protocols violate the control symmetry dictated by the combination of the Hamiltonian and the initial and target states; consequently, for $T>T_\text{sb}$, the search for optima in the landscape cannot be restricted to the space of symmetric control protocols.
We emphasize that, unlike for $T_c$ and $T_\text{QSL}$, the existence of control symmetry breaking at $T=T_\text{sb}$ cannot be deduced a priori from any property of the quantum control problem alone; instead, it is observed via direct exploration of the control landscape. This is a prime example for the manifestation of the physics underlying CLPTs.
Moreover, our analysis shows that the $T_\text{sb}$ transition is associated with a characteristic instability in the infidelity landscape, which occurs and can be observed right before the transition.
This instability is a direct manifestation of the symmetry-breaking transition and provides information about the shape of optimal control protocols beyond the transition, which, in turn, are important for experimental optimal control.
Namely, even though there exist two degenerate optimal control protocols, the two trajectories explore different parts of Hilbert space, and they are inequivalent in terms of physical observables \cite{beato2025_topological}. In this case, blindly optimizing the infidelity may produce a protocol of suboptimal performance in experiments.

In summary, the CLPT at $T_c$ and $T_\text{sb}$ signals changes in the shape of optimal protocols, and can help in making an informed choice when selecting the control protocol family in experiments.
The CLPT at $T_\text{QSL}$ signals a significant reorganization of the structure of local minima in the optimal level set and exhibits features which are independent of the controlled quantum system. These properties help locate the quantum speed limit and approximately identify the set of optimal control protocols beyond the transition.
Overall, we have shown that CLPTs carry a direct significance for both controlled quantum systems and control variables used to manipulate them, and can be used to reveal and understand their physical properties.

\subsection{Outlook}

By proposing a general framework for an analytical theory of quantum control landscapes, our work sheds new light on the theory of optimal control, which allows one to go beyond purely numerical studies. The perturbative expansions we develop for the effective landscape models reveal the control propagator as a central object in control landscape theory: indeed, the landscape of any specific control problem can be obtained by projecting this propagator onto the initial and target states, emphasizing its fundamental importance. Crucially, these expansions provide means to analytically derive controlled approximations for the quantum control landscape.

A key insight brought in by our theory is the reconciliation of seemingly disparate control problems under a unified analytical framework: e.g., the landscapes of fidelity maximization, energy minimization, and observable extremization, all admit a natural analytical description within the same formalism. 
Another interesting feature is the natural separation of quantum (i.e., physical) and control degrees of freedom, which makes the classical nature of control phase transitions evident. 
Whether these transitions are of equilibrium (i.e., thermodynamic) or nonequilibrium (i.e., spin-glass-like) character, remains an open question for many-body control problems; nevertheless, in the single- or two-qubit problem, we show that (i) CLPTs are associated with critical changes in the optimal level set and (ii) local traps are absent or can be easily avoided; therefore, in these two cases, the landscape does not possess glassy complexity.
In addition, our theory allows us to handle on equal footing continuous and discrete (e.g., bang-bang) control protocols: while the latter result in theoretically appealing $\mathbb{Z}_2$ Ising models, the former are often more relevant for experiments. Importantly, the coupling constants in these classical landscape models do not depend on the family of control protocols used.

To analyze the critical behavior of the control landscape in the vicinity of the quantum speed limit ($T_\text{QSL}$), we develop new techniques based on statistical field theory.
For bang-bang protocols of duration less than $T_\text{QSL}$, we show that multiple local minima can appear in the landscape as a result of the entropy associated with the number of different bang-bang approximations to a single continuous optimal protocol; hence, the bang-bang control landscape may be sensitive to the details of the time discretization, and exhibits features not necessarily representative of the behavior of the controlled physical system. Our theory circumvents this issue by providing a uniform description for bang-bang and continuous drives. This is expected to shed new light on the performance of state-of-the-art algorithms, such as the quantum approximate optimization algorithm~\cite{farhi2014quantum,matos2021quantifying,sack2023recursive,medina2024analytical}.

Beyond $T_\text{QSL}$, we overcome the difficulty of taking into account boundary constraints (e.g., boundedness, bang-bang, etc.) analytically by developing a novel path integral approach, inspired by techniques used to study spin glass physics. This allows us to obtain an analytical estimate of the critical scaling behavior of the protocol order parameter in the vicinity of $T_\text{QSL}$.
Our results pave the way for future applications of statistical field theory methods in the characterization of quantum control landscapes and their phase transitions. 

While we focused on two toy models where we computed the coefficients of the control landscape exactly, we expect our methodology to be broadly applicable. The formalism suggests straightforward extensions with little-to-no modifications to optimally control open few-qubit systems subject to dissipation or decoherence (Lindblad control), or exactly solvable many-body systems, such as the transverse-field Ising model, which admit a mapping to a collection of independent two-level systems. Moreover, by considering quantum models in which $H_{0,1}$ are random matrices drawn from a Gaussian unitary ensemble, the ensemble average of the effective landscape model may give rise to a classical spin-glass system already for a single-qubit control problem (cf.~Sec.~\ref{sec:expansions}).
When it comes to investigating the properties of optimal control landscapes in nonintegrable quantum many-body systems, we believe one can develop suitable low-rank approximations to the control propagator \cite{lerose2021influence}, and then apply our techniques. Hence, our work lays the foundations for a comprehensive theory of control phase transitions.

\begin{acknowledgments}
We would like to thank F.~Balducci, S.~Cacciatori, M.~Ciarchi, L.~Piroli, M.~Radice, M.~Serbyn and X.~Turkeshi for insightful discussions.
Funded by the European Union (ERC, QuSimCtrl, 101113633). Views and opinions expressed are however those of the authors only and do not necessarily reflect those of the European Union or the European Research Council Executive Agency. Neither the European Union nor the granting authority can be held responsible for them.
M.B.~was supported by the Marie Sk\l{}odowska-Curie grant agreement No 890711 (until 01.09.2022). 
Numerical simulations were performed on the MPIPKS HPC cluster.
\end{acknowledgments}

\section*{DATA AVAILABILITY}
The data and code that support the findings of this article are openly available \cite{data_availability}.

\appendix

\section{Infidelity expansions}   \label{app:expansions}

In this appendix, we discuss in more detail the analytical expansions introduced in Sec.~\ref{sec:expansions}.
In the first two subsections, we introduce the infidelity Taylor expansion and show how it connects with the Dyson expansion. 
In the last subsection, we comment on the radius of convergence of the Dyson and Magnus expansions.

We assume a generic Hamiltonian with the form in Eq.~\eqref{eq:H(t)}. The procedure can be generalized to the case with multiple control parameters.

\subsection{Taylor expansions centered around the arbitrary protocol \texorpdfstring{$s_0(t)$}{s0(t)}}

In what follows, we adopt the round and square brackets convention to distinguish between, respectively, the parametric dependence on the quantum duration $T$ and the functional dependence on the control protocol $s$. Consider the Taylor expansion for the infidelity functional $I[s](T)$:
\begin{align}
I[s](T) &= I[s_0](T) + \int_0^T\dd{t_1} \fdv{I}{s(t_1)}\eval_{s_0} \delta s(t_1) + \notag\\ 
        +&  \frac{1}{2!}\int_0^T\dd^2t\, \frac{\delta I}{\delta s(t_1) \delta s(t_2)}\eval_{s_0} \delta s(t_1)\delta s(t_2) + \dots \notag \\
        &=\sum_{i=0}^\infty \frac{1}{n!} \int_0^T \dd^nt\, \frac{\delta^n I}{\delta s(t_1) {\cdots} \delta s(t_n)}\eval_{s_0} \delta s(t_1) {\cdots} \delta s(t_n)
\end{align}
where $\delta s(t) {=} s(t){-}s_0(t)$ denotes deviation from protocol $s_0(t)$ at time $t$. The generic variational derivative 
\begin{equation*}
    \frac{\delta^n I}{\delta s(t_1) \dots \delta s(t_n)}\eval_{s_0}
\end{equation*}
 can be obtained explicitly from Eq.~\eqref{eq:infid-def}. The evolution operator $\hat U[s](T,0){=}\mathcal T e^{-i \int_0^T \hat H(t)dt}$, acting on the left, evolves quantum states from time $t{=}0$ to time $t{=}T$. Notice that $\mathcal T$ represents the time-ordering operator. From the definition we obtain
\begin{gather}
\frac{\delta^n \bra{\psi_*} \hat U [s](T,0) \ket{\psi_0}}{\delta s(t_1) \cdots \delta s(t_n)} \eval_{s_0} = \notag\\
(-i)^n \bra{\psi_*} \mathcal T \hat U[s_0](T,t_1)\, \partial_s \hat H(t_1)\, \hat U[s_0](t_1,t_2) \cdots \times \notag\\
\hat U[s_0](t_{n-1},t_{n})\, \partial_s \hat H(t_n) \hat U[s_0](t_{n},0) \ket{\psi_0},
\label{eq:Uexp_nth}
\end{gather}
where $\partial_s \hat H$ denotes the partial derivative of $\hat H$ with respect to the control parameter $s$. The time-ordering operator $\mathcal T$ contained in the evolution operator ensures the correct time-ordering of the time labels $\{t_n\}$. As a consequence, the functional derivative of the unitary operator is symmetric under permutations of the time labels. For example, for $n{=}1$ we have
\begin{gather}
    \frac{\delta \bra{\psi_*} \hat U [s](T,0) \ket{\psi_0}}{\delta s(t_1)} \eval_{s_0} \notag\\ = (-i)\bra{\psi_*} \hat U[s_0](T,t_1)\, \partial_s\hat H(t_1) \hat U[s_0](t_1,0) \ket{\psi_0}.
    \label{eq:Uexp_linear}
\end{gather}
For $n{=}2$,
\begin{gather}
     \frac{\delta^2 \bra{\psi_*} \hat U [s](T,0) \ket{\psi_0}}{\delta s(t_1) \delta s(t_2)} \eval_{s_0} \notag\\ = (-i)^2 \bra{\psi_*} \hat U[s_0](T,t_1)\, \partial_s\hat H(t_1) \times \notag\\ \hat U[s_0](t_1,t_2)
    \partial_s\hat H(t_2) \hat U[s_0](t_2,0) \ket{\psi_0}
    \label{eq:Uexp_quadratic}
\end{gather}
where we assumed $t_1{>}t_2$ (if $t_2{>}t_1$ one exchanges $t_1 {\leftrightarrow} t_2$ in the right-hand side to preserve the correct time-ordering).
The infidelity Taylor expansion can be easily written in terms of the variational derivative of the unitary operator $\hat U[s](T)$ given above. Introducing the variable $z {=} \bra{\psi_*} \hat U[s](T,0) \ket{\psi_0}$ and its complex conjugated $\bar z$ as shorthand notation, we have $I[s](T) = 1{-}\bar z z$ and
\begin{align}
    \frac{\delta I}{\delta s(t_1)}\eval_{s_0} &= -2 \Re \Biggl[ \bar z\frac{\delta z}{\delta s(t_1)}\eval_{s_0} \Biggl] 
    \label{eq:infexp_linear} \\
    \frac{\delta^2 I}{\delta s(t_2)\delta s(t_1)}\eval_{s_0} &= -2 \Re \Biggl[\bar z \frac{\delta^2 z}{\delta s(t_2)\delta s(t_1)}\eval_{s_0} + \notag \\
    & \frac{\delta \bar z}{\delta s(t_2)}\eval_{s_0} \frac{\delta z}{\delta s(t_1)}\eval_{s_0} \Biggr].
    \label{eq:infexp_quadratic}
\end{align}
where $\Re$ indicates the real part. 
Notice that in $t_1 {=} t_2$ (or in $t_i {=} t_j$ when $n{>}2$) the variational derivative has a removable discontinuity. Our analysis is not affected by these discontinuities: in the functional expansions the derivatives appear under integration and the discontinuities involve a zero-measure set. 

In Sec.~\ref{sec:expansions} we conveniently rewrite the infidelity in terms of the orthogonal evolution operator (cf.~Eq.~\eqref{eq:fid-real}). Let us now consider the Taylor expansion in the orthogonal operator space. Repeating the same steps as above we obtain
\begin{gather}
\frac{\delta^n I[s](T)}{\delta s(t_1) \dots \delta s(t_n)} \eval_{s_0} = \notag \\
-(1/2)\vec n_*\cdot  \mathcal T M[s_0](T,t_1)\, \partial_sm\, M[s_0](t_1,t_2) \dots \times \notag \\
M[s_0](t_{n-1},t_{n})\, \partial_sm\, M[s_0](t_n,0)\, \vec n_0. 
\label{eq:infexp_dual_nth}
\end{gather}
where $m(t)$ is the generator of the orthogonal evolution operator associated with the Hamiltonian $\hat H(t)$. By comparison with the same expansion in the unitary operator space, we see that the orthogonal operator space allows a considerable simplification of each term in the series: the $n$-th order coefficients of the infidelity expansion simply corresponds to the $n$-th functional derivative of the infidelity. In particular, in the orthogonal operator space each term is simply given by an alternating product of evolution operators $M[s_0](t_i,t_{i+1})$ and generators $\partial_s m(t_i)$. For example, for $n{=}1$
\begin{align}
    \frac{\delta I[s](T)}{\delta s(t_1)} \eval_{s_0} &= -(1/2)\vec n_* \cdot M[s_0](T,t_1)  \times
    \notag\\
    &\partial_sm(t_1) M[s_0](t_1,0) \vec n_0
    \label{eq:infexp_dual_linear}
\end{align}
and for $n{=}2$, 
\begin{gather}
    \frac{\delta^2 I[s](T)}{\delta s(t_1) \delta s(t_2)} \eval_{s_0} = \notag \\ -(1/2) \vec n_* \cdot M[s_0](T,t_1)\, \partial_sm(t_1) \times \notag \\
     M[s_0](t_1,t_2) \partial_sm(t_2) M[s_0](t_2,0)\, \vec n_0
    \label{eq:infexp_dual_quadratic}
\end{gather}
when $t_1{>}t_2$.

\subsection{Connection between Taylor and Dyson expansions}

As in Sec.~\ref{sec:expansions}, let us now perform the reference-frame transformation
\begin{align*}
\hat H(t) \mapsto \hat H'(t) &= \hat U_0^\dagger (t,0)[\hat H(t) -i \partial_t] \hat U_0(t,0) \\
\hat U[s](T,0) &\mapsto \hat U_0(T,0) \hat U'[s](T,0)
\end{align*}
 where
 \begin{align*}
     \hat U_0(T,0) &= \exp(-i T \hat H_0) \\
     \hat U'[s](T,0) &= \mathcal T e^{-i \int_0^T \dd t \hat H'(t)}.
 \end{align*}
We obtain $\hat H'(t){=}U_0^\dagger (t,0) \hat H_1(t) \hat U_0(t,0)$ so that the new Hamiltonian $\hat H'(t)$ is proportional to the protocol $s(t)$. Consequently, for $s{=}0$ the new evolution operator $\hat U'[s](T,0)$ reduces to the identity operator: $\hat U'[0](T,0){=}\mathbb 1$. 
 
Similarly, in the dual space we perform the reference-frame transformation,
\begin{align*}
m(t) \mapsto m'(t) &= M_0(T,0)^\mathrm{t} [m(t) - \partial_t] M_0(T,0) \\
M[s](T,0) &\mapsto M_0(T,0) M'[s](T,0)
\end{align*}
where 
\begin{align*}
    M_0(T,0) &= \exp(T m_0) \\
    M'[s](T,0) &= \mathcal T e^{\int_0^T\dd{t}\, m'(t)}.
\end{align*}
Again, the new generator $m'(t)$ in the orthogonal operator space is proportional to the protocol $s(t)$: $m'(t) {=} M_0(T,0)^\mathrm{t} m_1(t) M_0(T,0)$. Hence, for $s{=}0$ the evolution operator reduces to the identity operator: $M'[0](T,0){=}\mathbb 1$.

The properties $\hat U'[0](T,0){=}\mathbb 1$ and $M'[0](T,0){=}\mathbb 1$ considerably simplify the coefficients appearing in the infidelity expansion. From Eqs.~\eqref{eq:Uexp_nth} and ~\eqref{eq:infexp_dual_nth} we obtain,
\begin{gather}
\frac{\delta^n \bra{\psi_*} \hat U' [s](T,0) \ket{\psi_0}}{\delta s(t_1) \cdots \delta s(t_n)} \eval_{s_0} = \notag \\
(-i)^n\bra{\psi_*} \hat U_0(T,0) \mathcal T \partial_s \hat H'(t_1) \cdots \partial_s \hat H'(t_n) \ket{\psi_0}
\label{eq:Uexp0_nth}
\end{gather}
and
\begin{gather*}
\frac{\delta^n I[s](T)}{\delta s(t_1) \dots \delta s(t_n)} = \notag \\ 
-(1/2)\vec n_* \cdot M_0(T,0) \mathcal T \partial_sm'(t_1) \cdots \partial_sm'(t_n) \vec n_0
\end{gather*}
Notice that for $n{=}1,2$ this last result corresponds to the ``interaction terms" in Eq.~\eqref{eq:infid-exp-0-coeffs}; this expression generalizes the Dyson expansion terms presented in Sec.~\ref{sec:expansions} to an arbitrary order $n{\in}\mathbb N$.

\subsection{Convergence of the landscape expansions}

The convergence properties of the landscape expansion introduced in Sec.~\ref{sec:expansions} sensibly depend on the properties of the controlled quantum system and the space of allowed control protocols.
The quantum control problems introduced in Sec.~\ref{sec:model} are characterized by the following two properties: (i) they have a finite-dimensional Hilbert space and (ii) are defined for bounded control protocols, $\abs{s(t)}{\le}1,\,\forall t\in[0,T]$. As a consequence, in our case, the norm of the Hamiltonian is bounded at each time, $\max_{t}\norm{H(s(t))}<\infty$.

In this subsection, for the sake of generality, we can relax hypothesis (ii) and allow the more general class of control protocols satisfying
\begin{equation}
    C_H \equiv \int_0^T \dd t\,\norm{H(s(t))} < \infty.
\end{equation}
This second class contains the class of bounded control protocols, as in the latter case we have $C_H = T \max_{t}\norm{H(s(t))} < \infty$.

\subsubsection{Dyson and Taylor expansions}

Under the above assumptions, the Dyson (and Taylor) expansions are absolutely convergent for any fixed $T{<}\infty$.
To see this, consider the generic evolution operator
\begin{align*}
\hat U(T,0) &= \mathcal T_t \exp(-i\int_0^T \dd t\, \hat H(t))\qquad \mathrm{or}  \\
M(T,0) &= \mathcal T_t \exp(\int_0^T \dd t\, m(t)) 
\end{align*}
and observe that each term in the associated Dyson series is bounded. For example,
\begin{align*}
\norm{\mathcal T_t \exp(\int_0^T \dd t\, m(t))} &= 
\norm{\sum_{n=0}^\infty \frac{1}{n!} \int_0^T \dd^nt\, \prod_{i=0}^n m(t_i)} \\
&\le \sum_{n=0}^\infty \norm{\frac{1}{n!} \prod_{i=1}^n \int_0^T \dd t_i\, m(t_i)} \\
&\le \sum_{n=0}^\infty \frac{1}{n!} \qty(\int_0^T \dd t\, \norm{m(t_i)})^n.
\end{align*}
where we used the triangular inequality and the sub-multiplicative property of the operator (matrix) norm.
Thus, for any finite $T$ the Dyson and Taylor expansions are bounded in norm by the absolutely convergent numerical series
$$
\sum_{n=0}^\infty a_n = \sum_{n=0}^\infty \frac{(C_m)^n}{n!} = e^{C_m}.
$$

From this result, we extract (i) a \emph{conservative} estimate of the order of the expansion, $P$, after which higher-order terms decrease in norm, and (ii) a bound on the error associated with a finite order of truncation, $\Delta I_\text{err}$.

Let $P$ be the integer for which the numerical series $\{a_n\}$ achieves its maximum value, $P=\mathrm{argmax}_{n\in\mathbb N}\abs{a_n}$.
In particular, we estimate $P$ by imposing the condition $a_{P-1}{=}a_{P}$, that yields 
\begin{equation}
    P = C_m.
    \label{eq:P}
\end{equation}
Thus, $P$ grows linearly with $C_m$ (cf.~$C_m{=}T\max_{t}\norm{m(s(t))}$, for bounded protocols).
For example, in the single-qubit problem in Eq.~\eqref{eq:1q-m}, we have $\max_s\norm{m(s)}{\simeq}h_x{=}\sqrt5$ and $T_\text{QSL}{\simeq}2.5$ so that $P(T_\text{QSL}){\simeq}6$.
In the two-qubit case in Eq.~\eqref{eq:2q-m}, $\max_s\norm{m(s)}{\simeq}\sqrt2h_x{=}\sqrt{10}$ and $T_\text{QSL}{\simeq}3.0$ yield $P(T_\text{QSL}){\simeq}10$.
In both cases, the analytical estimate \eqref{eq:P} returns $P(T_\text{QSL})>3$, consistent with the numerical results in Fig.~\ref{fig:SD}.

Let $N\in\mathbb N$ be the finite order of truncation of the Dyson expansion. A bound on the error $\Delta I_\text{err}$ between the exact infidelity and the truncated Dyson expansion is given by the norm of the terms neglected by the truncation,
\begin{align*}
    \Delta I_\text{err} &= \norm{\sum_{n=N+1}^\infty \frac{1}{n!} \int_0^T \dd^nt\, \prod_{i=0}^n m(t_i)} \\
    &\le \sum_{n=N+1}^\infty a_n = \sum_{n=N+1}^\infty \frac{(TC)^n}{n!}.
\end{align*}

\subsubsection{Magnus expansion}

The well-known sufficient condition for the convergence of the Magnus expansion reads \cite{blanes2009magnus}
\begin{equation}
\int_0^{T} \dd t\, \norm{m(t)} \le \pi.
\label{eq:magnus_convergence_crit}
\end{equation}
In the case of Eq.~\eqref{eq:magnus_exp}, this yields a relatively short radius where convergence is guaranteed.
In particular, for the single- and two-qubit problem we have, respectively, $T{\lesssim}1.4$ and $T{\lesssim}1.0$ (to be compared with $T_\text{QSL}{\simeq}2.5$ and $T_\text{QSL}{\simeq}3.0$). 

However, the following remarks are in order. 
First, Eq.~\eqref{eq:magnus_convergence_crit} gives a \emph{sufficient} condition so that the Magnus expansion may still converge even when the bound is violated \cite{bukov2015universal}.
Second, the right-hand side of Eq.~\eqref{eq:magnus_convergence_crit} depends on the specific protocol $s(t)$ appearing in $m(t)$. Consequently, although not uniformly, the Magnus expansion may still converge in a subset of the protocol space. 
As a specific example, consider the single-qubit problem, where the optimal protocol $s_{\Delta_0(T)}$ for $T{\le}T_\text{QSL}$ yields (cf.~Sec.~\ref{sec:stability})
$$
\int_0^{T} \dd t\, \norm{m(t)[s_{\Delta_0(T)}]} = T_c h_x < \pi, \quad T{\le}T_\text{QSL}.
$$ 
Third, remember that the Magnus expansion can be centered around an arbitrary protocol $s_0(t)$ via reference-frame transformation. This property can be used to obtain a convergent Magnus expansion around a desired control protocol.

\subsubsection{Cumulants expansion}

We use the cumulants expansion to obtain, from the Magnus expansion, an infidelity expansion with scalar-valued coefficients; hence, the sufficient criterion \eqref{eq:magnus_convergence_crit} also applies in this case.

There are no other constraints on the radius of convergence set by the cumulants expansion.
To see this, observe that (i) the cumulants expansion is a Taylor expansion of the quantity $\log(1-I(T)[s])$, $I(T)[s]\in[0,1]$ and (ii) the radius of convergence of the Maclaurin series of the function $\log(1+x),x\in\mathbb R$ is unity.

\section{Stochastic Descent (SD) simulations}
\label{app:SD}

In this section we provide additional information on the Stochastic Descent simulations (SD) used in Sec.~\ref{sec:expansions} to sample locally-optimal bang-bang protocols in the quantum control landscape.

In the Stochastic Descent algorithm, the initial bang-bang protocol is drawn from a uniform distribution site-per-site: $s_i{=}\pm1$. At each iteration, a spin-flip $s_j {\mapsto} {-}s_j$ at random site $j$ is performed; the modified protocol is accepted whenever it leads to decrement in the infidelity; otherwise the spin-flip is undone. The algorithm terminates when a single spin-flip in all sites does not decrease infidelity. For each $T$ we collect 250 locally-optimal protocols from 250 independent SD runs. From the set of locally-optimal protocols, $\mathcal S$, we estimate the average protocol $\expval{s_i}_{\mathcal S}$ and compute $q_\text{BB}(T)$ (cf.~Eq.~\eqref{eq:q_BB}).  Notice that we do not observe any sensible change in the curve $q_\text{BB}(T)$ by doubling the number of locally-optimal protocols sampled. 
In the exact infidelity landscape, we let SD search in the space of bang-bang protocols with $N_\text{ex}{=}5000$ bang-bang steps. In the case of the approximated infidelity landscape, SD searches in the space of bang-bang protocols with $N{=}500$ bang-bang steps. The different numbers $N_\text{ex},N$ of bang-bang steps in the two infidelity landscape (exact and approximated, respectively) is motivated by the different scalings in the number of bang-bang steps exhibited by the numerical $q_\text{BB}(T)$ curve. Intuitively, the different scaling behavior is related to the different type of discretization: in the exact infidelity the bang-bang protocol directly enters the unitary operator; instead, the infidelity expansions are derived in the more general space of piecewise continuous protocols, and the bang-bang constraint is imposed after the expansion is performed.

\begin{figure}
    \centering
    \includegraphics[width=.237\textwidth]{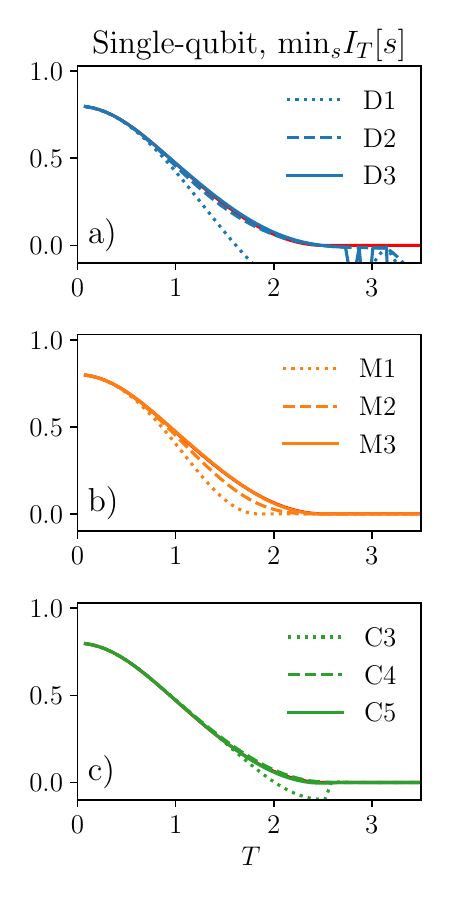}
    \includegraphics[width=.237\textwidth]{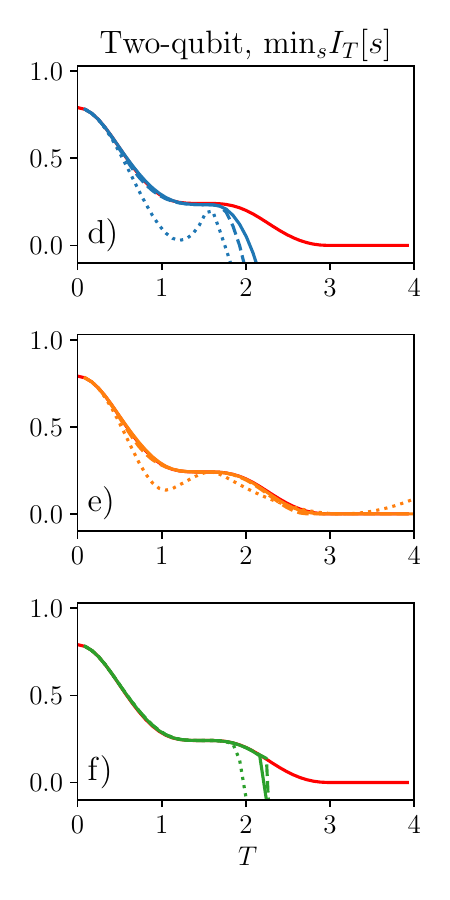}
    \caption{
    Stochastic Descent results complementing Fig.~\ref{fig:SD}.
    In the plot, we compare the minimum infidelity $\min_sI(T)[s]$ achieved in the approximated control landscapes with respect to the exact control landscape (showed in red). 
    Magnus is the only expansion preserving infidelity's domain $[0,1]$ and closely following the exact (red) curve. The Dyson and cumulants expansions break by construction the unitarity of the evolution operator and this leads to strong deviations in the corresponding curves (for $T{\gtrsim} 2.5$ and $T{\gtrsim}2.0$ in the single- and two-qubit problem, respectively). The letters D, M, and C in the legend stand for Dyson, Magnus, and Cumulant expansion while the adjacent integer specifies the order of truncation.
    }
    \label{fig:app_SD_inf}
\end{figure}

In Secs.~\ref{ssec:SD_1Q},\ref{ssec:SD_2Q} we showed the comparison between the curve $q_\text{BB}(T)$ obtained from SD performed over the exact infidelity landscape and the approximated infidelity landscape. It is also interesting to compare the minimum infidelity obtained. 
In the single-qubit case, from Fig.~\ref{fig:app_SD_inf} (left column) we see that for small $T$ all the expansions at the highest order have minimum infidelity close to the exact landscape result. For $T{\gtrsim}3.0$, the Dyson expansion (panel a) breaks the unitarity of the evolution operator and infidelity is not bounded within $[0,1]$. On contrary, the Magnus and cumulants expansions  (panel b and c) approximate the minimum infidelity curve in the whole interval $T{\in}[0,3.5]$.
In the two-qubit case, the Dyson and cumulant expansions  (panel d and e) break down beyond $T{\simeq}2$ while the Magnus expansion (panel f) reproduces the exact landscape minimum infidelity curve in the whole interval $T{\in}[0,4]$.

\begin{figure}
    \centering
    \includegraphics[width=.49\textwidth]{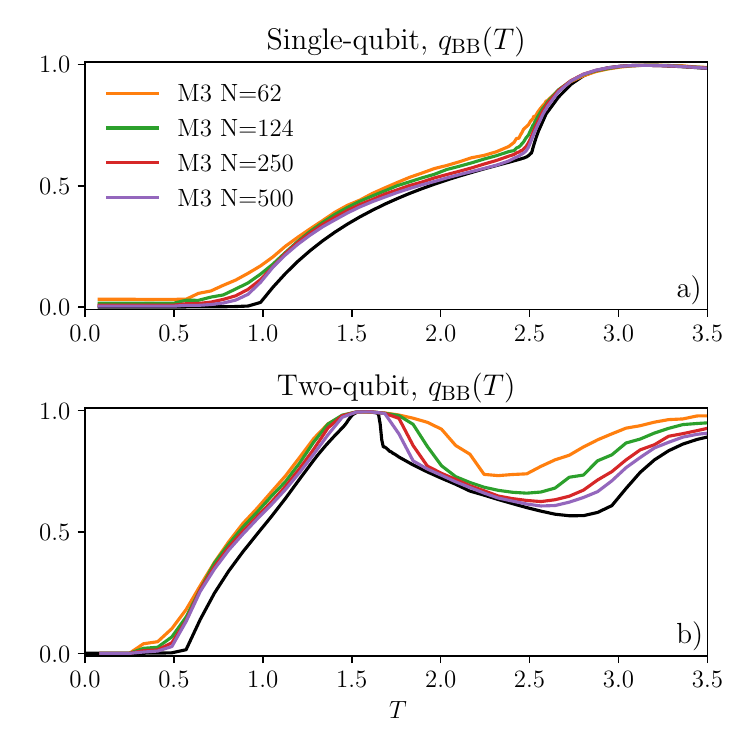}
    \caption{
    Stochastic Descent results complementing Fig.~\ref{fig:SD}.
    Finite-size scaling of the curve $q_\text{BB}(T)$ obtained via Magnus expansion, as the number of bang-bang steps $N$ is increased. For reference, in black we report the curve $q_\text{BB}(T)$ obtained in the exact control landscape (with $N_\text{ex}{=}5000$ bang-bang steps). In the legend, ``M3" stands for Magnus expansion truncated at third order.
    }
    \label{fig:app_SD_Nscaling}
\end{figure}

Let us now show how the number of bang-bang steps $N$ affects the order parameter curve $q_\text{BB}(T)$. In Fig.~\ref{fig:app_SD_Nscaling} the curve $q_\text{BB}(T)$ is plotted for the Magnus expansion truncated at third order for different numbers of bang-bang steps $N$.
As $N$ increases, the main change in $q_\text{BB}(T)$ occurs around the critical points: $T_c,T_\text{QSL}$ for the single-qubit  (panel a) and $T_\text{sb},T_\text{QSL}$ for the two-qubit problem  (panel b). Notice that in the two-qubit case, convergence of $q_\text{BB}(T)$ is significantly slower around the critical points $T{=}T_\text{QSL}$. Nevertheless, in both cases the qualitative behavior of the curve $q_\text{BB}(T)$ is already visible for $N {\gtrsim} 250$.

\begin{figure}
    \centering
    \includegraphics[width=.45\textwidth]{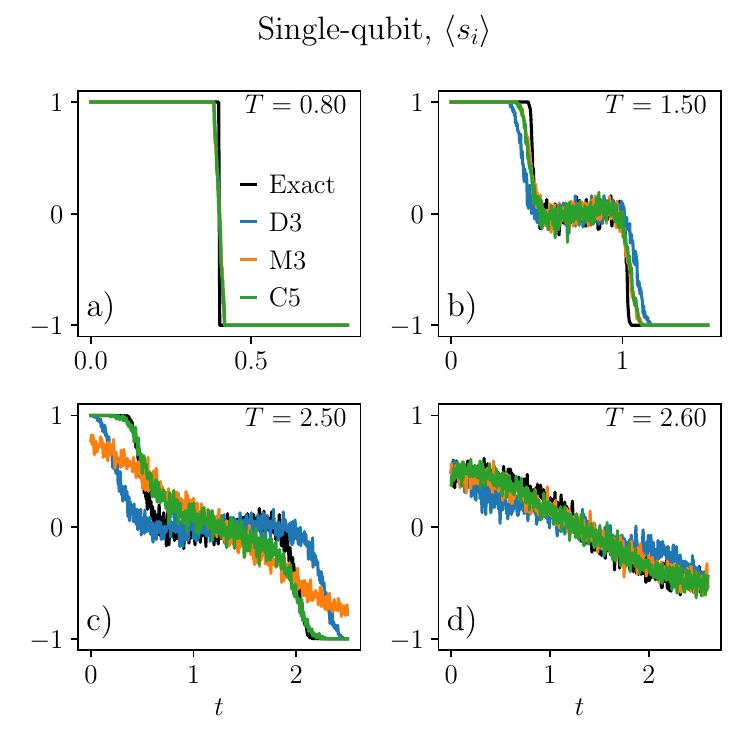}
    \caption{
    Single-qubit problem: average protocols $\expval{s_i}$ for different quantum durations $T{=}0.80,1.50,2.50,2.60$. The average is performed over 250 bang-bang protocols sampled by SD in the exact and approximated control landscapes -- D3: Dyson expansion truncated at third order; M3: Magnus expansion truncated at third order; C5: cumultants expansion truncated at fifth order (cf.~Sec.~\ref{sec:expansions}).
    Interestingly, from panel (\textbf{\textit{c}}) we observe that the shape of optimal protocols may qualitatively differ in the approximated and exact landscapes. This suggests some care when using the infidelity expansions to predict the shape of optimal protocols, even when they reproduce quantitatively well the behavior of other observables (such as $q_\text{BB}(T)$ in Fig.~\ref{fig:SD}).
    }
    \label{fig:app_SD_optprot}
\end{figure}

Finally, in Fig.~\ref{fig:app_SD_optprot} we directly compare the average protocols $\expval{s_i}_{\mathcal S}$ extracted from the different SD simulations in the single-qubit problem. Besides the order parameter $q(T)$ and fidelity $F(T)$, which effectively reduce the complexity of the system to a single value, in this plot we observe precisely how much the approximated landscapes reproduce the correct optimal protocol. Even though the Magnus expansion at third order better approximates the order parameter curve $q_\text{BB}(T)$, the average sampled protocols in Magnus do not always follow the qualitative behavior of the exact result (cf.~Fig.~\ref{fig:app_SD_optprot}c). On the contrary, the average protocols sampled by the Dyson and cumulants expansions possess the qualitative structure of the exact result. From this result, we understand that the quality of the approximation associated with the analytical expansions may sensibly depend on the features one is interested in comparing.

\section{Quadratic stability analysis} 
\label{app:quadratic_scaling}

In this section, we provide details about the quadratic stability analysis presented in Sec.~\ref{sec:stability}. The analysis is based on the numerical diagonalization of the Hessian operator $J_{t_1t_2}(T)$ appearing in Eqs.~\eqref{eq:infid-exp-0-coeffs} and \eqref{eq:hess-eigeq}. Numerically, we discretize the operator in an $L\times L$ matrix and account for the scaling prefactor $1/L$. 

\begin{figure}
    \centering
    \includegraphics[width=.5\textwidth]{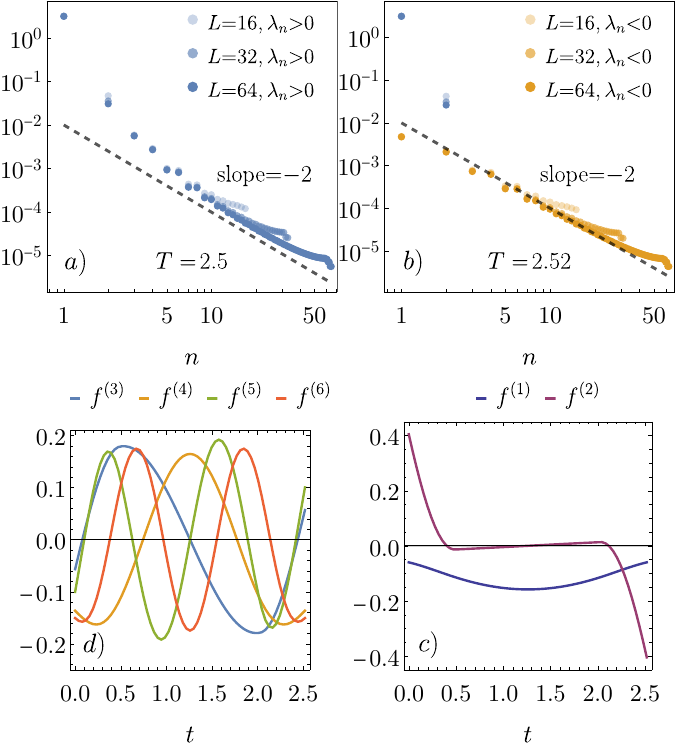}
    \caption{
    Single qubit-problem. 
    (\textbf{\textit{a}}),(\textbf{\textit{b}}) Spectrum of the discretized second-order operator in Eq.~\eqref{eq:infid-exp} relative to the protocol $s_{\Delta_0(T)}$ (cf.~Eq.~\eqref{eq:stab_s_Delta}), optimal for $T{\to}T_\text{QSL}^-{\simeq}2.51$. We observe the characteristic scaling of eigenvalues $\lambda_n{\sim}n^{-2}$ in the $L{\to}\infty$ limit.
    (\textbf{\textit{c}}),(\textbf{\textit{d}})
    Eigenfunctions associated with the two positive and the first four negative eigenvalues at $T{=}2.52$. Notice the close similarity to the Fourier basis.
    }
    \label{fig:app_quadratic_scaling}
\end{figure}

First, we show how the Hessian eigenvalues scale as a function of the discretization points. 
Let us consider explicitly the single-qubit problem and the quadratic operator relative to the protocol $s_{\Delta_0(T)}$, defined in Eq.~\eqref{eq:Delta} and optimal in the $[T_c,T_\text{QSL}]$ interval. In Fig.~\ref{fig:app_quadratic_scaling}a,b we show how eigenvalues scale as $L{=}16,32,64$ at $T{=}2.50,2.52$. The convergence in $L$ is relatively fast both for positive and negative eigenvalues. Notice the change in sign happening exactly after $T_\text{QSL}$: beyond this critical point, $L{-}2$ negative eigenvalues are present. It is also interesting to visualize the eigenfunctions associated with the numerical diagonalization. In Fig.~\ref{fig:app_quadratic_scaling}c,d we show the first six eigenfunctions of the quadratic operator for $L{=}64,T{=}2.52$. Except for the eigenfunction $f^{(2)}$, the eigenbasis $\{f^{(n)}\}$ appears to be a (relatively small) deformation of the standard Fourier basis on the $[0,T]$ domain. The role of the $f^{(2)}$ eigenfunctions becomes clear when considering deformations of $s_{\Delta_0(T)}$ along the directions of the eigenfunctions (see App.~\ref{app:optman_param}).

\begin{figure}
    \centering
    \includegraphics[width=.5\textwidth]{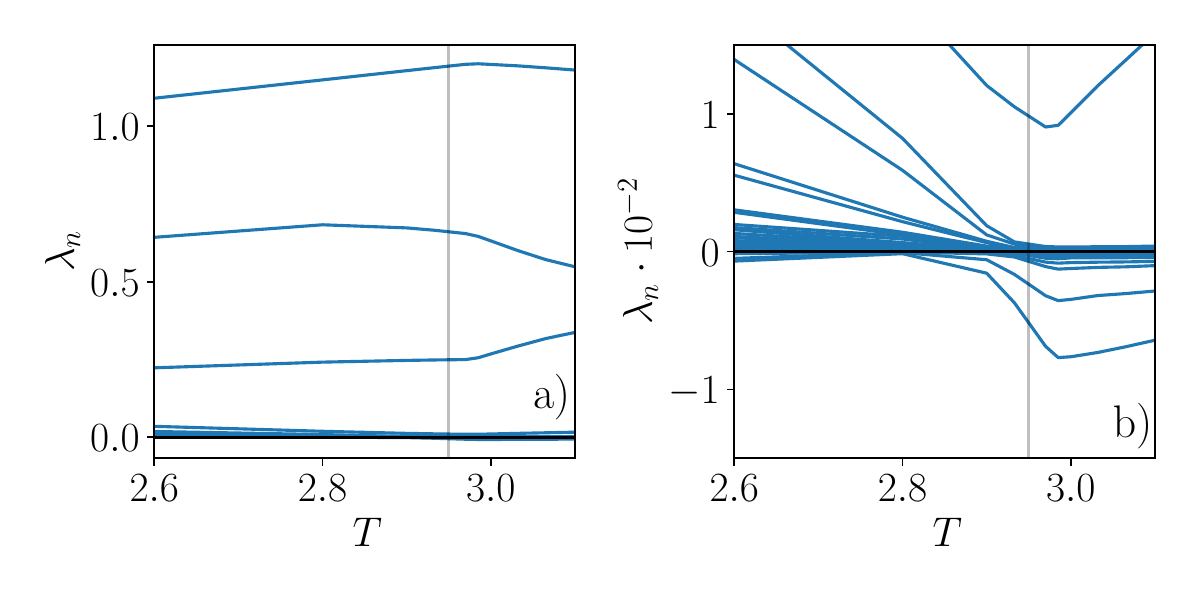}
    \caption{
    Two-qubit problem: spectrum of the Hessian operator around $T_\text{QSL}{\simeq}2.95$ (vertical gray line).
    The infidelity expansion is centered around one of the \emph{two} optimal protocols found by the LMC algorithm introduced in Sec.~\ref{sec:LMC} (the spectrum is identical for the two optimal protocols as discussed in App.~\ref{app:quadratic_scaling}).
    From the plot, we see $L-4$ eigenvalues vanishing for $T{\to}T_\text{QSL}$, up to deviations due to numerical errors.
    The result is analogous to the single-qubit case, shown in Fig.~\ref{fig:stab_1q_Delta}.
    }
    \label{fig:app_stab_2q_quad}
\end{figure}

Second, in Fig.~\ref{fig:app_stab_2q_quad} we show how eigenvalues of the Hessian operator change around the $T_\text{QSL}{\simeq}2.95$ in the two-qubit problem. 
In particular, the infidelity expansion is centered around one of the two isolated optimal protocols found numerically by the LMC algorithm, for each fixed $T$ (cf.~App.~\ref{sec:LMC}).
In LMC, the time interval $[0,T]$ is divided in $L{=}64$ uniform points and protocols are then piecewise constant on $L$ steps. 
To reduce finite size effects in the computation of the spectrum, the Hessian operator is instead discretized in $L'{=}256$ steps: the optimal protocol with $L'$ steps is obtained by interpolating the optimal protocol with $L$ steps found by LMC (using function \texttt{interp} from Python's package NumPy).
Notice that, for $T_\text{sb}{\le}T{\le}T_\text{QSL}$, the two isolated optimal protocols present in the landscape are related to each other via the control problem symmetry $s(t) {\leftrightarrow}{-}s(T{-}t)$ \cite{Bukov18_Broken}. As the infidelity is invariant under this symmetry transformation, the Hessian operators relative to the two optimal protocols possess the same spectrum.
The Hessian spectrum shows qualitatively similar behavior to the single-qubit case (cf.~Fig.~\ref{fig:stab_1q_Delta}a) except for the number of non-vanishing eigenvalues: four for the two-qubit, two for the single-qubit. 
Notice that, in the two-qubit case, the $L{-}4$ eigenvalues are not exactly zero for $T{\ge}T_\text{QSL}$, since the protocol chosen as the center of the infidelity expansion (at each fixed $T$) is optimal within numerical error; in this case, the error depends on the finite number of discretization steps, $L,L'$, and the finite value of the parameter $\beta$ appearing in the LMC update rule \eqref{eq:LMC_update_metropolis}.

\section{Gradient-free Langevin-Monte Carlo (LMC) simulations } 

In this appendix we present in more detail results from the LMC dynamics. We divide the discussion into three sections, respecting the three consecutive stages of LMC dynamics: thermal relaxation, diffusion along the optimal level set valleys and equilibrium sampling, as each one of the three stages is affected by CLPTs.

First, let us clarify the reason behind the simple dynamics of the algorithm. We discarded the gradient term, normally present in the Langevin update rule, because we are interested in exploring the optimal ($I(T)[s]{=}0$) level set, where the gradient term (as well as $L{-}n_+$ eigenvalues of the Hessian operator, with the positive integer $n_+$ dependent on the quantum control problem) are expected to vanish (cf.~Sec.~\ref{sec:stability}). Reinstating it results in a faster convergence to the optimal level set in the first part of the LMC simulations. However, its contribution during the level set exploration is expected to be less relevant.

Second, this algorithm uses purely stochastic moves to explore the optimal level set. Note that the update rule above is likely not the most efficient one: for example, one may exploit the Hessian analysis and restrict the stochastic move to the Hessian eigenspace associated with vanishing eigenvalues. This modification would probably result in a higher acceptance rate during the Metropolis rule (since in the update rule the infidelity would be preserved up to second-order corrections), albeit at the cost of requiring more computational time due to implementing an update in a rotated basis. Nevertheless, as a first approach, we prefer to avoid any bias in the numerical exploration of the optimal level set at the cost of a less efficient algorithm.

We tested LMC dynamics varying parameters in the following intervals: $T {\in} [0,3]$, $\sigma{\in}[10^{-4},10^{-3/2}]$ and $\beta{\in}[10^2,10^8]$. At the beginning of each LMC run, the initial protocol configuration $s_0=(s_{0,1},\dots,s_{0,L})$ is chosen randomly by drawing each component $s_{0,i}$ from the uniform distribution in $[-1,1]$. We define a single LMC iteration as $L$ local attempted moves.

\subsection{Thermal relaxation stage} 
\label{app:LMC_T}

\begin{figure}
\centering
\includegraphics[width=.45\textwidth]{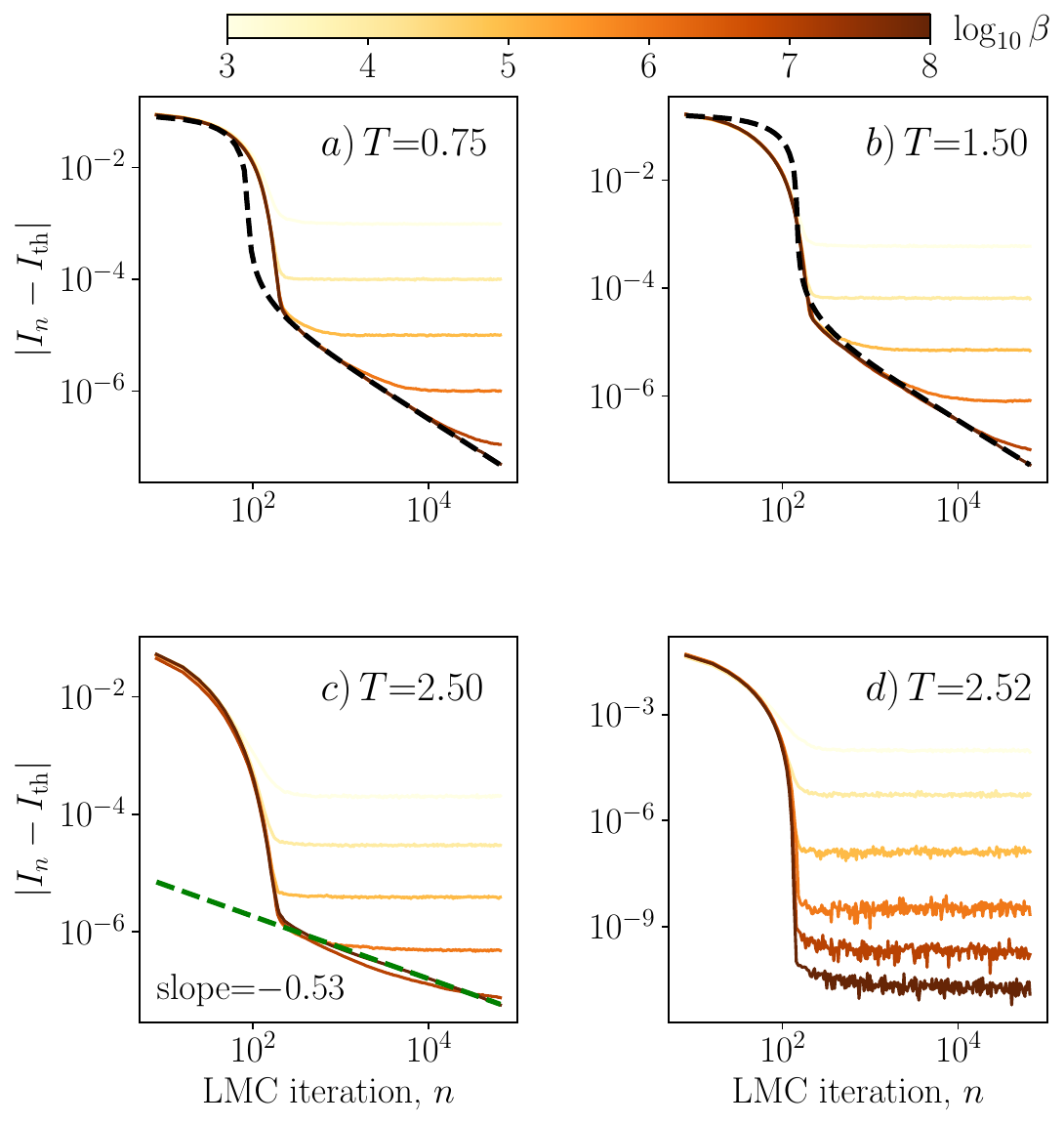}
\caption{
Single-qubit problem. The plots show the infidelity during thermal relaxation, averaged over ten independent LMC runs. The parameter $T$, different in the four panels, is changed to investigate the effects of CLPTs (cf.~$T_c{\simeq}0.98,T_\text{QSL}{\simeq}2.51$). Each curve refers to a different Monte Carlo inverse-temperature $\beta$ while $\sigma{=}10^{-3/2}$, $L{=}512$ are fixed. In particular, we plot the absolute value of the difference between the infidelity and the exact optimal infidelity, analytically known in the single-qubit case (but not in the two-qubit case, cf.~Fig.~\ref{fig:app_LMC2_T}).
Interestingly, in each panel, the six curves lie on top of each other until thermalization occurs. A simple effective one-dimensional toy model (black-dashed line in panels (\textbf{\textit{a}}) and (\textbf{\textit{b}}), cf.~\eqref{eq:toymod-ode}) reproduces the typical scaling observed in the numerical data when $T$ is sufficiently different from the critical value $T_\text{QSL}{\simeq}2.51$. At $T{=}2.50$ (panel (\textbf{\textit{c}}); just before $T_\text{QSL}$), the ${\simeq}1/2$ slope differs from the ${=}1$ slope characteristic of the late-time scaling of the one-dimensional model in Eq.~\eqref{eq:toymod-scaling2}. On the contrary, at $T{=}2.52$ (panel (\textbf{\textit{d}}); just beyond $T_\text{QSL}$), LMC dynamics thermalizes much faster. 
}
\label{fig:app_LMC1_T}
\end{figure}

\begin{figure}
\centering
\includegraphics[width=.45\textwidth]{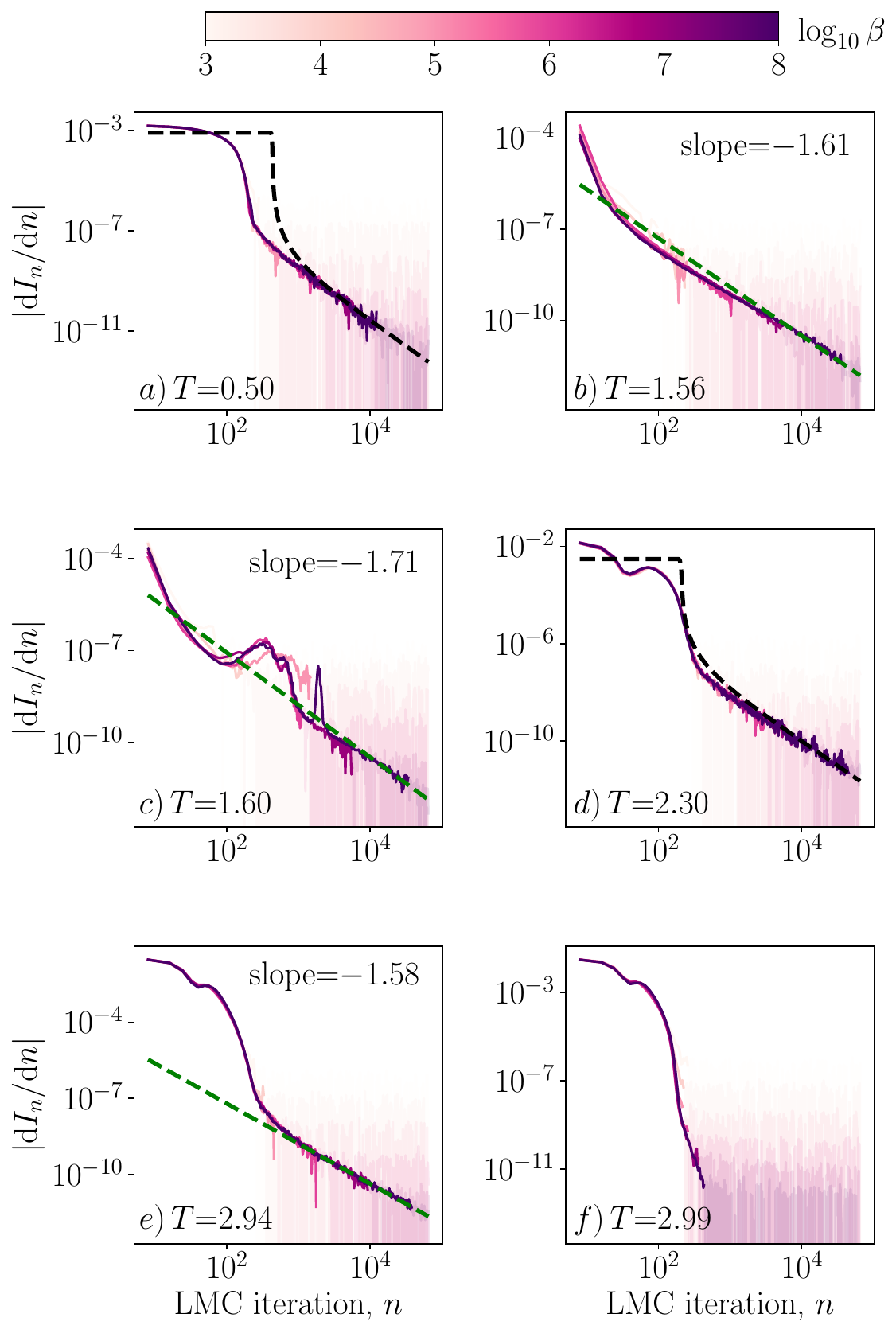}
\caption{
Two-qubit problem. The plots show the discrete \emph{derivative} of infidelity, $\dd I_n/\dd n$ during thermal relaxation, averaged over ten independent LMC runs. The parameter $T$, different in the four panels, is changed to investigate the effects of CLPTs (cf.~$T_\text{sb}{\simeq}1.57,T_\text{QSL}{\simeq}2.95$). Each curve refers to a different Monte Carlo inverse-temperature $\beta$ while $\sigma{=}10^{-3/2}$, $L{=}512$ are fixed.
As in the single-qubit system, a simple effective one-dimensional toy model (black-dashed line in panels (\textbf{\textit{a}}) and (\textbf{\textit{d}}); cf.~\eqref{eq:toymod-ode}) can approximate the typical scaling observed in data away from the critical values. The change in slope (from ${\simeq}1$ to ${\simeq}1/2$) observed just before $T_\text{QSL}$ in the single-qubit case (cf. Fig.~\ref{fig:app_LMC1_T}) is also visible here around $T_\text{sb}$ (panels (\textbf{\textit{b}}) and (\textbf{\textit{c}})) and just before $T_\text{QSL}$ (panel (\textbf{\textit{e}})).
The discrete numerical derivative is computed with the function \texttt{gradient} from Python's package NumPy. We decreased the opacity of each curve to 10\% after thermalization occurred.
}
\label{fig:app_LMC2_T}
\end{figure}

In Fig.~\ref{fig:app_LMC1_T} and \ref{fig:app_LMC2_T} we show how infidelity changes as the LMC dynamics evolves the protocol for $\beta=10^3,10^4,\dots,10^8,\sigma=10^{-3/2}$, in the single- and two-qubit problem, respectively. 
In the single qubit case, we plot the infidelity difference between the LMC and the theoretical optimal value computed from the exact solution of the problem \cite{Bukov18_Reinforcement}.
In the two-qubit case, the exact solution is not known and we plot the discrete \emph{derivative} of the infidelity as a function of the LMC iteration.
In both cases, we average over a set of ten independent LMC runs.
When the Monte Carlo inverse-temperature $\beta$ is large enough, the average motion in the thermal relaxation stage does not depend sensibly on it. In this case, the infidelity (derivative) curve displays in general two consecutive transient behaviors: the first one characterizes the dynamics at early iterations whereas the second one originates with a sudden transition (occurring around LMC iteration $n{=}10^2$ for $\sigma{=}10^{-3/2}$) and terminates once the system reaches the thermal equilibrium in the low-infidelity region. As visible from the plots, these two sub-stages possess distinct characteristic scaling behaviors independent of $\beta$. 

The characteristic scaling behaviors observed during the LMC thermal relaxation stage for $\beta {\gtrsim} 10^3$ are captured to some extent by an effective one-dimensional toy model. Consider a one-dimensional system $x_n{\in}\mathbb R$ evolving according to LMC dynamics in Eq.~\eqref{eq:LMC_update_update} in the landscape $I(x){=}\alpha \abs{x {-} x_\text{min}}$ from a random initial position $x_0{\ne}x_\text{min}$. 
The transition probability for a single iteration is then
\begin{equation}
    P(x_{n+1}|x_n) = \mathcal N_{x_{n+1}}(x_n,\sigma)p_\text{acc}(\Delta I_n).
\end{equation}
where $\mathcal N_x(x_n,\sigma)$ is a Gaussian distribution centered at $x_n$ with variance $\sigma^2$, $p_\text{acc}(\Delta I_n){=}\max(1,e^{-\beta \Delta I_n})$ is the Metropolis acceptance probability and $\Delta I_n = I(x_{n+1})-I(x_n)$ is the change in infidelity during the iteration. On average, the final state is then
\begin{equation*}
\mathbb E(x_{n+1}|x_n) = \int_{\mathbb R} \dd x_{n+1}\,x_{n+1}P(x_{n+1}|x_n).
\end{equation*}
For $\beta{\to}\infty$, thermal effects play no role since $p_\text{acc}(\Delta I_n){\to}\mathrm{sgn}(\Delta I_n)$.
In this case, the dynamics only depends on $\sigma$ and the relative distance $\ell_n {=} \abs{x_n {-} x_\text{min}}$ between $x_n$ and the minimum of the infidelity at $x_{\text{min}}$. Through a change of coordinate we fix $x_n{=}0$ and $x_\text{min}{>}0$. Then, in the limit $\beta{\to}\infty$ we obtain
\begin{align*}
    \mathbb E(x_{n+1}|x_n{=}0)
    &= \int_{\mathbb R} \dd x_{n+1}\,x_{n+1} \mathcal N_{x_n+1}(0,\sigma) \mathrm{sgn}(\Delta I_n) \\
    &= \int_{0}^{2\ell_n} \dd x_{n+1}\,x_{n+1} \mathcal N_{x_n+1}(0,\sigma) \\
    &= \sqrt{\frac{\sigma^2}{2\pi}} (1-e^{-(1/2)(2 \ell_n)^2/\sigma^2}).
\end{align*}
Thus, while averaging over different LMC realizations, we obtain the differential equation for $\ell_n$
\begin{align}
    \dv{\ell_n}{n} &\approx -\mathbb E(x_{n+1}|x_n{=}0),
    \label{eq:toymod-ode}
\end{align}
where we promoted $n$ to a continuous variable assuming sufficiently small changes in $\ell_n$ over single LMC iterations.
This equation admits simple closed-form solutions in the two limiting cases
\begin{align}
    \ell_n/\sigma &\gg 1: & \ell_n &\sim - \alpha_1 n 
    \label{eq:toymod-scaling1} \\
    \ell_n/\sigma &\ll 1: & \ell_n &\sim \alpha_2 n^{-1}
    \label{eq:toymod-scaling2}
\end{align}
where the two prefactors $\alpha_{1,2}$ depends on $\sigma$.
These two analytical limiting cases are related with the two characteristic scaling behaviors visible in Figs.~\ref{fig:app_LMC1_T} and \ref{fig:app_LMC2_T}. In the single-qubit system, the prediction in Eq.~\eqref{eq:toymod-ode} fits qualitatively well the whole numerical curves away from the $T_\text{QSL}{\simeq}2.51$ transition (black dashed curve in Fig.~\ref{fig:app_LMC1_T}a,b); in the two-qubit problem, the same is true away from the $T_\text{sb}{\simeq}1.57$ transition and again before the $T_\text{QSL}{\simeq}2.95$ transition (black dashed curve in Fig.~\ref{fig:app_LMC2_T}a,d). 
In this way, we gain intuition about the two characteristic scaling behaviors visible during LMC thermal relaxation stage: the first (second) transient behavior characterizes the motion when the protocol $\vb s_n$ is sufficiently far away from (close to) the low-infidelity region with respect to the average jump-size $\sigma$ of the Langevin update rule.

Interestingly, the scaling behavior changes across the CLPT at the quantum speed limit. 
For example, in the single-qubit system compare $T{=}2.50$ with $T{=}2.52$ (cf.~$T_\text{QSL}{\simeq}2.51$) or in the two-qubit system $T{=}2.94$ with $T{=}2.97$ (cf.~$T_\text{QSL}{\simeq}2.95$).
We observe that the early-time characteristic behavior remains qualitatively unaffected by the $T_\text{QSL}$ transition: we deduce that there are no significant changes around $T_\text{QSL}$ in the infidelity landscape in the high-infidelity region. On the contrary, the $T_\text{QSL}$ transition heavily affects the scaling behavior in the late-time thermal relaxation stage. Observe that the late-time characteristic scaling at the two sides of the $T_\text{QSL}$ transition has opposite behavior: it slows down when $T {\to} T_\text{QSL}^-$ and it becomes fast when $T {\to} T_\text{QSL}^+$. 
On a qualitative level, the different behavior at the two sides of the $T_\text{QSL}$ transition can be understood as follows. In the limit $T {\to} T_\text{QSL}^-$, the stochastic search for the unique optimal protocols is complicated by an increasing number of sub-optimal protocols having similar infidelity as the true optimal protocol. In the other case, for $T {\to} T_\text{QSL}^+$, there is an extensive number of optimal protocols and the stochastic dynamics easily converges to one of the many possible minima. 
More in detail, at $T{=}T_\text{QSL}^-$, the late-time characteristic scaling is $n^{-1/2}$ (notice the scaling $n^{-1}$ observed away from the $T_\text{QSL}$ transition). This different scaling behavior is phenomenologically explained by the one-dimensional effective model by assuming an effective potential $V(\abs{x}) {\propto} \abs{x}^{1/2}$ in the $\ell_n{\ll}\sigma$ case. 
On the other side, at $T{=}T_\text{QSL}^+$, the second transient is not visible at all: all curves converge approximately simultaneously to the low-infidelity region shortly after the end of the first transient. This other behavior is due to the presence of many optimal protocols and is effectively described in the one-dimensional model by an extended flat valley $V{=}0$ with a length greater than the $\sigma$.

The two-qubit problem presents additional features not observed in the single-qubit case. Around the $T_\text{sb}{\simeq}1.57$ transition, the early-time scaling behavior is different than what is observed for other values of $T$. In particular, for $T{=}1.56$ the scaling during the thermal relaxation phase is approximately algebraic (with a slightly different slope in the early- and late-time thermal relaxation substages). For $T{=}1.60$, the same scaling persists but an additional ``bump" in the infidelity curve is now separating the two sub-stages. 
Although a more detailed analysis is needed to better understand this other behavior, intuitively the ``bump" may be related to a saddle-point present in the landscape, slowing down the thermal relaxation dynamics.
Finally, the characteristic late-time scaling, $n^{1/2}$, observed in the single-qubit problem just before the $T_\text{QSL}$ transition, appears also in the two-qubit problem not only before its $T_\text{QSL}$ transition but also around the $T_\text{sb}$ transition.

Last, let us point out that in all our simulations in the single-qubit landscape, LMC always thermalizes around a global optimal protocol without getting trapped in any local minimum.
On the contrary, in the two-qubit problem for $L{=}16,32,64$ we sometimes found the LMC dynamics trapped in a local minimum of the landscape for $T{\in}[2.0,3.3]$, approximately. In this case, a Simulated Annealing protocol combined with a post-selection among a set of five thermal relaxation runs is enough to guarantee the convergence of LMC to the optimal level set in all of our simulations.

\subsection{Diffusion stage}
\label{app:LMC_D}

In order to develop some intuition about the diffusion stage, it is convenient to consider the theoretical limit $\sigma{\to}0$ at fixed $\beta{\gg}\sigma^2$. In this limit, the acceptance probability $P_\text{acc}$ approaches unity and the dynamics becomes effectively a free random walk in $L$ dimensions. In practice, we consider a small but finite $\sigma$ so that $P_\text{acc} \simeq 1$ only during the very early diffusion stage. As $\vb s_n$ moves, $P_\text{acc}$ suppresses the motion in the directions associated with increasing infidelity. As long as the dynamics involves a neighborhood of the starting point, the motion of $\vb s_n$ can be decomposed along the local flat directions (\emph{tangent} directions) and the local infidelity increasing directions (\emph{orthogonal} directions): the tangential motion is approximately free whereas the motion along orthogonal directions is suppressed by $P_\text{acc}$. Hence, LMC becomes effectively a random walk in $L$ dimensions restricted by the infidelity landscape to move along the valley's flat directions. Asymptotically, $\vb s_n$ diffuses through the valleys and explores a global portion of the optimal level set.

Using the random walk as a theoretical limiting behavior, we characterize the diffusive motion along the optimal level set by considering the covariance matrix associated with a protocol's fluctuations. More precisely, we sample protocols at different iterations $\mathcal D_M {=} \{\vb s_{n_1} \dots \vb s_{n_M}\}$ and compute the covariance-matrix
\begin{align}
    \sigma_{M,ij} = \expval{s(t_i) s'(t_j)}_M - \expval{s(t_i)}_M \expval{s'(t_i)}_M
    \label{eq:covmat}
\end{align}
where the average is performed over the set $\mathcal D_M$. As we discussed above, the diffusion stage is expected to have three different time scales, each associated with some properties in the covariance matrix \eqref{eq:covmat}.
\begin{enumerate}
    \item Microscopic time-scale. At early iterations, the dynamics can be approximated by a free random walk in $L$-dimensions. In this regime, the growth of covariance matrix eigenvalues $\lambda_i$ has a diffusive scaling behavior $\lambda_i \sim t$.
    \item Mesoscopic time-scale. When the motion involves a neighborhood of the starting point the dynamics can be decomposed into a free random walk along the tangent directions and a bounded fluctuating motion in the orthogonal directions. Upon diagonalization of the covariance matrix, we expect the tangent/orthogonal ($\parallel/\perp$) subspace decomposition $\sigma_n {\cong} D_{\parallel,n} \oplus D_{\perp}$ where $D_{\perp}$ is a diagonal matrix (approximately) independent from $n$ and $D_{\parallel,n}$ is a diagonal matrix whose eigenvalues exhibits diffusive scaling behavior.
    \item Macroscopic time-scale. Asymptotically, the dynamics involves the whole optimal level set. In this regime, covariance matrix eigenvectors specify the principle axis decomposition (of the Gaussian approximation) of the distribution of optimal protocols, while the eigenvalues specify the lengths of each semi-axis. 
\end{enumerate}

\begin{figure}
\centering
\includegraphics[width=.45\textwidth]{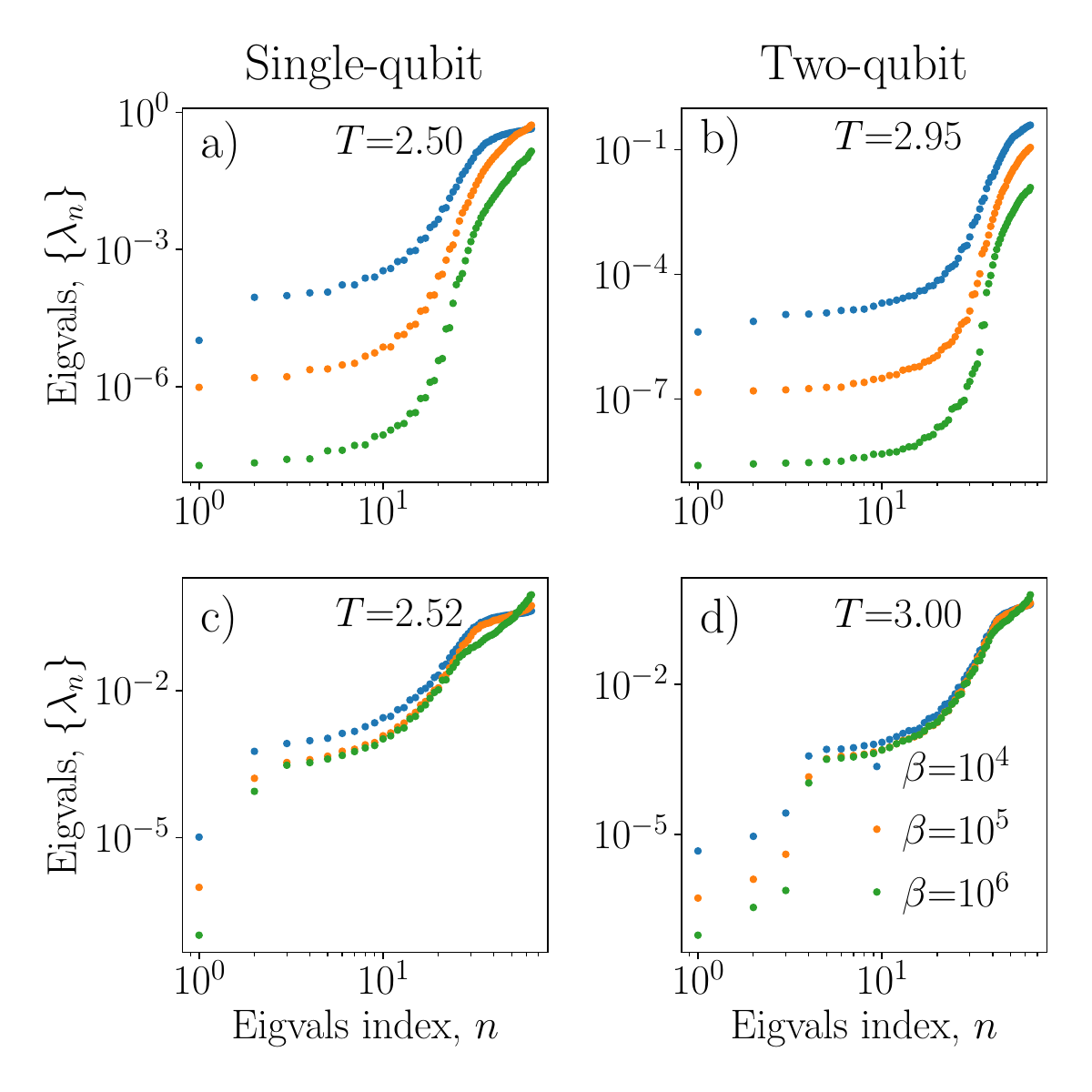}
\caption{
LMC diffusion stage, covariance-matrix asymptotic eigenvalues for a single LMC run. The eigenvalues capture fluctuations in the protocol dynamics and depend in general on $\beta$, the Monte Carlo effective inverse-temperature in Eq.~\eqref{eq:LMC_update_metropolis}. For $T{>}T_\text{QSL}$ (panels (\textbf{\textit{c}}) and (\textbf{\textit{d}})), protocol fluctuations along the optimal level set become independent from $\beta$ ($T_\text{QSL}{\simeq}2.51$, $2.98$ in single- and two-qubit problems, respectively). 
Notice the different number $n_+$ of $\beta$-dependent eigenvalues for $T{>}T_\text{QSL}$ in the two problems: the optimal level set expands in $L{-}2$ ($L{-}4$) directions of the $L$ dimensional protocol space in the single-qubit (two-qubit) problem.
}
\label{fig:app_LMC_D_eigvals}
\end{figure}

Global properties of the optimal level set can be extracted by letting LMC run on a macroscopic time scale. LMC works at finite $L$ and $\beta$ so we eventually need to study how results scale as $L,\beta {\to} \infty$. To this end, we run LMC at fixed $L$ and $\beta$ and let the covariance matrix eigenvalues stabilize to constant values. In Fig.~\ref{fig:app_LMC_D_eigvals}, the asymptotic eigenvalues of the covariance matrix are shown for $\beta{=}10^4,10^5,10^6$. The asymptotic independence of $(L{-}n_+){>}0$ eigenvalues from $\beta$ for $T{>}T_\text{QSL}$ gives a direct numerical confirmation of the presence of an optimal level set of continuously connected optimal protocols. 
In particular, we see that $n_+{=}2,4$ in the single- and two-qubit case, respectively. We notice that the eigenvalue $\lambda_2$ ($\lambda_4$) in the single- (two-) qubit case shows a weak dependence with $\beta$ (cf.~Figs.~\ref{fig:app_LMC_D_eigvals}c,d): this is compatible with the observed magnitude of the smallest positive eigenvalue of the Hessian, visible in Figs.~\ref{fig:stab_1q_Delta}b and \ref{fig:app_stab_2q_quad}b.
Notice that the presence of $n_+$ $\beta$-dependent eigenvalues means that along these directions (in the $L$-dimensional space) the protocol fluctuations still depend on $\beta$. Geometrically, this implies that the optimal level set lies in a $n_+$-dimensional hyperplane of the $L$ dimensional space.

From the LMC diffusive stage analysis, we estimate the number of LMC iterations $\Delta n_\text{dec}$ required during diffusion to explore a global portion of the optimal level set. In particular, we define $\Delta n_\text{dec}$ as the number of iterations required for half of covariance matrix eigenvalues to stabilize to a constant value. Hence, the set of protocols $\mathcal E {=} \{\vb s_{n_1} \dots \vb s_{n_M}\}$ sampled during the LMC equilibrium sampling stage are separated by $\Delta n_\text{dec}$ iterations (namely, $n_{i+1}{-}n_i {=} \Delta n_\text{dec}$). Based on our observations, we set $\Delta n_\text{dec} {=} 2^{12}$ in the single-qubit problem and $\Delta n_\text{dec} {=} 2^{14}$ in the two-qubit problem.

\subsection{Equilibrium sampling stage}
\label{app:LMC_E}

Eventually, in the LMC equilibrium sampling stage we collect $r$ sets of protocols $\mathcal{E}_a,\,a=1,2,\dots,r$ from $r$ independent LMC runs. 

\begin{figure}
\centering
\includegraphics[width=.5\textwidth]{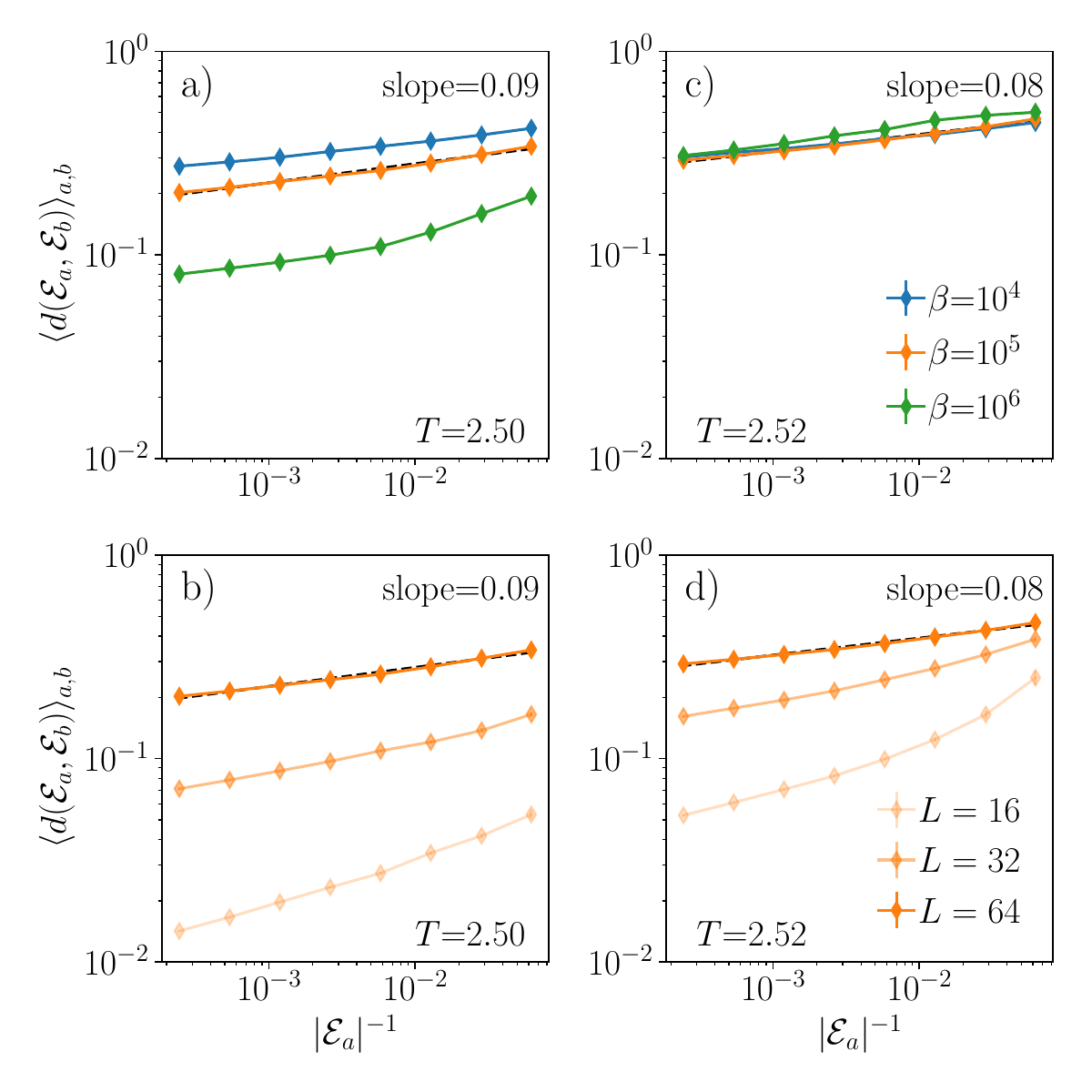}
\caption{
Single qubit problem, LMC equilibrium sampling stage. The average distance between pairs of LMC runs (out of a total of 20) is shown as a function of $\abs{\mathcal E_a}^{-1}=\abs{\mathcal E_b}^{-1}$, namely the inverse number of protocols contained in each run. The same slope at the two sides of the $T_\text{QSL}{\simeq}2.51$ transition (compare panels (\textbf{\textit{a}}) with (\textbf{\textit{c}}) and (\textbf{\textit{b}}) with (\textbf{\textit{d}})) suggests the presence of a single connected component in the optimal level set for $T{\simeq}T_\text{QSL}$.
}
\label{fig:app_LMC_E_d_scaling}
\end{figure}

LMC allows to estimate the number of connected components in the optimal level set. To this end, we consider the Euclidean distance between two protocols defined by $$d(s_1,s_2)=\sqrt{\frac1T \int_0^T \dd t\,(s_1(t)-s_2(t))^2}.$$ 
The distance $d(\mathcal{E}_a,\mathcal{E}_b)$ between sets $\mathcal{E}_a,\mathcal{E}_b$ is defined as the minimum distance found between any two of their protocols:
\begin{equation}
    d(\mathcal{E}_a,\mathcal{E}_b) = \min_{\vb s \in \mathcal{E}_a\, \vb s' \in \mathcal{E}_b}{d(\vb s, \vb s')}.
    \label{eq:LMC_run_distance_set}
\end{equation}
The presence of multiple disconnected components can be detected from the (asymptotic) distribution of distances between pairs of LMC runs. If only one connected component is present, we expect to observe a single-peaked distribution of distances with a peak location that shifts to zero as a larger number of protocols are considered within each set. Otherwise, when multiple disconnected components are present, the distribution of distances is expected to possess multiple peaks while its average value converges to a non-zero value as more protocols are considered. Here, we assume a single LMC run to sample from a single connected component via (approximately) homotopic transformations, for sufficiently large $\beta$ (cf.~Eq.~\eqref{eq:LMC_update_update}). 
For the single qubit system, we show how the mean value $\expval{d(\mathcal{E}_a,\mathcal{E}_b)}_{a,b}$ scales as the number of protocol samples is increased in Figs.~\ref{fig:app_LMC_E_d_scaling}. Across the $T_\text{QSL}$ transition, the minimum distance curves show a similar behavior. For $T{>}T_\text{QSL}$, distances are independent from $\beta$ and decrease with the same slope as in $T{<}T_\text{QSL}$, thus suggesting the presence of a single connected component in the optimal level set. The situation in the two-qubit problem is richer and it is discussed in Ref.~\cite{beato2025_topological}. In general, we expect different quantum control problems to have different numbers of disconnected components.

\begin{figure}
\centering
\includegraphics[width=.4\textwidth]{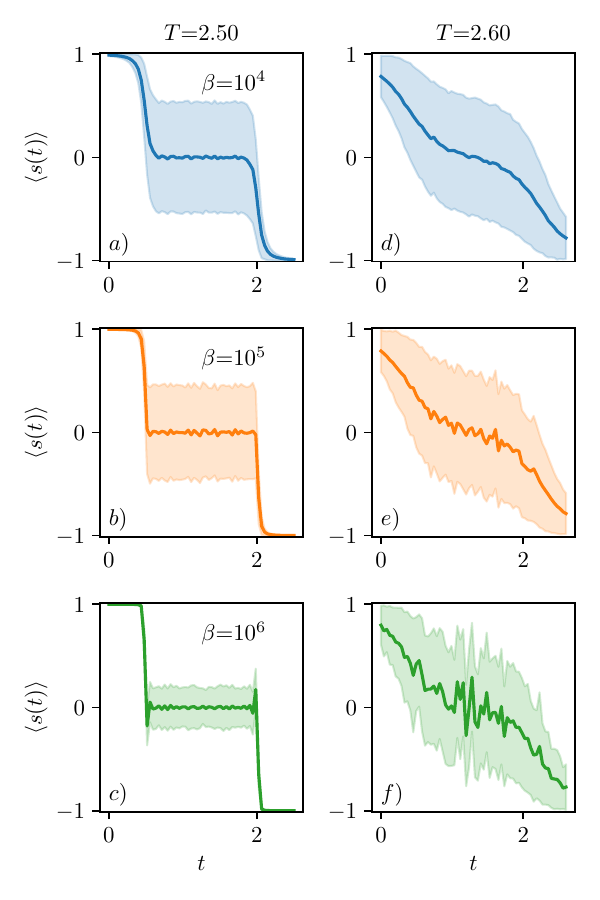} 
\caption{
Single qubit problem. Average protocol resulting from a single LMC run during the equilibrium sampling stage. The colored area shows fluctuations generated by the LMC dynamics in the optimal level set. Notice that the size of fluctuations depends on $\beta$ only for $T{<}T_\text{QSL}{\simeq2.51}$ (panels (\textbf{\textit{a}}),(\textbf{\textit{b}}) and (\textbf{\textit{c}})). It is interesting to compare this result with the average protocols in SD, in Fig.~\ref{fig:app_SD_optprot}.
}
\label{fig:app_LMC_E_OptProt}
\end{figure}

On a different note, in Fig.~\ref{fig:app_LMC_E_OptProt} we report the average protocols obtained from LMC equilibrium sampling at different Monte Carlo inverse-temperature $\beta$ in the single-qubit problem. The average shapes of the optimal protocols collected from LMC match the average optimal protocols obtained with SD (cf.~Fig.~\ref{fig:app_SD_optprot}) in the exact infidelity landscape. 
Notice the protocol fluctuations, represented by the colored area in the plot, affected by $\beta$ only for $T{=}2.50{<}T_\text{QSL}$.

\begin{figure}
\centering
\includegraphics[width=.48\textwidth]{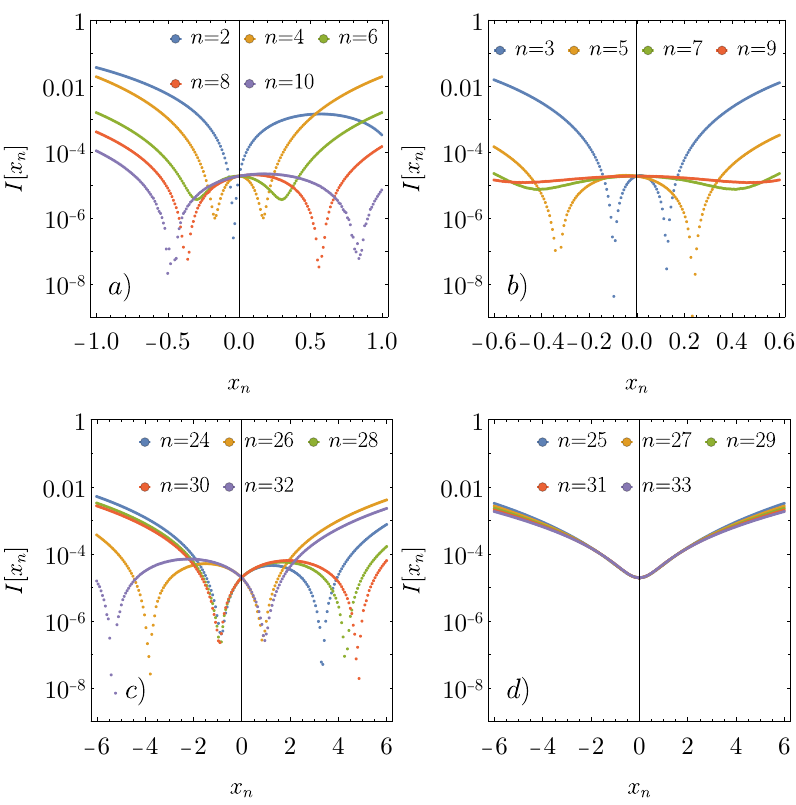}
\caption{
Single-qubit problem. Infidelity evaluated for different families of protocols $s_n(x_n),\, n{\in}\mathbb N$ in Eq.~\eqref{eq:optman_param_deform}, parametrized by the real variable $x_n$, at fixed $T{=}2.52>T_\text{QSL}$.
According to the infidelity expansion truncated at second-order, along each direction $n{>}2$ the infidelity should intersect the $I(T)[s]{=}0$ hyperplane at two values $x_n^+{>}0$ and $x_n^-{<}0$. From the plot we observe that this is not always the case in the exact infidelity landscape: for example, for $n{=}6,7,9$, $I(x_n^\pm){\gtrsim}I(0)$ (panel (\textbf{\textit{b}})) and for odd $n{\gg}1$, $x_n^\pm{=}0$ (panel (\textbf{\textit{d}})). 
We conclude that, in some cases, the infidelity expansion truncated at second-order may fail to correctly approximate (even at a qualitative level) the exact infidelity landscape; if so, higher-order terms are important for quantitative comparison.
}
\label{fig:app_optman_param_deform}
\end{figure}

\section{Optimal level set parametrization} 
\label{app:optman_param}

In this section, we test the analytical parametrization proposed in Sec.~\ref{sec:Tqslcritical} for the optimal level set after the quantum speed limit, $T_\text{QSL}$.
In particular, we consider the single-qubit problem defined in Sec.~\ref{sec:model}, which possesses a single optimal control protocol $s_{\Delta_0(T)}$ up to the quantum speed limit (cf.~Eq.~\eqref{eq:stab_s_Delta}).

We consider the parameterization of the optimal level set obtained from the infidelity expansion centered at $s_{\Delta_0(T)}$ and truncated at second-order. 
We explicitly evaluate the infidelity for a continuous family of protocol deformations of the form 
\begin{align}
    s_n(x_n) &= s_{\Delta_0(T)} + x_n f^{(n)}
    \label{eq:optman_param_deform}
\end{align}
where $n{\in}\mathbb N$ labels orthogonal eigenfunctions and $x_n{\in}\mathbb R$ controls the magnitude of the deformation; $f^{(n)}(t)$ are the eigenfunctions of the Hessian operator in the infidelity expansion centered in $s_{\Delta_0(T)}$ (cf.~Fig.~\ref{fig:app_quadratic_scaling}).

In the analytical parametrization of the optimal level set truncated at second-order, along each direction $n$ associated with a negative Hessian eigenvalue $\lambda_n{<}0$, the infidelity eventually crosses the zero-infidelity hyperplane. Since fourth-order terms are neglected, we use these intersection points to estimate the locations of global minima in the exact infidelity landscape.

We test the extent to which this approximation holds by computing how the exact infidelity behaves under the deformations in \eqref{eq:optman_param_deform}. At fixed $x_n$, we numerically solve the Schr\"odinger equation associated with the quantum control problem. In Fig.~\ref{fig:app_optman_param_deform}, we show the infidelity as a function of $x_n$ for small and large values of the integer $n$, for fixed $T{=}2.52$. At fixed $n$, the infidelity curve exhibits in general two minima $I(x_n^\pm)$, located in the two half-planes $x_n^-{<}0,x_n^+{>}0$. We notice that not every minimum is numerically indistinguishable from zero: for example, for $n{=}3,5$ the infidelity at the minima is ${\lesssim}10^{-8}$ whereas for $n{=}7,9$ it reaches a value ${\sim}10^{-5}$. In fact, for odd $n$, infidelity evaluated at the minima locations increases as $n$ grows. That is, for odd $n$'s deformations $s_n$ are not able to reach zero infidelity, except for $n{=}3,5$. On the other hand, for even $n$, infidelity evaluated at the minima locations remains small (${\lesssim}10^{-8}$) as $n$ grows. Thus, we observe an important limitation of the infidelity second-order approximation: some predicted optimal protocols in the approximated landscape (at second-order) have non-vanishing infidelity in the exact landscape. 

As we saw in App.~\ref{app:quadratic_scaling}, parity of $n$ is related to the parity of the $f^{(n)}$ eigenfunction: odd (even) $n$ corresponds to an even-parity (odd-parity) eigenfunction with respect to the transformation $f(t){\mapsto}{-}f(T{-}t)$. Hence, for small $n$, protocols obtained by deforming $s_{\Delta_0}(t)$ along even-parity $f^{(n)}$ are (within numerical precision) optimal and violate the symmetry of the quantum control problem (cf.~Sec.~\ref{sec:model}). Therefore, for $T{>}T_\text{QSL}$ the quantum control symmetry (respected for $T{<}T_\text{QSL}$) is broken in the optimal level set. 
Interestingly, deformations along eigenfunctions with even-parity (\textit{i.e.} violating the quantum control problem's symmetry) are exactly the ones strongly affected by higher-order corrections in the infidelity expansion.

Let us also comment on the infidelity curve along the $n{=}2$ direction, which is highly non-symmetric with respect to axis $x_2{=}0$ (see Fig.~\ref{fig:app_optman_param_deform}). To understand this, remember that $n{=}2$ is associated with one of the two \emph{positive} eigenvalues visible in the Hessian spectrum (the other being $n{=}1$, not shown in the four plots). Hence, the minimum visible in $x_2 {\simeq} {-}0.05$ is due to the linear-order term in the infidelity expansion.

\begin{figure}
\centering
\includegraphics[width=.45\textwidth]{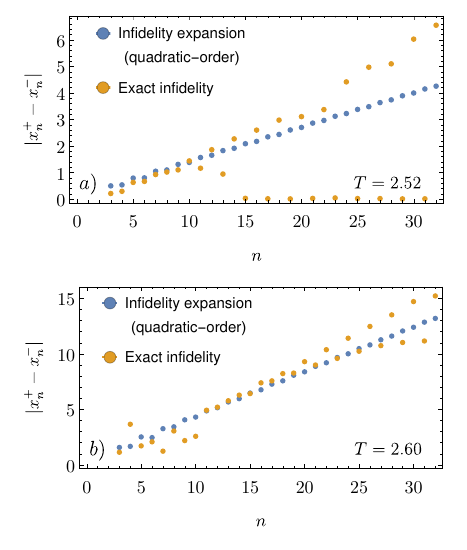}
\caption{
Single-qubit problem. Optimal level set parametrization shortly after the quantum speed limit, $T_\text{QSL}{\simeq}2.51$. 
The plot shows the separations $\abs{x_n^+{-}x_n^-}$ between the two minima $x_n^+,x_n^-$ in the infidelity curves $I[x_n]$ of Fig.~\ref{fig:app_optman_param_deform}, for different directions $n{>}2$.
Due to the Hessian eigenvalues, negative for $n{>}2$, minima are in general present pairwise. 
Here, we compare the separation estimated by the infidelity expansion (truncated at second-order; orange points) with the separation resulting from the exact landscape (obtained from numerical integration of the Schr\"odinger equation; blue points).
At $T{=}2.52$, we observe good \emph{quantitative} agreement between the exact and approximated infidelity for $n{\lesssim}12$; for $n{\gtrsim}12$, the agreement is good at a \emph{qualitative} level for even $n$ only.
Comparing $T{=}2.52,2.60$ (in panels (\textbf{\textit{a}}) and (\textbf{\textit{b}}), respectively), we observe that for large $n$ there is a better agreement between the exact infidelity and the second-order expansion results. We relate this behavior to the quantum control symmetry $s(t){\leftrightarrow}{-}s(T{-}t)$, constraining optimal protocols for $T{\le}T_\text{QSL}$.
}
\label{fig:app_optman_param_minima}
\end{figure}

As a second comparison between the second-order approximation and the exact infidelity, in Fig. \ref{fig:app_optman_param_minima} we compare the \emph{location} of minima $x_n^\pm$ in the exact infidelity landscape with the prediction obtained from the second-order expansion of the infidelity, for $T{=}2.52$ and $T{=}2.60$. For $n {\lesssim} 12$, there is a good quantitative agreement between the exact (blue points) and approximated locations (orange points). As $n$ increases, higher-order corrections in the infidelity expansion become important and the two sets of points deviate from each other.
In the exact landscape, the location of \emph{odd} $n$ minima eventually moves to zero, for $n{\gtrsim}15$: infidelity does not decrease along these directions and higher-order terms in the infidelity expansion are important for a qualitative comparison with the exact infidelity landscape. 
Conversely, the second-order expansion of the infidelity underestimates the location of \emph{even} $n$ minima; in this case, higher-order corrections mildly change the locations of the minima: the second-order expansion and the exact landscape produce the same qualitative result.
As a final remark, we notice that, as $T{\to
}T_\text{QSL}$, $\abs{x_n^+},\abs{x_n^-}{\to}0$ for all \emph{odd} $n$ (corresponding to even-parity eigenfunctions). This is consistent with the quantum control symmetry $f(t){\mapsto}{-}f(T{-}t)$ respected by the isolated optimal protocol for $T{\le}T_\text{QSL}$.
We deduce that the relatively large contribution of higher-order corrections to the optimal level set parametrization is a direct consequence of the symmetry of the quantum control problem we consider.

\section{Calculation of the order parameter \texorpdfstring{$q(T)$}{q(T)}}
\label{app:q}

In this appendix, we provide details of the calculation of the order parameter $q(T)$ discussed in Sec.~\ref{ssec:partfunc_q}.

The starting point is the partition function in Eq.~\eqref{eq:partfunc_approx}.
Here and in Apps.~\ref{app:steepest_descent},\ref{app:qBB}, for notational convenience, we redefine coefficients $b_n,\lambda_n,\dots$ appearing in the infidelity expansion in Eq.~\eqref{eq:infidelity-expansion-n} and absorb the prefactors $T,T^2,\dots$.
This operation does not affect the critical scaling of the coefficient as $T \to T_\text{QSL}$ for $\Delta T \to 0$.

\subsection{The generating functional \texorpdfstring{$G[k]$}{G[k]}}

Our goal is to evaluate the moments $\expval{s_n},\,\expval{s_n^2}$, appearing in the order parameter $q(T)$ in Eq.~\eqref{eq:q_nspace}.

More in general, the moments $\expval{s_n^\alpha},\alpha{\in}\mathbb N$ are conveniently obtained from the generating functional
\begin{equation*}
    G[k] = Z[e^{\int_0^T \dd t\,k_t s_t/T}]
\end{equation*}
via derivatives, $\expval{s_n^\alpha}=\partial_{k_n}^\alpha G[k] \eval_{k = 0}/G[0]$. 

Therefore, let us evaluate the following quantity
\begin{align}
    G[k] &= \int_{\mathbb R} \prod_n \dd s_n\int_{\mathbb R} \dd z
    e^{\kappa L F[s] + s \cdot k}
    \label{eq:partfunc-G-1} \\
    F[s] &= - s^2/2 + i z I[s].
    \notag
\end{align}
This expression follows from Eq.~\eqref{eq:partfunc_approx} with the change of variables $s_t = \sum_{n=1}^L s_n f_t^{(n)}$ and the Fourier representation of the Dirac delta distribution, 
\begin{equation*}
    \delta(I) = \frac{\kappa L}{2\pi} \int_{\mathbb R}\dd z\,\exp[\kappa L(i I z)].    
\end{equation*}
In addition, for later convenience, we use the parameter $\kappa\in(0,\infty)$ in place of the numerical prefactor $3$ (see App.~\ref{app:qBB}) and we introduce the shorthand notation 
\begin{align*}
    s_1 \cdot s_2 &= \frac1T \int_0^T\dd t\, s_1(t)s_2(t) = \sum_n s_{1n} s_{2n}, \\
    s^2 &= s \cdot s.
\end{align*}

\subsection{Integration over control protocol variables \texorpdfstring{$s_n$}{sn}}

As a first step for the explicit evaluation of $G[k]$, we perform the shift of control protocol variables, $s_n \mapsto s_{0n} + \delta s_n$, and let $\delta s_n$ be fixed by the stationarity condition
\begin{equation}
    \frac{\delta F[s_0+\delta s]}{\delta(\delta s_n)}\eval_{\delta s = \delta \bar s} = 0,
    \label{eq:F-spa}
\end{equation}
to be solved up to a given order in $\delta s$.
Notice that the solution of Eq.~\eqref {eq:F-spa} depends on the Lagrange multiplier $z$, $\delta \bar s{=}\delta \bar s(z)$.
This operation centers the functional $F[s]$ at one of its stationary points and leaves us with the functional $\bar F[\Delta] = F[s_0 {+} \delta \bar s {+} \Delta]$, where the field $\Delta$ describes fluctuations around the stationary point $s_0 {+} \delta \bar s {=} \bar s_0$.

We notice that, when $F[s]$ is truncated to second-order, the stationarity condition has a unique solution (geometrically, the vertex of a paraboloid). Truncating at a higher order may introduce multiple solutions that can be distinguished by the signs of eigenvalues of the associated Hessian operator \cite{dominy2014characterization, nicolaescu2007invitation}.
For the moment, we keep the discussion general and postpone the choice of the order of truncation. 
Eventually, we will see that a second-order truncation is sufficient to understand the critical behavior of the order parameter $q(T)$.
In contrast, in Sec.~\ref{ssec:partfunc_qBB}, we will show that quartic order terms are necessary to explain the critical behavior of the order parameter $q_\text{BB}(T)$.

After the shift $s_0 \mapsto s_0 + \delta \bar s$, we obtain (ignoring the overall prefactor)
\begin{align}
    G[k] &\approx
    \int_{\mathbb R} \dd z\, 
    e^{-\kappa L \bar s_0^2/2 + k \cdot \bar s_0}
    \int_{\mathbb R} \prod_n \dd \Delta_n\,
    e^{\kappa L \bar F[\Delta]}
    \notag\\
    \bar F[\Delta] &=
    iz \bar c 
    + k \cdot \Delta / (\kappa L)
    + \Delta \cdot \bar \Pi(z) \cdot \Delta/2
    + \dots
    \notag
\end{align}
where the second-order term
\begin{equation}
    \Delta \cdot \bar \Pi(z) \cdot \Delta = \sum_{n,m} \Delta_n (\delta_{nm} - iz \bar J_{nm}) \Delta_m,
    \label{eq:Pi}
\end{equation}
contains the new Hessian operator 
\begin{equation}
\bar J_{nm} = \delta_{nm}\lambda_n+\sum_k d_{nmk}\delta \bar s_k+\frac12\sum_{kl} g_{nmkl} \delta \bar s_k \delta \bar s_l + \dots\,,
\label{eq:hessian-renorm}
\end{equation}
that includes corrections from higher order terms $d_{nmk},g_{nmkl},\dots$ caused by the shift $\delta \bar s$ (analogously for the new constant term $\bar c$).

Next, we go to the diagonal basis of the Hessian operator, $\{\bar f_t^{(p)}\}$, with eigenvalues $\{\bar \lambda_p\}$. Neglecting $\order{\Delta^3}$ terms, we perform the Gaussian integrals over the variables $\{\Delta_p\}$ and obtain
\begin{align}
    G[k] &\approx \int_{\mathbb R} \dd z e^{\kappa L \Omega_z[k]} 
    \label{eq:partfunc-G-3} \\
    \Omega(z)[k] &=
   -\bar s_0^2/2 + iz \bar c + \bar s_0 \cdot k / (\kappa L)
    \notag\\
    &+ (1/2) k \cdot \bar \Pi^{-1} \cdot k / (\kappa L)^2
    \notag\\
    &- (1/2) \Tr \log \bar \Pi / (\kappa L).
    \label{eq:partfunc-Omega}
\end{align}
In the following, we ignore logarithmic corrections in $G[k]$ as they will not affect $q(T)$.

\subsection{Integration over Lagrange multiplier \texorpdfstring{$z$}{z}}

In the $L{\to}\infty$ limit, the integral in Eq.~\eqref{eq:partfunc-G-3} is dominated by the saddle-point $z_*$ of the exponent $\Omega(z)[k]$, satisfying 
\begin{equation}
    \pdv{\Omega(z)[k]}{z}\eval_{z=z_*} = 0.
    \label{eq:Omega-spa}
\end{equation}
For $z {\in} \mathbb R$, the exponent is a complex number so that valid saddle-points must satisfy additional conditions (cf.~App.~\ref{app:steepest_descent}).
Although there is no valid saddle point for $z {\in} \mathbb R$, following the steepest-descent method we deform the integration contour in the complex plane and intersect a complex-valued saddle-point, $z_* {\in} \mathbb C$. 
We notice that each positive (negative) eigenvalue $\bar \lambda_p$ corresponds to a branch point located on the negative (positive) imaginary semi-axis, $z = i\bar\lambda_p^{-1}$. The deformation of the integration contour is valid provided we do not cross any branch point during the process \cite{donaldson2011riemann}

\begin{figure}
\centering
\includegraphics[width=.5\textwidth]{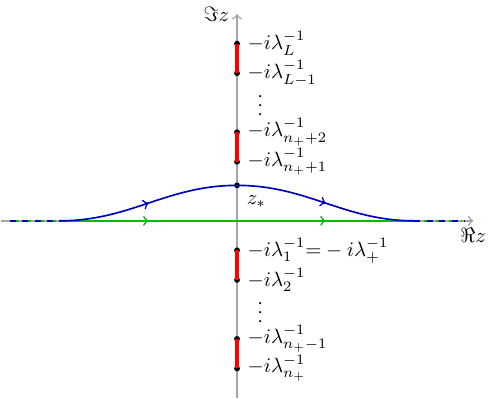}
\caption{
Schematic representation in the complex plane $z{\in}\mathbb C$ of the simple poles $\{-i\lambda_n^{-1}\}$ and the saddle-point $z_*$ of the complex function $\Omega(z)$ defined in Eq.~\eqref{eq:partfunc-Omega}. $\Omega(z)$ appears in the calculation of the generating functional $G[k]$ (cf.~Eq.~\eqref{eq:partfunc-G-3}).
The spectrum $\{\lambda_n\}$ of the second-order term in the infidelity expansion determines the poles' location. The number of positive eigenvalues $n_+$ depends on the particular controlled quantum system.
The green path represents the initial integration path along the real axis appearing in $G[k]$.
We perform a steepest-descent approximation by deforming the green into the blue path passing through the saddle-point $z_*$ of $\Omega(z)$ in the direction of steepest descent.
The red segments represent branch cuts created by the square root appearing in the integrand \eqref{eq:partfunc-Omega}.
}
\label{fig:steepest_descent}
\end{figure}

The saddle-point conditions in Eqs.~\eqref{eq:F-spa},\eqref{eq:Omega-spa} have to be solved self-consistently to obtain a valid pair of solutions $\delta \bar s, z_*$. In general, due to the nonlinearity of the infidelity $I[s]$, this is a difficult task. 
In App.~\ref{app:steepest_descent} we show that, within a second-order approximation in the infidelity expansion $I[\delta \bar s]$, a unique purely imaginary saddle point $z_* = iy_*,\,y_*{\in}\mathbb R$, always exists in the interval
\begin{align}
\Im(z_*) = y_* &\in (-\lambda_+^{-1}, -\lambda_-^{-1})
\label{eq:y}
\end{align}
for $T{\to}T_\text{QSL}^+$.
Here, we indicate with $\lambda_{+,-}$ the positive and negative eigenvalues among $\{\lambda_p\}$ with largest absolute value (cf.~Fig.~\ref{fig:steepest_descent}).

In conclusion, following the steepest-descent method, the cumulant-generating functional reads
\begin{align}
    \log G[k] &\approx \kappa L\, \Omega(z_*)[k]
    \label{eq:partfunc-logG}
\end{align}
with $\Omega(z)[k]$ defined in Eq.~\eqref{eq:partfunc-Omega}. 

From $\log G[k]$, we obtain $q_i(T)$ by deriving two times with respect to the source terms $k$. 
This operation is cumbersome, as the quantities $\bar s_0, \bar c, \bar \Pi_p$ depend on $\delta \bar s_0(y_*)$ and $y_*$ depends on $k$ through the saddle-point condition \eqref{eq:Omega-spa}, $y_*=y_*[k]$.
Nevertheless, as shown in the next section, we obtain an informative result by truncating the infidelity expansion in Eq.~\eqref{eq:infidelity-expansion-n} to second-order.

\subsection{The order parameter \texorpdfstring{$q(T)$}{q(T)}}
\label{subapp:q-calc}

In this section, we truncate the infidelity expansion in Eq.~\eqref{eq:infidelity-expansion-n} to second-order and show that the only non-vanishing contribution to $q_i(T)$ in the limit $L\to\infty$ comes from the second-to-last term of Eq.~\eqref{eq:partfunc-Omega}.
In other words, we demonstrate that the dependence $y_*=y_*[k]$ does not affect the behavior of $q(T)$, in the $L\to\infty$ limit.
Consequently, we obtain the result discussed in Sec.~\ref{ssec:partfunc_q},
\begin{align*}
    q_i(T) &\sim \frac{\Tr(\Pi^{-1})}{\kappa L} = \frac{1}{\kappa L}\sum_n \frac{1}{1+y_*\lambda_n}, &  L &\to \infty.
\end{align*}

The starting point is the cumulant generating function $\log G[k]$ in Eq.~\eqref{eq:partfunc-logG}.
For convenience, we introduce the compact partial derivative notation 
$$
\partial_n f[k] |_0 \equiv \partial_{k_n} f[k] \eval_{k=0}.
$$
We remember that $\delta \bar s,\, y_*$ are determined by the stationarity conditions in Eqs.~\eqref{eq:F-spa},\eqref{eq:Omega-spa}.

Evaluating $\partial_n^2 \Omega(y_*)[k]|_0$ we obtain the terms
\begin{align}
    \partial_n^2 \bar s_0(y_*)^2 |_0 &= 2[\bar s_0 \cdot \partial_n^2 \delta \bar s + (\partial_n \delta \bar s )^2 ]|_0
    \notag\\
    \partial_n^2 k \cdot \bar s_0 |_0 &= 2 \partial_n \delta \bar s_n |_0
    \label{eq:partfunc-k-derivatives}
    \\
    \partial_n^2 y_* \bar I |_0 &= [ (\partial_n^2 y_* ) \bar I + 2 \partial_n y_*  \partial_n \bar I  + y_* \partial_n^2 \bar I ]|_0
    \notag
\end{align}
where we used $\bar s_0(y_*) {=} s_0 {+} \delta \bar s(y_*)$ and omitted the dependence of $\bar s_0, \delta \bar s, \bar I$ on $y_*$ for notational convenience.
Let us explicitly write the terms appearing in Eq.~\eqref{eq:partfunc-k-derivatives}.
From the infidelity expansion \eqref{eq:infidelity-expansion-n}, we have
\begin{align*}
    \partial_n I|_0     &= 
    [
    b \cdot \partial_n \delta \bar s + 
    \delta \bar s \cdot \Lambda \cdot \partial_n \delta \bar s]|_0
    \\
    \partial_n^2 I|_0   &= [
    b \cdot \partial_n^2 \delta \bar s + 
    \partial_n \delta \bar s \cdot \Lambda \cdot \partial_n \delta \bar s +
    \delta \bar s \cdot \Lambda \cdot \partial_n^2 \delta \bar s
    ]|_0.
\end{align*}
and from the stationarity condition \eqref{eq:F-spa},
\begin{align*}
    \partial_n \delta \bar s_m |_0 &= 
    (\partial_{y_*} \delta \bar s_m) (\partial_n y_*) |_0
    \\
    \partial_n^2 \delta \bar s_m |_0 &= 
    (\partial_{y_*}^2 \delta \bar s_m)(\partial_n y_*) |_0^2 + (\partial_{y_*} \delta \bar s_m)(\partial_n^2 y_*) |_0
\end{align*}
with
\begin{align*}
    \partial_{y_*} \delta \bar s_m &= (s_{0m} - b_m/\lambda_m) \lambda_m \Pi_m^{-2} \\
    \partial_{y_*}^2 \delta \bar s_m &= -2(s_{0m} - b_m/\lambda_m) \lambda_m^2 \Pi_m^{-3} \\
    \Pi_m &= 1 + y_* \lambda_m.
\end{align*}
Finally, we obtain $\partial_n y_* |_0$ and $\partial_n^2 y_* |_0$ from the stationarity condition \eqref{eq:Omega-spa}, using the implicit function theorem:
\begin{align*}
    \partial_n y_* |_0 &= - \frac{F_n}{G} L^{-1}
    \\
    \partial_n^2 y_* |_0 &= - \frac{A (\partial_n y_*|_0)^2 + B_n (\partial_n y_*|_0) + C_n}{G} L^{-2}
\end{align*}
with
\begin{align*}
    F_n &= \kappa^{-1} \partial_{y_*} \delta \bar s_n |_0
    &
    A &= \partial_{y_*}^3 [ - \bar s_0^2 - y_*\bar I ] |_0
    \\
    G &= \partial_{y_*}^2 [ - \bar s_0^2 - y_*\bar I  ]|_0
    &
    B_n &= \kappa^{-1} \partial_{y_*}^2 \delta \bar s_n |_0
    \\
    & 
    & 
    C_n &= \kappa^{-2} \Pi_n^{-1}.
\end{align*}

From this expression, we deduce that terms in Eqs.~\eqref{eq:partfunc-k-derivatives} are of order $\order{L^{-2}}$. 
Therefore, their contribution to $q(T)$ is of the form $L^{-1} \sum_{n=1}^\infty a_n$, where the numerical sequence $\{a_n\}$   depends on the details of the landscape expansion. 
A sufficiently fast convergence to zero of the gradient components $\{b_n\}$ and the Hessian eigenvalues $\{\lambda_n\}$ guarantees that the sum $\sum_{n=1}^L a_n$ converges to a finite value, as $L\to\infty$, and therefore does not contribute to $q(T)$.

\section{Steepest-descent method and saddle-point of \texorpdfstring{$\Omega(z)$}{Omega(z)}}
\label{app:steepest_descent}

In this appendix, we discuss the existence of a saddle-point $z_*$ of the complex function $
\Omega(z)=\Omega(z)[k]\eval_{k=0}$ in Eq.~\eqref{eq:partfunc-Omega}, valid for the steepest-descent method applied to Eq.~\eqref{eq:partfunc-G-3}.

We show that, within a second-order truncation of the infidelity expansion, the integration domain $z\in\mathbb R$ can always be deformed to pass through a saddle-point $z_*$ lying on the imaginary axis, $\Re(z){=}0$.

First, we separate the real and imaginary parts $z=x{+}iy,\,x,y{\in}\mathbb R$.
Notice that $\Im \Omega(z)$ vanishes identically on the imaginary axis, $x{=}0$; the function $\Omega(z)$ restricted to the imaginary axis is the real function
\begin{align}
    \Omega(y) &=     
    - \frac12 \sum_n \bar s_{0n}^2 - y \bar c(y).
    \label{eq:Omega_y}
\end{align}
Truncating the infidelity expansion at second-order, we obtain from Eqs.~\eqref{eq:infidelity-expansion-n},\eqref{eq:F-spa}
\begin{align*}
    \bar c(y) &= c + \sum_n b_n \delta \bar s_n + \frac12 \sum_n \lambda_n \delta \bar s_n^2
    \\
    \bar s_{0n} &= s_{0n} + \delta \bar s_n
    \\
    \delta \bar s_n &= - \frac{s_{0n}+y b_n}{1+y \lambda_n}.
\end{align*}
Substituting in Eq.~\eqref{eq:Omega_y}, we obtain eventually
\begin{align*}
    \Omega(y) &= -\frac12 \sum_n s_{0n}^2 - yc + \frac12\sum_n \frac{(s_{0n} + y b_n)^2}{1+y \lambda_n}.
\end{align*}
From this expression, we deduce that $\Omega(y)$ possesses a minimum in
$$
y \in (-\lambda_+^{-1},-\lambda_-^{-1}) \equiv A,
$$
with $\lambda_{+,-}$ the positive and negative eigenvalues of the Hessian operator $J_{nm}$ with largest absolute value. 
To prove this, it is sufficient to observe that $\Omega(y)$ is continuous for $y\in A$ and
\begin{align*}
    \Omega(y) &\to +\infty & y\lambda_\pm &\to (-1)^+.
\end{align*}
Thus, there exists $y_*\in A$ such that $y_*$ is a stationary point (minimum) of the real function $\Omega(y)$.
Finally, as $\Omega(z)$ is holomorphic in a neighborhood of $z_*{=}iy_*$, $z_*$ is a also stationary point of the complex function $\Omega(z)$. 

To prove that $y_*$ is the unique saddle-point of $\Omega(z)$ restricted to $x{=}0,y\in A$, we observe that the real function $\Omega(y)$ is convex for $y\in A$, \textit{i.e.}, we observe $d_y^2\Omega(y){>0}$ for $y\in A$.

As a last step, we prove that $z_*$ is a \emph{valid} stationary point for the steepest-descent method \cite{murray2012asymptotic}. In particular, we prove that the integration path in Eq.~\eqref{eq:partfunc-G-3} may be deformed to cross the saddle-point $z{=}iy_*$ in the tangent direction $z(x)$, for which (i) the imaginary part is constant and (ii) the second derivative of the real part is negative.
The above two properties follow from the holomorphicity of $\Omega(z)$ in $z_*$. In particular, holomorphicity implies that $\Im \Omega(z)$ is constant around the point $y_*$ along the tangent direction $z(x){=}x{+}iy_*$ and that $d_x^2\Omega(z_*) {<} 0$ \cite{murray2012asymptotic}.
This completes the proof that $z_*{=}iy_*$ is a valid point for the steepest-descent method applied to Eq.~\eqref{eq:partfunc-G-3}.

\section{Computing \texorpdfstring{$q_\text{BB}(T)$}{qBB(T)}}
\label{app:qBB}

In this appendix we compute the critical scaling of $q_\text{BB}(T)$, as $\Delta T = T-T_\text{QSL}{\to}0^+$, predicted by the partition function in Eq.~\eqref{eq:partfunc-exact-BB}.

Following the same steps as in App.~\ref{app:q}, we arrive at the generating functional
\begin{align}
    G_\text{BB}[k] &= \sum_{i} G_\text{BB}[s_0^{(i)};k]
    \notag\\
    G_\text{BB}[s_0;k] &\approx \int_{\mathbb R} \prod_{n=1}^L \dd s_n\, \int_{\mathbb R} \dd z\, e^{\alpha N F_\text{BB}[s_0;s]} 
    \label{eq:partfunc_G_BB_0} \\
    F_\text{BB}[s_0;\delta s] &=
    - s^2/2 
    + i z I[s_{0n} + \delta s] + k \cdot s /(\alpha N)
    \notag
\end{align}
where $N$ is the number of bang-bang steps of the ``microscopic" protocol and $L$ is the number of steps of the ``mesoscopic" protocols, obtained through the coarse-graining average. 

Contrary to Sec.~\ref{ssec:partfunc_q}, here we do not separate contributions from different locally optimal control protocols $\{s_0^{(i)}\}$.
This choice is motivated by our interest in comparing with numerical results of Sec.~\ref{fig:phase_diags}, where the computation of the average $\expval{\cdot}$ does not separate contributions from disconnected cluster of solutions.
 
We notice that $G_\text{BB}[s_0;k]$ has the same structure as $G[k]$ in Eq.~\eqref{eq:partfunc-G-1} with the identification $\kappa {=} \alpha (N/L)$. 
Hence, within a second-order approximation, the Shannon entropy originating from the coarse-graining average in Eq.~\eqref{eq:shannon} has the same effect as the hard-boundary constraint, $\abs{s_t}\le1$.

From Eqs.~\eqref{eq:q_BB} and \eqref{eq:partfunc-G-3}, we have
\begin{equation}
    q_\text{BB}(T) = 1 - \sum_n (\partial_n \log G[0])^2
    \label{eq:qBB_xn_1}
\end{equation}
and
\begin{align*}
    G[k] &\approx \sum_{i} \exp (\kappa L \Omega(y_*)[s_0^{(i)}, k]).
    \notag
\end{align*}
Here, we highlighted the dependence $\Omega(y_*)=\Omega(y_*)[s_0^{(i)}, k]$ on the center of expansion $s_0^{(i)}$.

We observe that, in the $\kappa L\to\infty$ limit, the dominant contributions in
\begin{align*}
    \partial_n \log G[0] &= 
    \frac{
        \sum_{i} \kappa L \partial_n \Omega[s_0^{(i)},0] \exp(\kappa L \Omega[s_0^{(i)},0])
    }{
        \sum_{i} \exp(\kappa L \Omega[s_0^{(i)},0])
    }
\end{align*}
arise from protocols in $\{s_0^{(i)}\}$ minimizing the quantity $\Omega[s_0^{(i)},0]$. 
Therefore, denoting this subset with $\{s_*^{(j)}\}$, we are left with the evaluation of the quantity
\begin{align*}
    \partial_n \log G[0] &= 
        \frac{1}{N_*}\sum_{j=1}^{N_*} \partial_n \Omega[s_*^{(j)},0].
\end{align*}

For the evaluation of $\partial_n \Omega[s_*^{(j)},0]$, it is necessary to consider the dependence of the quantities $\bar s_*, \bar I, y_*$ on the source terms $k$ (through the stationarity conditions in Eqs.~\eqref{eq:F-spa},\eqref{eq:Omega-spa}).
However, using the results from Sec.~\ref{subapp:q-calc}, we verified that this dependence does not affect the critical scaling of $q_\text{BB}(T)$, within a quadratic order truncation of the infidelity. 
Therefore, in the following discussion, we will neglect this dependence.

In this case, the order parameter $q_\text{BB}(T)$ solely depends on the terms $k \cdot \bar s_*^{(j)}/\kappa L$ with $\bar s_*^{(j)} {=} s_*^{(j)} {+} \delta \bar s_*^{(j)}$.
Therefore, we obtain
\begin{align*}
    \partial_n \log G[0] &\approx 
        \frac{1}{N_*}\sum_j \bar s_{*p}^{(j)} = \frac{1}{N_*}\sum_j (s_{*p}^{(j)} + \delta \bar s_{*p}^{(j)}),
\end{align*}
with $\delta \bar s_*$ determined by Eq.~\eqref{eq:F-spa}, and
\begin{align}
    \Delta q_\text{BB}(T) &= q_\text{BB}(T) - q_{\text{BB},*}(T)
    \notag \\
    &\approx - \frac{1}{N_*^2} \sum_{ j,j' } 
    \qty( s_*^{(j)} \cdot \delta \bar s_*^{(j')} + \delta \bar s_*^{(j)} \cdot \delta \bar s_*^{(j')}) 
    \label{eq:DeltaqBB}
    \\
    q_{\text{BB},*}(T) &= 1 - \biggl( \frac{1}{N_*} \sum_{j} 
    s_*^{(j)} \biggr)^2. \notag
\end{align}
Here, $q_{\text{BB},*}(T)$ is the contribution to $q_\text{BB}(T)$ given by the locally optimal protocols $\{s_*^{(j)}\}$, found for $T\le T_\text{QSL}$. 

The non-analytic behavior of $q_\text{BB}(T)$ at the quantum speed limit $T_\text{QSL}$ is contained in the deviation $\Delta q_\text{BB}(T) = q_\text{BB}(T) - q_{\text{BB},*}(T)$.
In particular, we extract its critical behavior for $\Delta T\to0$ from the protocol deviation $\delta \bar s_*$.

Truncating the infidelity expansion at second-order, we obtain,
\begin{align*}
    \delta \bar s_{*n} 
    &= -\frac{s_{*n} + y_* b_n}{1+y_*\lambda_n} 
    \notag\\ 
    &\sim 
    \begin{cases}
        b_{n} & n\le n_+ \\
        s_{*n} & n> n_+
    \end{cases}
    \sim
    \begin{cases}
        \Delta T & n\le n_+ \\
        1 & n> n_+
    \end{cases}    
\end{align*}
where we used Eqs.~\eqref{eq:scalings-lambda},\eqref{eq:scalings-rest}.
Therefore, within a second-order approximation of the infidelity, we obtain the result in Eq.~\eqref{eq:qBB-partfunc-quadratic}, for which $\Delta q_\text{BB}(T)$ is discontinuous at the quantum speed limit. 

We observe that the jump-discontinuity of $\Delta q_\text{BB}(T)$ is an artifact of the second-order truncation of the infidelity expansion.
Intuitively, for $\Delta T \to 0^+$ the ``entropic potential'', $\int_0^T\dd t\, s(t)^2$, dominates over the infidelity constraint, since the second-order term of the infidelity expansion, $\int_0^T \dd^2t\, J_{t_1t_2} s_{t_1} s_{t_2}$, possesses a number $n_+$ of vanishing eigenvalues, $\{\lambda_n\}_{n>n_+}$. 
Therefore, considering terms up to fourth order in the infidelity expansion is crucial for determining the critical behavior of $q_\text{BB}(T)$.

\begin{figure}
    \centering
    \includegraphics[width=0.999\linewidth]{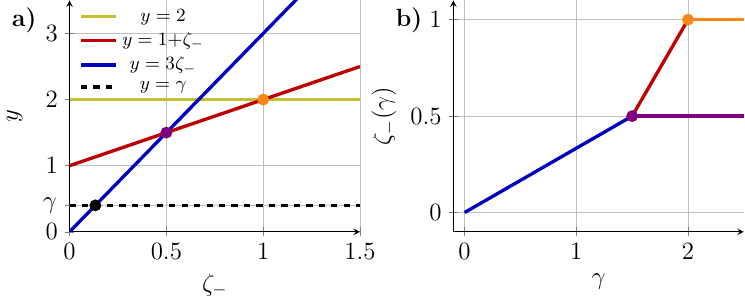}
    \caption{
        Geometrical representation of the dimensional analysis of Eq.~\eqref{eq:F-spa-fourth}, used to extract the critical behavior of the order parameter $q_\text{BB}(T)$, for $T\to T_\text{QSL}^+$. 
        (\textbf{\textit{a}}) Plot of the dimensions of relevant terms in Eq.~\eqref{eq:dimensional-analysis}, as a function of $\zeta_-$ and $\gamma$. 
        Depending on the value of the parameter $\gamma$, the intersections between the different curves change. 
        For each $\gamma$, $\zeta_-$ is fixed such that (i) at least two curves intersect at $\zeta_-$ and (ii) there are no other curves below the intersection point at $\zeta_-$.
        (\textbf{\textit{b}}) Plot of $\zeta_-=\zeta_-(\gamma)$ determined from panel (\textbf{\textit{a}}) by the conditions (i) and (ii).
    }
    \label{fig:dimensional-analysis}
\end{figure}

Truncating the infidelity expansion at fourth order, we obtain,
\begin{align}
    0
    &= -s_{*n} - \delta \bar s_{*n}
    \notag\\
    &+ y_* 
    \Biggl[
    b_n + \lambda_n \delta \bar s_{*n} + \frac{1}{2} \sum_{m,k} d_{nmk} \delta \bar s_{*m}\delta \bar s_{*k} 
    \notag\\
    &+ \frac{1}{3!} \sum_{m,k,l} g_{nmkl} \delta \bar s_{*m}\delta \bar s_{*k}\delta \bar s_{*l}
    \Biggr]
    \label{eq:F-spa-fourth}
\end{align}
From this equation, we extract the critical scaling in $\Delta T$ of $\delta \bar s$ by dimensional analysis.
In particular, we use the scaling behavior in Eqs.~\eqref{eq:scalings-lambda},\eqref{eq:scalings-rest}, assume $y_* \sim \Delta T^{-\gamma}$ with $\gamma>0$ unknown and let
\begin{equation*}
    \delta \bar s_{*n} \sim 
    \begin{cases}
    \Delta T^{\zeta_+} & n\le n_+ \\
    \Delta T^{\zeta_-} & n> n_+
    \end{cases},
\end{equation*}
with $\zeta_{+,-}$ to be determined consistently with Eq.~\eqref{eq:F-spa-fourth}.

As a working hypothesis, let us first assume $\zeta_-\le\zeta_+$. In this case, we start the analysis of Eq.~\eqref{eq:F-spa-fourth} for $n>n_+$, where we obtain the dimensional equality
\begin{align}
    0 &\sim \Delta T^\gamma + \Delta T^{\gamma+\zeta_-}
    \notag
    \\
    &+ \Delta T^2 + \Delta T^{1+\zeta_-} + \Delta T^{1+2\zeta_-} + \Delta T^{3 \zeta_-}.
    \label{eq:dimensional-analysis}
\end{align}
This dimensional equality fixes $\zeta_-$ as a function of $\gamma$.
Namely, for each $\gamma$, $\zeta_-$ is fixed such that: (i) there are at least two terms in the right hand side with the same dimension (\textit{i.e.}, have the same exponent); (ii) the other terms are subdominant (\textit{i.e.}, have a larger exponent). These conditions are necessary for the leading terms of Eq.~\eqref{eq:F-spa-fourth} to cancel.
Imposing these conditions yields (cf.~Fig.~\ref{fig:dimensional-analysis})
\begin{align}
    \zeta_-(\gamma) =
    \begin{cases}
        \gamma/3 & 0\le\gamma\le3/2 \\
        1/2,\gamma-1 & 3/2\le\gamma\le2 \\
        1/2,1 & 2\le\gamma
    \end{cases}.
    \label{eq:betam-gamma}
\end{align}
Now, considering Eq.~\eqref{eq:F-spa-fourth} for $n<n_+$ yields
\begin{align}
    0 &\sim \Delta T^\gamma + \Delta T^{\gamma+\zeta_+} \\
    +& \Delta T^1 + \Delta T^{\zeta_+} + \Delta T^{1+2\zeta_-} + \Delta T^{3 \zeta_-}
    \notag.
\end{align}
We substitute $\zeta_-(\gamma)$ from Eq.~\eqref{eq:betam-gamma} and obtain
\begin{align}
    \zeta_+(\gamma) =
    \begin{cases}
        \gamma & 0\le\gamma\le1 \\
        1 & 1\le\gamma
    \end{cases}.
    \label{eq:betap-gamma}
\end{align}
These results for $\zeta_-(\gamma),\zeta_+(\gamma)$ are consistent with the initial assumption $\zeta_-\le\zeta_+$. 
On the contrary, we notice that repeating the same procedure with the alternative hypothesis $\zeta_->\zeta_+$, does not yield consistent results. 

The allowed analytical range for $\zeta_-$ can be further narrowed, as we now show. In particular, we observe that $\zeta_->1/2$ is inconsistent with $\gamma>1$.
To see this, we need to remember the constraint $y_* < {-}\bar\lambda_-^{-1}$, with $\{\bar\lambda_n\}$ eigenvalues of the Hessian operator $\bar J_{nm}$ defined in Eq.~\eqref{eq:hessian-renorm} and $\bar\lambda_-$ the negative eigenvalue with largest absolute value. 
Hence, substituting $\delta \bar s_{n} \sim \Delta T^{\zeta_-}$ in Eq.~\eqref{eq:hessian-renorm}, we can observe that corrections from third and fourth order terms, 
\begin{align*}
d_{nmk}\delta \bar s_{k} &\sim \Delta T^{1+\zeta_-} &
g_{nmkl}\delta \bar s_{k}\delta \bar s_{l} &\sim \Delta T^{2\zeta_-},
\end{align*}
are subdominant, when $\zeta_->1/2$, with respect to the lowest order contribution, $\lambda_n$ (cf.~Eq.~\eqref{eq:scalings-lambda}).
As a consequence, when $\zeta_->1/2$, eigenvalues $\{\bar\lambda_n\}$ of the Hessian operator $\bar J_{nm}$ necessarily possess the same scaling as $\{\lambda_n\}$ (\textit{i.e.}, $\lambda_n \sim \Delta T$) and we have $\gamma\le1$.
In the other case, $\zeta_-\le1/2$, fourth-order corrections can change the qualitative behavior of negative eigenvalues, thus allowing for $\gamma\ge1$.

Finally, substituting the scaling behavior of $\delta \bar s_{*n}$ in Eq.~\eqref{eq:DeltaqBB}, we obtain the result in Eq.~\eqref{eq:qBB-partfunc-quartic}.

In conclusion, we note that the key difference between the second- and fourth-order results is that there is now a continuous set of potential solutions, $\zeta_-\in[0,1/2]$. 
Here, the precise value of $\zeta_-$ (and $\gamma$) is fixed by solving the stationary conditions in Eqs.~\eqref{eq:F-spa},\eqref{eq:Omega-spa}.
On the contrary, within a second-order truncation, we obtain the unique solution $\zeta_-=0$ (and $\gamma=1$).

\section{Scaling collapse of \texorpdfstring{$q_\text{BB}(T)$}{qBB(T)} at the quantum speed limit}
\label{app:qBB-scalcoll}

In this appendix, we discuss details regarding the scaling collapse shown in Fig.~\ref{fig:q_BB_critical}.
The scaling collapse is performed on the curves $q_\text{BB}(T)$ shown in Fig.~\ref{fig:phase_diags}, while varying the number of bang-bang steps $N$ in the Stochastic Descent simulations.
We refer to Refs.~\cite{melchert2009,sorge2015} for details regarding the curve-fitting procedure.

An important preliminary step, necessary for the scaling collapse, is the estimation of the numerical behavior of $q_{\text{BB},N}(T)$ in the $N\to\infty$ limit and for $T\le T_\text{QSL}$.
The procedure we adopt to estimate the limiting curve $\lim_{N\to\infty}q_{\text{BB},N}(T\le T_\text{QSL})$ consists of two steps. 
First, for a fixed $T_0<T_\text{QSL}$, we interpolate $q_{\text{BB},N}(T_0)$ with the curve $f(N) = AN^{-B}+C$ ($A,B,C$: fitting parameters). In the limit $N\to\infty$, we estimate the limiting value $q_{\text{BB},\infty}(T_0)$ from $C$.
Second, we repeat the first step for different values of $T<T_\text{QSL}$ and obtain the points $q_{\text{BB},\infty}(T)$. Then, 
we interpolate the points $q_{\text{BB},\infty}(T)$ with a linear function in $T$, $q_{\text{BB},0}(T)=q+mT$ ($q,m$: fitting parameters).
Eventually, we perform the scaling collapse on the shifted data $\Delta q_{\text{BB},N}(T) = q_{\text{BB},N}(T) - q_{\text{BB},0}(T)$ and estimate the critical exponent $\Delta q_\text{BB}(\Delta T) \sim \Delta T^{\zeta_n}$.

The estimate presented in Sec.~\ref{ssec:partfunc_qBB}, $\zeta_n\approx0.7$,  has an uncertainty of order $\sim 0.1$. 
Here, the main source of error originates in the estimation of $q_{\text{BB},0}(T)$, and, more specifically, in the particular choice of the interval of $T$ used to estimate $q_{\text{BB},0}(T)$.
To produce Fig.~\ref{fig:q_BB_critical}, we extrapolated $q_{\text{BB},0}(T)$ using $T\in[2.3,2.45],[2.8,2.95]$ in the single- and two-qubit problem, respectively.
Different choice of these intervals leads to slightly different values of the critical exponent within the range $[0.6,0.8]$, in both problems.

\bibliography{bibliography}

\end{document}